\documentclass[onecolumn,aps,prc,amsfonts,preprintnumbers,superscriptaddress,nofootinbib]{revtex4-1}
\pdfoutput=1
\usepackage{graphics}
\usepackage{amsmath}
\usepackage{amssymb}
\usepackage{psfrag}
\usepackage{epsfig}
\usepackage{epsf}
\usepackage{float}
\usepackage{slashed}
\usepackage{hyperref}
\usepackage{cleveref}
\allowdisplaybreaks
\usepackage{physics}
\usepackage{xcolor}
\usepackage{graphicx}
\usepackage{tabularx} 
\usepackage{siunitx}
\usepackage{array,mathtools,booktabs}

\newcommand{\beq}{\begin{equation}}
	\newcommand{\eeq}{\end{equation}}
\newcommand{\beqa}{\begin{eqnarray}}
	\newcommand{\eeqa}{\end{eqnarray}}
\newcommand{\nn}{\nonumber \\ }

\newcommand{\Mp}{M_\pi}

\renewcommand\theequation{\arabic{equation}}

\def\Xint#1{\mathchoice
	{\XXint\displaystyle\textstyle{#1}}%
	{\XXint\textstyle\scriptstyle{#1}}%
	{\XXint\scriptstyle\scriptscriptstyle{#1}}%
	{\XXint\scriptscriptstyle\scriptscriptstyle{#1}}%
	\!\int}
\def\XXint#1#2#3{{\setbox0=\hbox{$#1{#2#3}{\int}$}
		\vcenter{\hbox{$#2#3$}}\kern-.5\wd0}}

\def\dashint{\Xint-}
\begin{document}

\title{Chiral $3\pi$-exchange potential using the method of unitary transformation}
	
\author{Victor Springer}
\email[]{victor.springer@rub.de}
\affiliation{Ruhr-Universit\"at Bochum, Fakult\"at f\"ur Physik und Astronomie,
Institut f\"ur Theoretische Physik II, 
D-44780 Bochum, Germany}
	
\author{Hermann Krebs}
\email[]{hermann.krebs@rub.de}
\affiliation{Ruhr-Universit\"at Bochum, Fakult\"at f\"ur Physik und Astronomie,
Institut f\"ur Theoretische Physik II, 
D-44780 Bochum, Germany}
	
\author{Evgeny Epelbaum}
\email[]{evgeny.epelbaum@rub.de}
\affiliation{Ruhr-Universit\"at Bochum, Fakult\"at f\"ur Physik und Astronomie,
Institut f\"ur Theoretische Physik II, 
D-44780 Bochum, Germany}
\date{\today}
	
\begin{abstract}
  Nuclear potentials are known to exhibit a considerable degree of scheme dependence. For one- and two-pion exchange nucleon-nucleon (NN) potentials, unitary ambiguities start showing up at the level of the leading relativistic corrections to the dominant static contributions. However, for the three-pion exchange potential, scheme-dependent contributions are expected to appear already at the static level. 
Here, we analyze the leading and subleading chiral $3\pi$-exchange NN potentials using the method of unitary transformation. In line with the expectations, our results for selected classes of contributions differ from those obtained by Kaiser using S-matrix matching. We present analytical expressions for the $3\pi$-exchange potential, which are off-shell consistent with the interactions used by the Bochum group, and discuss the numerical importance of the observed differences.
\end{abstract}
	
\maketitle
	
\section{Introduction}
\def\theequation{\arabic{section}.\arabic{equation}}

Chiral effective field theory (EFT) has established itself as an efficient approach to derive nuclear interactions and current operators and to study low-energy nuclear structure and dynamics, see Refs.~\cite{Epelbaum:2008ga,Machleidt:2011zz} for review articles. Initiated by Weinberg in the early 1990s as a method to explore the implications of the spontaneously broken approximate chiral symmetry of QCD for nuclear physics \cite{Weinberg:1990rz, Weinberg:1991um}, this approach has proven phenomenologically successful and capable of providing precision results \cite{Epelbaum:2019kcf}. 
Two-nucleon (NN) potentials have been worked out completely to fifth chiral order $Q^5$ (N$^4$LO) in the framework of heavy-baryon chiral perturbation theory (ChPT) with pions and nucleons as the only active degrees of freedom (using dimensional regularization or equivalent methods to deal with the loop integrals)  \cite{Ordonez:1995rz,Kaiser:1997mw,Kaiser:1998wa,Epelbaum:1998ka,Epelbaum:1999dj, Kaiser:1999ff,Kaiser:1999jg,Kaiser:2001dm,Kaiser:2001pc,Kaiser:2001at,Entem:2014msa}. Here, $Q \in \{|\vec p \, |/\Lambda_{\rm b}, \, M_\pi/\Lambda_\chi \}$ denotes the expansion parameter with $|\vec p\, | \sim M_\pi$ referring to external three-momenta of the nucleons in the center-of-mass frame and $\Lambda_{\rm b}$ being the breakdown scale of chiral EFT. Three- and four-nucleon forces have been worked out completely up through the fourth chiral order $Q^4$ (N$^3$LO) in Ref.~\cite{vanKolck:1994yi,Epelbaum:2002vt,Epelbaum:2005bjv,Epelbaum:2007us,Ishikawa:2007zz,Bernard:2007sp,Bernard:2011zr}, and most of the N$^4$LO contributions to the three-nucleon force have been derived in Refs.~\cite{Girlanda:2011fh,Krebs:2012yv,Krebs:2013kha}. Isospin-breaking contributions to the two- and three-nucleon interactions have been analyzed using the same theoretical framework in Refs.~\cite{vanKolck:1997fu, Friar:1999zr,Friar:2003yv,Friar:2004ca,Epelbaum:2004xf,Friar:2004rg,Kaiser:2003ty,Kaiser:2006ck,Kaiser:2006na,Reinert:2020mcu}. Furthermore, electroweak and scalar nuclear current operators were calculated through N$^3$LO\footnote{Here and in what follows, we use the standard power counting of chiral perturbation theory based on naive dimensional analysis, except for the nucleon mass $m$, which is assumed to be of the order of $m \sim \Lambda_{\rm b}^2/M_\pi$ \cite{Weinberg:1991um} as often done in the few-nucleon sector.}  in Refs.~\cite{Park:1993jf,Park:1995pn,Kolling:2009iq,Kolling:2011mt,Krebs:2016rqz, Krebs:2019aka, Krebs:2020plh,Pastore:2008ui,Pastore:2009is,Pastore:2011ip,Baroni:2015uza}, see Ref.~\cite{Krebs:2020pii} for a review article. Finally, for similar studies of the parity- and time-reversal-violating nuclear interactions see Ref.~\cite{deVries:2020iea} and references therein.  

The extensive work on the derivation of the NN force in Refs.~\cite{Kaiser:1997mw,Kaiser:1998wa,Epelbaum:1998ka,Epelbaum:1999dj, Kaiser:2001pc,Kaiser:2001at,Entem:2014msa}, coupled with the novel approach for extracting the $\pi$N low-energy constants (LECs) from the solution of the $\pi$N  Roy-Steiner equations \cite{Hoferichter:2015hva,Hoferichter:2015tha,Siemens:2016jwj} and the improved regularization methods \cite{Epelbaum:2014efa} have resulted in the development of accurate and precise N$^4$LO$^+$ NN potentials \cite{Reinert:2017usi,Reinert:2020mcu}, which lead to a statistically perfect description of the neutron-proton and proton-proton scattering data up to the pion production threshold, see Refs.~\cite{Epelbaum:2022cyo,Reinert:2022jpu} for details. For a related work using a different regularization scheme see Ref.~\cite{Entem:2017gor}. Importantly, the semi-locally regularized potentials of Refs.~\cite{Epelbaum:2014efa,Epelbaum:2014sza,Reinert:2017usi} were found to provide  a clear evidence of the chiral two-pion (2$\pi$) exchange, which is predicted by the chiral symmetry of QCD and the empirical information on the pion-nucleon system needed to pin down the relevant LECs \cite{Epelbaum:2024gfg}. In particular, it was shown in Refs.~\cite{Epelbaum:2014efa,Epelbaum:2014sza,Reinert:2017usi} that the description of the NN data improves considerably when going from the order $Q^2$ (NLO) to $Q^3$ (N$^2$LO) as well as from the order  $Q^4$ (N$^3$LO) to $Q^5$ (N$^4$LO). In both cases, the  only isospin-invariant contributions to the NN potential are given by the parameter-free contributions to the 2$\pi$-exchange potential. For a related recent work focused on the role of chiral symmetry in peripheral neutron-$\alpha$ scattering see Ref.~\cite{Yang:2025mhg}. 

\begin{figure}[t!]
	\begin{center}
          \includegraphics[width=0.7\textwidth]{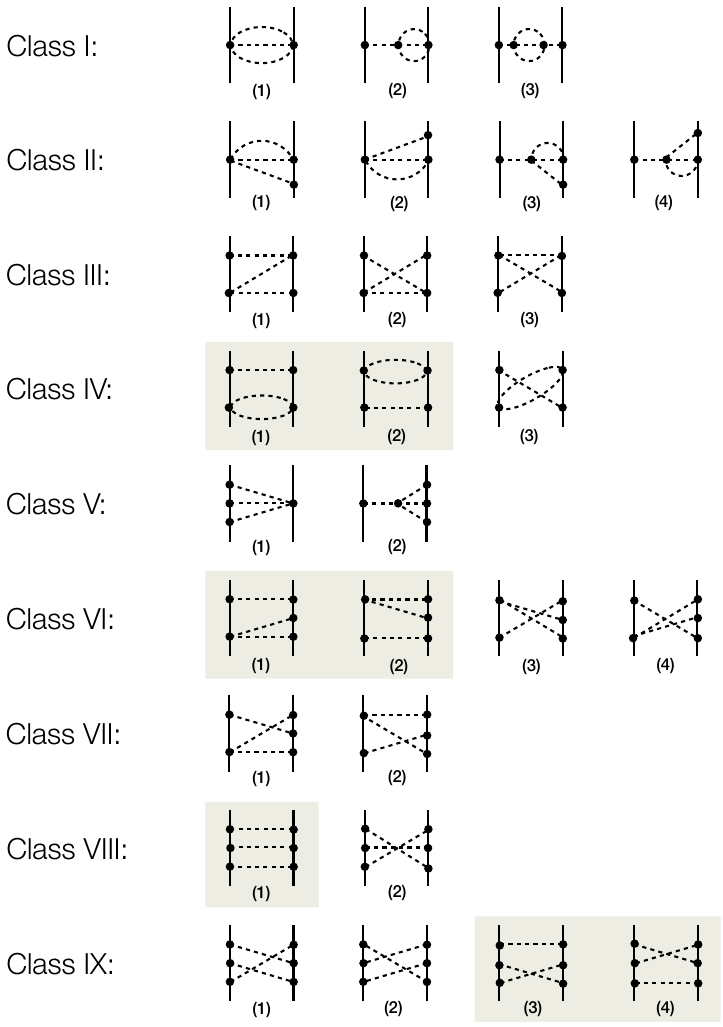}
          \end{center}
	\caption{Diagrams contributing to the leading $3\pi$-exchange NN potential at N$^3$LO. Solid and dashed lines refer to nucleons and pions, respectively, while solid dots denote the lowest-order vertices from the effective Lagrangians $\mathcal{L}_{\pi N}^{(1)}$ and $\mathcal{L}_{\pi}^{(2)}$. Diagrams resulting from the interchange of the nucleon lines are not shown (except for the class IX). Light-shaded gray areas mark reducible-like contributions which are expected to be scheme-dependent.}
	\label{fig:3pi_N3LO}
\end{figure}

\begin{figure}[t!]
	\begin{center}
          \includegraphics[width=0.7\textwidth]{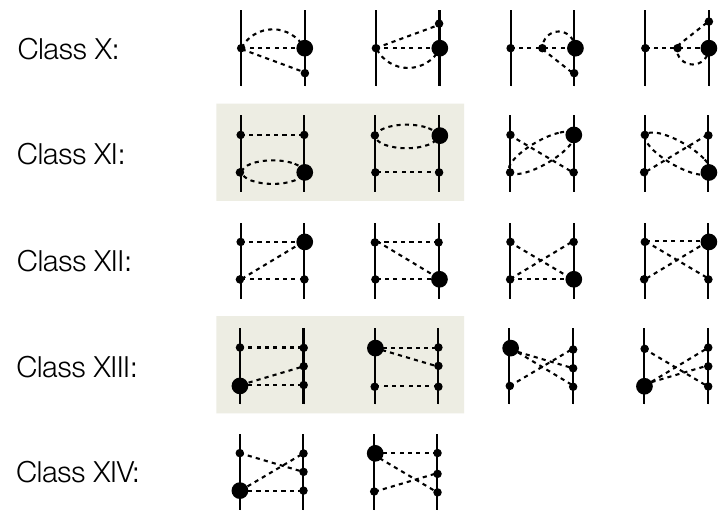}
          \end{center}
	\caption{Diagrams contributing to the subleading $3\pi$-exchange NN potential at N$^4$LO. Filled circles denote the subleading pion-nucleon vertices from the effective Lagrangian $\mathcal{L}_{\pi N}^{(2)}$. For remaining notation see Fig.~\ref{fig:3pi_N3LO}. }
	\label{fig:3pi_N4LO}
\end{figure}

While the phenomenological importance of the 2$\pi$-exchange can be regarded as well established, the situation with the three-pion (3$\pi$) exchange potential, which also comes out as a parameter-free prediction within chiral EFT, is more subtle due to its shorter range and the additional suppression by two powers of the expansion parameter $Q$ as compared to the 2$\pi$-exchange. 
The leading and subleading contributions to the chiral 3$\pi$-exchange stemming from diagrams shown in Figs.~\ref{fig:3pi_N3LO} and \ref{fig:3pi_N4LO}, respectively, have been worked out by Kaiser in Refs.~\cite{Kaiser:1999ff,Kaiser:1999jg,Kaiser:2001dm}, see also Ref.~\cite{Pupin:1999ba} for an earlier work along this line. 
Instead of directly calculating the complicated two-loop diagrams, Kaiser used the Cutkosky cutting rules to express the corresponding spectral functions as an integral of the on-shell $\bar{\rm N} {\rm N} \to \pi\pi\pi$ amplitudes over the $3\pi$ phase space. Reducible contributions to the $\rm NN \to NN$ scattering amplitude stemming from iterations of the Lippmann-Schwinger equation, which  
do not correspond to the potential, were identified by performing $1/m$-expansion of the scattering amplitude in a close analogy to what is done for the  $2 \pi$-exchange in Ref.~\cite{Kaiser:1997mw}. However, in contrast to the S-matrix, nuclear potentials are defined off-the-energy-shell and thus exhibit a large degree of scheme dependence, see also Ref.~\cite{Epelbaum:2025aan} for a related recent discussion. To illustrate this point, consider the on-shell NN scattering amplitude $\mathcal{A}_{2\pi}^{g_A^4}$, where $g_A$ denotes the nucleon axial-vector coupling, stemming from the two-pion exchange planar- and crossed-box diagrams.  The corresponding on-shell amplitude is defined unambiguously and can be obtained by calculating the two Feynman graphs. On the other hand, the same on-shell amplitude must come out of the iterations of the Lippmann-Schwinger equation, 
\beq
\label{Matching'2pi}
\mathcal{A}_{2\pi}^{g_A^4}  = \Big[ \hat V_{1\pi}^{g_A^2} \, \hat G_0 \,  \hat V_{1\pi}^{g_A^2} 
+   \hat V_{2\pi}^{g_A^4} \Big]_{\rm on-shell}\,,  
\eeq
where $\hat G_0$ is the two-nucleon resolvent operator while $V_{1\pi}^{g_A^2}$ and  $V_{2\pi}^{g_A^4}$ refer to the one- and two-pion exchange potentials, respectively. Here, $\hat{X}$ indicates that the corresponding quantity should be understood as an operator. Notice that the four-dimensional (three-dimensional) loop integrals appearing on the left-hand (right-hand) side of the above equation are ultraviolet divergent and require introducing the corresponding counterterms. We are, however, interested here in the long-range contributions to the amplitude, which are not affected by the counterterms. It is instructive to expand both sides of Eq.~(\ref{Matching'2pi}) around the static limit $m \to \infty$. Given that $\hat G_0 \sim \mathcal{O} (m)$, the dominant contribution to the amplitude $\sim m$ is generated solely by the iterated static one-pion exchange potential. When working with energy-independent and Hermitian potentials, the first relativistic corrections to the $1\pi$-exchange appear at order $m^{-2}$. Accordingly, the static $2\pi$-exchange potential is unambiguously determined (on-the-energy-shell) by the order-$\mathcal{O}(1)$ contribution to the amplitude $\mathcal{A}_{2\pi}^{g_A^4}$. Indeed, the static expressions for the non-polynomial parts of the $2\pi$-exchange potential in the energy-independent formulation calculated using different methods come out identical to each other \cite{Friar:1994zz,Kaiser:1997mw,Epelbaum:1998ka}. On the other hand, the leading relativistic corrections to $V_{2\pi}^{g_A^4}$ of order $\mathcal{O}(m^{-1})$ are scheme-dependent due to the unitary ambiguity of the order-$m^{-2}$
contributions to $V_{1\pi}^{g_A^2}$ and different schemes for treating relativistic effects in the framework of the Schr\"odinger equation (which affect the order-$m^{-1}$ corrections to $G_0$), see Ref.~\cite{Friar:1999sj} for details. 

The above considerations can be extended to the three-pion exchange. Consider, for example, the amplitude 
$\mathcal{A}_{3\pi}^{g_A^6}$ corresponding to the diagram (1) of class VIII, which is defined unambiguously and can be computed from the Feynman graph or by iterating  the Lippmann-Schwinger equation  
\beq
\mathcal{A}_{3\pi}^{g_A^6}  = \Big[ \hat V_{1\pi}^{g_A^2} \, \hat G_0 \,  \hat V_{1\pi}^{g_A^2} \, \hat G_0 \hat V_{1\pi}^{g_A^2}
+  \hat V_{1\pi}^{g_A^2} \, \hat G_0 \,  \hat V_{2\pi}^{g_A^4} +  \hat V_{2\pi}^{g_A^4} \, \hat G_0 \,  \hat V_{1\pi}^{g_A^2} +  \hat V_{3\pi}^{g_A^6} \Big]_{\rm on-shell}\,,
\eeq
where the two- and three-pion exchange potentials are understood to include only planar box diagrams. The non-relativistic expansion of $\mathcal{A}_{3\pi}^{g_A^6}$ starts with the order $\mathcal{O}(m^2)$, see the first term in the square brackets, followed by the order-$m$ corrections from the second and third terms. The last term in the square brackets denotes the genuine static (i.e., order $\mathcal{O} (1)$ in the $1/m$-expansion) three-pion exchange potential we are interested in here. Contrary to the unambiguously defined\footnote{The term ``unambiguous'' refers to the considered class of unitary transformations.} static two-pion exchange potential, the 
ambiguities in the order-$m^{-2}$ (order-$m^{-1}$) corrections to $V_{1\pi}^{g_A^2}$ ($V_{2\pi}^{g_A^4}$) and in the treatment of relativistic corrections to $G_0$ show that already the dominant static expressions for the three-pion exchange $ V_{3\pi}^{g_A^6}$ are scheme-dependent.
The above considerations suggest that  three-pion exchange potentials generated by reducible-like diagrams, which are  highlighted with light-shaded gray areas in Figs.~\ref{fig:3pi_N3LO} and \ref{fig:3pi_N4LO}, are ambiguously defined. Accordingly, the expressions derived in Refs.~\cite{Kaiser:1999ff,Kaiser:1999jg,Kaiser:2001dm} can potentially be inconsistent with the nuclear forces and currents from Refs.~\cite{Epelbaum:1998ka,Epelbaum:1999dj,Epelbaum:2002vt,Epelbaum:2005bjv,Epelbaum:2007us, Bernard:2007sp,Bernard:2011zr,Krebs:2012yv,Krebs:2013kha,Epelbaum:2004xf,Reinert:2020mcu,Kolling:2009iq,Kolling:2011mt,Krebs:2016rqz, Krebs:2019aka, Krebs:2020plh} derived using the method of unitary transformation (MUT). For this reason and assuming that  effects of the three-pion exchange can, for the employed cutoff values, be accurately represented by contact interactions, $3\pi$-exchange contributions were not explicitly taken into account in the NN potentials of Refs.~\cite{Epelbaum:2014efa,Epelbaum:2014sza,Reinert:2017usi,Reinert:2020mcu}. 

The purpose of this work is twofold. First, we rederive all scheme-independent contributions to the leading three-pion exchange potential using the same methods as developed and applied by Kaiser in Refs.~\cite{Kaiser:1999ff,Kaiser:1999jg,Kaiser:2001dm}. We provide a detailed description of the performed calculational steps, including intermediate results, and independently verify the expressions obtained in those papers. Secondly, for scheme-dependent contributions stemming from reducible-like diagrams, we perform calculations  using both the approach of Kaiser and the MUT. In line with the expectations, we find different results for class-VI, VIII and IX contributions. On the other hand, the class-IV, XI and XIII potentials calculated using the MUT turn out to coincide with those obtained in Refs.~\cite{Kaiser:1999ff,Kaiser:1999jg,Kaiser:2001dm}.

Our paper is organized as follows. In sec.~\ref{SecII}, we present a detailed description of various techniques used to derive the three-pion exchange potential including the MUT, the Cutkosky cutting rules for calculating the corresponding spectral functions and the alternative approach for directly computing the potentials in coordinate space. In sec.~\ref{SecIII}, we describe the derivation of the N$^3$LO three-pion exchange potential stemming from the class-I-IX diagrams shown in Figs.~\ref{fig:3pi_N3LO}, while the N$^4$LO contributions are considered in sec.~\ref{SecIV}. The main results of our paper are summarized in sec.~\ref{SecV}, where we also
compare the different potentials obtained using these methods in coordinate space. Finally, the calculation of various integrals relevant for this work is detailed in appendices \ref{Appendix_Principle Value Integrals}-\ref{subsec_SolvingImVcClassIX}.

\section{Anatomy of the calculation}
\setcounter{equation}{0}
\def\theequation{\arabic{section}.\arabic{equation}}
\label{SecII}

Our general strategy for deriving the three-pion exchange potential can be summarized as follows:
\vspace{-0.25cm}
\begin{itemize}
\item Instead of directly calculating the potentials by solving two-loop integrals that appear in the expressions for the Feynman or irreducible time-ordered-like diagrams, we focus on the imaginary part of the amplitude for $q = i \mu + 0^+$, where $q \equiv |\vec q \, |$ and $\vec q$ denotes the nucleon momentum transfer. The calculation of the imaginary part is facilitated by the application of the Cutkosky cutting rules and amounts to integrating the on-shell $\bar N N \to 3 \pi$ scattering amplitudes over the $3\pi$ phase space \cite{Kaiser:1999ff,Kaiser:1999jg,Kaiser:2001dm}, see sec.~\ref{Subsec:Frame1} for details.
Using dispersive representation, the calculated imaginary parts can be used to reconstruct the (non-polynomial part of the) potential in momentum space by integrating over the mass spectrum $\mu$ of the exchanged pions as described in sec.~\ref{subsecDisp}. Alternatively, the spectral representation can be used to obtain the (large-distance behavior of the) corresponding coordinate-space potentials. \\[-13pt]
\item
For certain class-VIII and class-IX contributions, the spectral function obtained using the Cutkosky cutting rules as described in sec.~\ref{Subsec:Frame1}  turns out to be ill-defined \cite{Kaiser:1999jg,Kaiser:2001dm}. In such cases, we follow the alternative approach proposed in Ref.~\cite{Kaiser:2001dm} and directly calculate the potentials in coordinate space. This is achieved by first Wick-rotating the pion loop momenta in the original expressions and subsequently performing the Fourier transform to coordinate space as explained in sec.~\ref{sec:Wick}. For the considered classes of contributions, the integrals over the three-momenta of exchanged pions factorize and, therefore, can be easily carried out. We have verified for selected cases that both methods lead to the same coordinate-space expressions. \\[-13pt]
\item
  For reducible types of diagrams highlighted with the light-shaded gray areas in Figs.~\ref{fig:3pi_N3LO} and \ref{fig:3pi_N4LO}, one needs to subtract out iterative contributions to the amplitude, which are scheme-dependent. Here, we perform calculations in two ways. The first approach, which we refer to as the S-matrix method (SMM), is the one that has been used by Kaiser in Refs.~\cite{Kaiser:1999ff,Kaiser:1999jg,Kaiser:2001dm}. Starting from the two-loop expressions for the corresponding Feynman diagrams written in terms of the relativistic nucleon propagators, we first perform the integrations over the $0$-th components of the loop momenta. This step essentially corresponds to switching from the covariant perturbation theory to the old-fashioned time-ordered perturbation theory. We then expand the result in inverse powers of the nucleon mass $m$, remove the contributions involving positive powers of $m$ which are associated with iterations of the Lippmann-Schwinger equation and keep only the static terms by taking the limit $m \to \infty$. After this step, the integrands are given in terms of rational functions of the pion energies $\omega_i = \sqrt{\vec{l}_i^{\; 2} + M_\pi^2}$, where $M_\pi$ is the pion mass. These functions can also be obtained by summing over the energy denominators of the corresponding time-ordered diagrams. To facilitate the calculation of the remaining integrals, it is advantageous to rewrite these expressions in the covariant form, i.e.~as loop integrals over four-momenta with the integrands given in terms of the usual Feynman propagators for pions and heavy-baryon propagators for nucleons\footnote{Such matching can be done straightforwardly using the residue theorem, see sec.~\ref{SecIII} for explicit expressions.}. The resulting covariant expressions, which have been ``purified'' to remove the iterative contributions, can be treated using the methods outlined above. This way we will reproduce the three-pion exchange potentials obtained using the SMM in Refs.~\cite{Kaiser:1999ff,Kaiser:1999jg,Kaiser:2001dm} (with one exception). The main focus of this work is, however, to derive the three-pion exchange potential that would be 
off-shell consistent with the nuclear forces and currents of Refs.~\cite{Epelbaum:1998ka,Epelbaum:1999dj,Epelbaum:2002vt,Epelbaum:2005bjv,Epelbaum:2007us, Bernard:2007sp,Bernard:2011zr,Krebs:2012yv,Krebs:2013kha,Epelbaum:2004xf,Reinert:2020mcu,Kolling:2009iq,Kolling:2011mt,Krebs:2016rqz, Krebs:2019aka, Krebs:2020plh} using the MUT. In sec.~\ref{sec:MUT}, we briefly explain the MUT and provide the relevant Fock-space expressions for the nuclear Hamiltonian in terms of the pion-nucleon vertices and energy denominators. We then evaluate the two-body momentum-space matrix elements of these operators and combine together the corresponding energy denominators. For the class-VI, VIII and IX contributions, the resulting rational functions of $\omega_i$ turn out to be different from those found using the SMM. The remaining derivation proceeds in the same way as outlined above using the SMM. 
\end{itemize}
\vspace{-0.4cm}
In the next subsections, we provide a detailed description of the above-mentioned calculational steps and techniques.

\subsection{Method of Unitary Transformation}
\label{sec:MUT}

Similarly to time-ordered perturbation theory, the MUT relies on the Hamiltonian $H$, which for the case at hand can be straightforwardly obtained from the effective pion-nucleon Lagrangian using the canonical formalism, see Ref.~\cite{Weinberg:1990rz, Weinberg:1991um}. To derive nuclear forces, one needs to decouple the purely nucleonic subspace of the Fock space from the rest. This can be achieved by acting on the pion-nucleon Hamiltonian $H$ with a suitable unitary transformation $U$. Let $\eta$ and $\lambda$ denote the projection operators onto the purely nucleonic subspace of the Fock space and the rest, respectively. Following Okubo \cite{Okubo:1954zz}, the ``minimal'' form of the unitary operator, $U_{\rm Okubo}$, is given by
\beq
U_{\rm Okubo} = \left( \begin{array}{ll} \eta (1 + A^\dagger A )^{-1/2} & - A^\dagger (1 + AA^\dagger )^{-1/2} \\
                         A (1 + A^\dagger A)^{-1/2} & \lambda  (1 + A A^\dagger )^{-1/2} \end{array}\right),
\eeq
where the operator $A$ has only mixed nonvanishing components: $A = \lambda A \eta$. The requirement $\eta U_{\rm Okubo}^\dagger H U_{\rm Okubo} \lambda = \lambda U_{\rm Okubo}^\dagger H U_{\rm Okubo} \eta = 0$ leads to the nonlinear decoupling equation for the operator $A$
\beq
\lambda ( H + [ H, \, A] - A H A ) \eta = 0\,.
\eeq
The solution of this equation and the calculation of the nuclear forces $V_{\rm MUT} = \eta U_{\rm Okubo}^\dagger H U_{\rm Okubo} \eta - \eta H_{\rm kin}$, where  $H_{\rm kin}$ refers to  the kinetic energy of the nucleons, can be carried out perturbatively by means of the chiral expansion \cite{Epelbaum:1998ka} as described in detail in Refs.~\cite{Epelbaum:2007us,Epelbaum:2019kcf}. 

The resulting expression for the nuclear Hamiltonian relevant for the class-III and IV diagrams is given by \cite{Epelbaum:2007us}
\begin{align}
  \label{MUTClass34}
	V^{\rm III,  IV}_{\rm MUT}&= \eta \left[ \frac{1}{2} H^1_{21} \frac{\lambda^1}{E_\pi} H^1_{21}\eta H^2_{22}\frac{\lambda^2}{E_\pi^2}H^2_{22} + \frac{1}{2} H^1_{21} \frac{\lambda^1}{E_\pi^2} H^1_{21}\eta H^2_{22}\frac{\lambda^2}{E_\pi}H^2_{22} - H^1_{21} \frac{\lambda^1}{E_\pi} H^1_{21}\frac{\lambda^2}{E_\pi}H^2_{22}\frac{\lambda^2}{E_\pi}H^2_{22}\right.\nonumber\\
	&\quad\left. - H^1_{21} \frac{\lambda^1}{E_\pi} H^2_{22}\frac{\lambda^1}{E_\pi}H^1_{21}\frac{\lambda^2}{E_\pi}H^2_{22} - \frac{1}{2} H^1_{21} \frac{\lambda^1}{E_\pi} H^2_{22}\frac{\lambda^1}{E_\pi}H^2_{22}\frac{\lambda^1}{E_\pi}H^1_{21} - H^1_{21} \frac{\lambda^1}{E_\pi} H^2_{22}\frac{\lambda^3}{E_\pi}H^1_{21}\frac{\lambda^2}{E_\pi}H^2_{22}\right.\nonumber\\
	&\quad\left. - \frac{1}{2} H^1_{21} \frac{\lambda^1}{E_\pi} H^2_{22}\frac{\lambda^3}{E_\pi}H^2_{22}\frac{\lambda^1}{E_\pi}H^1_{21} - \frac{1}{2} H^2_{22} \frac{\lambda^2}{E_\pi} H^1_{21}\frac{\lambda^3}{E_\pi}H^1_{21}\frac{\lambda^2}{E_\pi}H^2_{22} - \frac{1}{2} H^2_{22} \frac{\lambda^2}{E_\pi} H^1_{21}\frac{\lambda^1}{E_\pi}H^1_{21}\frac{\lambda^2}{E_\pi}H^2_{22}\right.\nonumber\\
	&\quad\left. + H^1_{21} \frac{\lambda^1}{E_\pi} H^1_{21}\frac{\lambda^2}{E_\pi}H^4_{42} + \frac{1}{2} H^1_{21} \frac{\lambda^1}{E_\pi} H^4_{42}\frac{\lambda^1}{E_\pi}H^1_{21}\right]\eta + \textrm{h.c.} \,,
\end{align}
where h.c.~means Hermitean conjugated. Here, $\lambda^n$ denotes the projection operator on the Fock-space states with $n$ pions, while $E_\pi = \sum_i \omega_i$ is the pion energy in the corresponding intermediate state. For the pion-nucleon vertices, we use the notation $H_{ab}^\kappa$, where $a$ and $b$ refer to the number of nucleon and pion field operators, respectively, while $\kappa$ is the inverse mass dimension of the coupling constant. For example, $H_{21}^1$ refers to the leading pion-nucleon vertex proportional to the axial-vector coupling $g_A$, while 
 $H_{22}^2$ denotes the Weinberg-Tomozawa interaction. 
For every term in Eq.~(\ref{MUTClass34}), the power $\nu$ of the chiral expansion parameter $Q$ can be easily obtained using $\nu = -2 + \sum_i \kappa_i$, where the sum goes through all vertices.  Accordingly, all operators in the above equation contribute at order $Q^4$ or N$^3$LO.  
Notice that in addition to the $g_A$ and Weinberg-Tomozawa vertices, some of the operators in Eq.~(\ref{MUTClass34}) involve the derivativeless  interaction $H_{42}^4$  with four nucleon and two pion fields, which at first sight seems to violate the chiral symmetry. Such a vertex is, in fact, absent in the effective Lagrangian.  It is generated in the Hamiltonian through the application of the canonical formalism and is caused by the time derivative entering the Weinberg-Tomozawa interaction \cite{Weinberg:1992yk}. We further emphasize that the operator in Eq.~(\ref{MUTClass34}) gives rise not only to two-nucleon potentials, but also to  three- and four-nucleon forces. 

In a similar way, one obtains the expression for the nuclear Hamiltonian made out of four $g_A$-vertices $H_{21}^1$ and a single Weinberg-Tomozawa interaction $H_{22}^2$, which can be used to calculate the class-VI and VII three-pion exchange potentials \cite{Epelbaum:2007us}: 
\begin{align}
 	V^{\rm VI,  VII}_{\rm MUT} &= \eta\left[-\frac{1}{2}H_{21}^{1}\frac{\lambda^{1}}{E_\pi}H_{21}^{1}\;\eta\; H_{21}^1\frac{\lambda^{1}}{E_\pi}H_{21}^{1}\frac{\lambda^{2}}{E_\pi^2}H_{22}^{2}-\frac{1}{2} H_{21}^{1}\frac{\lambda^{1}}{E_\pi}H_{21}^{1} \;\eta\; H_{21}^1\frac{\lambda^{1}}{E_\pi}H_{22}^{2}\frac{\lambda^{1}}{E^2_\pi}H_{21}^{1}\right.\nonumber\\
	&\quad\left.-\frac{1}{2}H_{21}^{1}\frac{\lambda^{1}}{E_\pi}H_{21}^{1}\;\eta\; H_{22}^2\frac{\lambda^{2}}{E_\pi}H_{21}^{1}\frac{\lambda^{1}}{E^2_\pi}H_{21}^{1}-\frac{1}{2}H_{21}^{1}\frac{\lambda^{1}}{E_\pi}H_{21}^{1}\;\eta\; H_{21}^1\frac{\lambda^{1}}{E^2_\pi}H_{21}^{1}\frac{\lambda^{2}}{E_\pi}H_{22}^{2}\right.\nonumber\\
	&\quad\left.-\frac{1}{2}H_{21}^{1}\frac{\lambda^{1}}{E_\pi}H_{21}^{1} \;\eta\; H_{21}^1\frac{\lambda^{1}}{E^2_\pi}H_{22}^{2}\frac{\lambda^{1}}{E_\pi}H_{21}^{1}-\frac{1}{2}H_{21}^{1}\frac{\lambda^{1}}{E_\pi}H_{21}^{1} \;\eta\; H_{22}^2\frac{\lambda^{2}}{E^2_\pi}H_{21}^{1}\frac{\lambda^{1}}{E_\pi}H_{21}^{1}\right.\nonumber\\
	&\quad\left.-\frac{1}{2}H_{21}^{1}\frac{\lambda^{1}}{E^2_\pi}H_{21}^{1} \;\eta\; H_{21}^1\frac{\lambda^{1}}{E_\pi}H_{21}^{1}\frac{\lambda^{2}}{E_\pi}H_{22}^{2}-\frac{1}{2}H_{21}^{1}\frac{\lambda^{1}}{E^2_\pi}H_{21}^{1}\;\eta\; H_{21}^1\frac{\lambda^{1}}{E_\pi}H_{22}^{2}\frac{\lambda^{1}}{E_\pi}H_{21}^{1}\right.\nonumber\\
                                   &\quad\left.
-\frac{1}{2}H_{21}^{1}\frac{\lambda^{1}}{E^2_\pi}H_{21}^{1} \;\eta\; H_{22}^2\frac{\lambda^{2}}{E_\pi}H_{21}^{1}\frac{\lambda^{1}}{E_\pi}H_{21}^{1}
                                     +\frac{1}{2}H_{21}^{1}\frac{\lambda^{1}}{E_\pi}H_{21}^{1} \frac{\lambda^{2}}{E_\pi}H_{22}^{2}\frac{\lambda^{2}}{E_\pi}H_{21}^{1}\frac{\lambda^{1}}{E_\pi}H_{21}^{1}\right.\nonumber\\
                                  &\quad\left.
                                    + H_{21}^{1}\frac{\lambda^{1}}{E_\pi}H_{21}^{1} \frac{\lambda^{2}}{E_\pi}H_{21}^{1}\frac{\lambda^{1}}{E_\pi}H_{21}^{1}\frac{\lambda^{2}}{E_\pi}H_{22}^{2}+ H_{21}^{1}\frac{\lambda^{1}}{E_\pi}H_{21}^{1} \frac{\lambda^{2}}{E_\pi}H_{21}^{1}\frac{\lambda^{1}}{E_\pi}H_{22}^{2}\frac{\lambda^{1}}{E_\pi}H_{21}^{1}\right.\nonumber\\
                                 &\quad\left.
                                    + H_{21}^{1}\frac{\lambda^{1}}{E_\pi}H_{21}^{1} \frac{\lambda^{2}}{E_\pi}H_{21}^{1}\frac{\lambda^{3}}{E_\pi}H_{21}^{1}\frac{\lambda^{2}}{E_\pi}H_{22}^{2}+ H_{21}^{1}\frac{\lambda^{1}}{E_\pi}H_{21}^{1} \frac{\lambda^{2}}{E_\pi}H_{21}^{1}\frac{\lambda^{3}}{E_\pi}H_{22}^{2}\frac{\lambda^{1}}{E_\pi}H_{21}^{1}\right]\eta  + {\rm h.c.}\,.\label{3pioperatorclass2}
\end{align}
As already mentioned above, the Okubo Ansatz $U_{\rm Okubo}$ considered so far represents a minimal possible unitary transformation that is capable of decoupling the pion states. After decoupling, it is always possible to perform additional transformations on the $\eta$-subspace, which affect the off-shell behavior of the corresponding potentials. For the case at hand, one can write down three anti-Hermitean operators made out of two $g_A$- and a single Weinberg-Tomozawa vertex  \cite{Epelbaum:2007us}:
\beqa
S_3 &=& \eta \bigg[ H_{21}^1 \frac{\lambda^1}{E_\pi^2} H_{22}^2 \frac{\lambda^1}{E_\pi} H_{21}^1  - {\rm h.c.} \bigg]\eta\,, \nonumber \\
S_4 &=& \eta \bigg[ H_{22}^2 \frac{\lambda^2}{E_\pi^2} H_{21}^1 \frac{\lambda^1}{E_\pi} H_{21}^1  - {\rm h.c.} \bigg]\eta\,, \nonumber \\
S_5 &=& \eta \bigg[ H_{22}^2 \frac{\lambda^2}{E_\pi} H_{21}^1 \frac{\lambda^1}{E_\pi^2} H_{21}^1  - {\rm h.c.} \bigg]\eta\,.
\eeqa
The corresponding $\eta$-space unitary transformation $U = \exp(\alpha_3 S_3 + \alpha_4 S_4 + \alpha_5 S_5)$ depends on three dimensionless phases $\alpha_3$,  $\alpha_4$ and $\alpha_5$ and induces additional contributions to the Hamiltonian via
\beq
\delta V_{\rm MUT} = U^\dagger (V_{\rm MUT} + \eta H_{\rm kin})U - V_{\rm MUT} - \eta H_{\rm kin}\simeq
[V_{\rm MUT} + \eta H_{\rm kin}, \; \alpha_3 S_3 + \alpha_4 S_4 + \alpha_5 S_5]\,. 
\eeq
When the commutator in the above equation is evaluated with the one-pion exchange potential $-\eta H_{21}^1 \frac{\lambda^1}{E_\pi} H_{21}^1 \eta$ entering $V_{\rm MUT}$, the resulting terms provide additional class-VI and VII contributions 
\begin{align}
  \label{MUTClass67add}
  \delta V^{\rm VI,  VII}_{\rm MUT} &=
                                      -\alpha_3\eta \left[H_{21}^{1}\frac{\lambda^{1}}{E_\pi}H_{21}^{1}\;\eta\; H_{21}^1\frac{\lambda^{1}}{E_\pi^2}H_{22}^{2}\frac{\lambda^{1}}{E_\pi}H_{21}^{1} - H_{21}^{1}\frac{\lambda^{1}}{E_\pi}H_{21}^{1}\;\eta\; H_{21}^1\frac{\lambda^{1}}{E_\pi}H_{22}^{2}\frac{\lambda^{1}}{E_\pi^2}H_{21}^{1}\right]\eta \nonumber \\
	& \quad -\alpha_4\eta \left[H_{21}^{1}\frac{\lambda^{1}}{E_\pi}H_{21}^{1}\;\eta\; H_{22}^2\frac{\lambda^{2}}{E_\pi^2}H_{21}^{1}\frac{\lambda^{1}}{E_\pi}H_{21}^{1} - H_{21}^{1}\frac{\lambda^{1}}{E_\pi}H_{21}^{1}\;\eta\; H_{21}^1\frac{\lambda^{1}}{E_\pi}H_{21}^{1}\frac{\lambda^{2}}{E_\pi^2}H_{22}^{2}\right]\eta \nonumber\\ 
	& \quad -\alpha_5\eta \left[H_{21}^{1}\frac{\lambda^{1}}{E_\pi}H_{21}^{1}\;\eta\; H_{22}^2\frac{\lambda^{2}}{E_\pi}H_{21}^{1}\frac{\lambda^{1}}{E_\pi^2}H_{21}^{1} - H_{21}^{1}\frac{\lambda^{1}}{E_\pi}H_{21}^{1}\;\eta\; H_{21}^1\frac{\lambda^{1}}{E_\pi^2}H_{21}^{1}\frac{\lambda^{2}}{E_\pi}H_{22}^{2}\right]\eta + {\rm h.c.}\,.
\end{align}
Notice that the considered unitary transformation also affects $1/m$ corrections to the two-pion exchange when acting on $\eta E_{\rm kin}$, cf.~the discussion in the Introduction, and it also generates contributions to the three- and four-nucleon forces and current operators.  It was shown in Refs.~\cite{Epelbaum:2007us} that in order to obtain renormalized three-nucleon force at N$^3$LO, the phases $\alpha_i$ must be restricted to
\beq
\label{alpha345}
\alpha_3 = -\alpha_5 , \quad \alpha_4 = \frac{1}{2}+ 2 \alpha_5\,.
\eeq

Finally, we give the expression for the nuclear Hamiltonian after performing the Okubo-transformation, which is relevant for the class-VIII and IX contributions: 
\begin{align}
 	V^{\rm VIII,  IX}_{\rm MUT} &=\frac{1}{2} \eta\left[H^{1}_{21}\frac{\lambda^{1}}{E_\pi}H^{1}_{21} \eta H^{1}_{21}\frac{\lambda^{1}}{E_\pi}H^{1}_{21}\frac{\lambda^{2}}{E_\pi}H^{1}_{21}\frac{\lambda^{1}}{E^2_\pi}H^{1}_{21}+H_{21}^{1}\frac{\lambda^{1}}{E_\pi}H_{21}^{1} \eta H^{1}_{21}\frac{\lambda^{1}}{E_\pi}H^{1}_{21}\frac{\lambda^{2}}{E^2_\pi}H^{1}_{21}\frac{\lambda^{1}}{E_\pi}H^{1}_{21}\right.\nonumber\\
		&\quad\left.+H^{1}_{21}\frac{\lambda^{1}}{E_\pi}H^{1}_{21} \eta H^{1}_{21}\frac{\lambda^{1}}{E^2_\pi}H^{1}_{21}\frac{\lambda^{2}}{E_\pi}H^{1}_{21}\frac{\lambda^{1}}{E_\pi}H^{1}_{21}+H^{1}_{21}\frac{\lambda^{1}}{E^2_\pi}H^{1}_{21} \eta H^{1}_{21}\frac{\lambda^{1}}{E_\pi}H^{1}_{21}\frac{\lambda^{2}}{E_\pi}H^{1}_{21}\frac{\lambda^{1}}{E_\pi}H^{1}_{21}\right.\nonumber\\
		&\quad\left.-H^{1}_{21}\frac{\lambda^{1}}{E_\pi}H^{1}_{21} \eta H^{1}_{21}\frac{\lambda^{1}}{E_\pi}H^{1}_{21}\eta H^{1}_{21}\frac{\lambda^{1}}{E^3_\pi}H^{1}_{21}
		-\frac{1}{4}H^{1}_{21}\frac{\lambda^{1}}{E^2_\pi}H^{1}_{21} \eta H^{1}_{21}\frac{\lambda^{1}}{E_\pi}H^{1}_{21}\eta H^{1}_{21}\frac{\lambda^{1}}{E^2_\pi}H^{1}_{21}\right.\nonumber\\
		&\quad\left.-\frac{3}{4}H^{1}_{21}\frac{\lambda^{1}}{E_\pi}H^{1}_{21} \eta H^{1}_{21}\frac{\lambda^{1}}{E^2_\pi}H^{1}_{21}\eta H^{1}_{21}\frac{\lambda^{1}}{E^2_\pi}H^{1}_{21}
		-H^{1}_{21}\frac{\lambda^{1}}{E_\pi}H^{1}_{21} \frac{\lambda^{2}}{E_\pi}H^{1}_{21}\frac{\lambda^{1}}{E_\pi}H^{1}_{21}\frac{\lambda^{2}}{E_\pi}H^{1}_{21}\frac{\lambda^{1}}{E_\pi}H^{1}_{21}\right.\nonumber\\
		&\quad\left.-H^{1}_{21}\frac{\lambda^{1}}{E_\pi}H^{1}_{21} \frac{\lambda^{2}}{E_\pi}H^{1}_{21}\frac{\lambda^{3}}{E_\pi}H^{1}_{21}\frac{\lambda^{2}}{E_\pi}H^{1}_{21}\frac{\lambda^{1}}{E_\pi}H^{1}_{21}\right]\eta + \mathrm{h.c.}\, .\label{3pioperator_gA^6}
\end{align}
Similarly to the previously considered case, one can perform the additional unitary transformation $U = \exp(\alpha_1 S_1 + \alpha_2 S_2)$
on the $\eta$-space, where the anti-Hermitean generators $S_1$ and $S_2$ are given by \cite{Epelbaum:2007us}
\beqa
\label{alpha12}
S_1 &=& \eta \bigg[ H_{21}^1 \frac{\lambda^1}{E_\pi} H_{21}^1 \eta H_{21}^1\frac{\lambda^1}{E_\pi^3} H_{21}^1  - {\rm h.c.} \bigg]\eta\,, \nonumber \\
S_2 &=& \eta \bigg[ H_{21}^1 \frac{\lambda^1}{E_\pi} H_{21}^1 \frac{\lambda^2}{E_\pi} H_{21}^1  \frac{\lambda^1}{E_\pi^2}  H_{21}^1- {\rm h.c.} \bigg]\eta\,.
\eeqa
The relevant additional terms in the Hamiltonian have the form 
\begin{align}
  \label{ga6corterm}
		\delta V^{\rm VIII,  IX}_{\rm MUT} =&-\alpha_1\eta\left[H^1_{21}\frac{\lambda^1}{E_\pi}H^1_{21}\eta H^1_{21}\frac{\lambda^1}{E_\pi}H^1_{21}\eta H^1_{21}\frac{\lambda^1}{E^3_\pi}H^1_{21}
		-H^1_{21}\frac{\lambda^1}{E_\pi}H^1_{21}\eta H^1_{21}\frac{\lambda^1}{E^3_\pi}H^1_{21}\eta H^1_{21}\frac{\lambda^1}{E_\pi}H^1_{21}\right]\eta\\
		&-\alpha_2\eta\left[H^1_{21}\frac{\lambda^1}{E_\pi}H^1_{21}\eta H^1_{21}\frac{\lambda^1}{E_\pi}H^1_{21}\frac{\lambda^2}{E_\pi}H^1_{21}\frac{\lambda}{E^2_\pi}H^1_{21}-H^1_{21}\frac{\lambda^1}{E_\pi}H^1_{21}\eta H^1_{21}\frac{\lambda^1}{E^2_\pi}H^1_{21}\frac{\lambda^2}{E_\pi}H^1_{21}\frac{\lambda^2}{E_\pi}H^1_{21}\right]\eta+ \mathrm{h.c.}\, .\nonumber
	\end{align}
        Again, the renormalizability of the three-nucleon force leads to constraints for the phases $\alpha_1$, $\alpha_2$, which have the form\footnote{There was a misprint for the value of $\alpha_2$ in the original publication  \cite{Epelbaum:2007us}.}
        \beq
\alpha_1 = - 2 \alpha_2 = - \frac{1}{2}\,.
        \eeq
The values for the phases $\alpha_i$ specified in Eqs.~(\ref{alpha345}),  (\ref{alpha12}) were used in the derivation of the three- and four-nucleon forces in Refs.~\cite{Epelbaum:2005bjv,Epelbaum:2007us, Bernard:2007sp,Bernard:2011zr} and the exchange current operators in Refs.~\cite{Kolling:2009iq,Kolling:2011mt,Krebs:2016rqz, Krebs:2019aka, Krebs:2020plh} and will also be adopted in this work. We further emphasize that while the phase $\alpha_5$ is not constrained through the renormalizability requirement, the above-mentioned results for the nuclear interactions and currents turn out to be independent of $\alpha_5$. As will be shown below, this also holds true for the three-pion exchange NN potential.   

	\subsection{Dispersive representation of the potential}
	\label{subsecDisp}
        
	The leading and subleading contributions to the NN $3\pi$-exchange give rise to the central, spin-spin and tensor potentials that depend
	only on the nucleon momentum transfer $\vec q = \vec p \, ' - \vec p$,
	\beq
	\label{VMomSpace}
	V_{3 \pi} (\vec q) = V_C (q) + \vec \tau_1 \cdot \vec \tau_2 W_C (q)  + \vec \sigma_1 \cdot \vec \sigma_2 \big[  V_S (q) + \vec \tau_1 \cdot \vec \tau_2 W_S (q) \big] + \vec \sigma_1 \cdot \vec q \, \vec \sigma_2 \cdot \vec q \big[  V_T (q) + \vec \tau_1 \cdot \vec \tau_2 W_T (q) \big]\,, 
	\eeq
	where $\vec \sigma_i$ and $\vec \tau_i$ refer to the spin and isospin Pauli matrices of nucleon $i$. Throughout this paper, we are only interested in the non-polynomial (in momenta) contributions to the potentials, which can be reconstructed using the dispersion relations
	\beqa
	\label{SpectralMomSpace}
	V_{C,S} (q) &=& -\frac{2}{\pi} \int_{3 \Mp}^{\infty} d \mu \mu \frac{\rho_{C,S}(\mu)}{\mu^2 +q^2} + \ldots , \qquad 	V_{T} (q) = -\frac{2}{\pi} \int_{3 \Mp}^{\infty} d \mu \mu \frac{\rho_{T}(\mu)}{\mu^2 +q^2} + \ldots, \nonumber \\
	W_{C,S} (q)& =&-\frac{2}{\pi} \int_{3 \Mp}^{\infty} d \mu \mu \frac{\eta_{C,S}(\mu)}{\mu^2 +q^2} + \ldots ,\qquad W_{T} (q) = -\frac{2}{\pi} \int_{3 \Mp}^{\infty} d \mu \mu \frac{\eta_{T}(\mu)}{\mu^2 +q^2} + \ldots, 
	\eeqa
	where $q \equiv | \vec q \, |$ and the ellipses denote subtraction terms needed to render the spectral integral finite. These subtraction conributions possess no left-hand cut and thus can be absorbed into a redefinition of the contact interactions. The spectral functions $\rho_i$ and $\eta_i$ are related to the potential via $\rho_i (\mu) = - \textrm{Im}V_i(i \mu)$ and $\eta_i(\mu) = - \textrm{Im}W_i(i \mu)$. Notice that to facilitate the comparison with the earlier results obtained by Kaiser in Refs.~\cite{Kaiser:1999ff,Kaiser:1999jg,Kaiser:2001dm}, the convention for the spectral functions considered here is chosen to coincide with the one used in those papers. It differs by the overall minus sign from the convention used, e.g., in Refs.~\cite{Epelbaum:2003gr,Epelbaum:2003xx,Epelbaum:2004fk,Epelbaum:2014efa,Epelbaum:2014sza,Reinert:2017usi,Reinert:2020mcu}.

        The spectral representation is particularly convenient for obtaining the coordinate-space expressions 
	\beqa
	\label{VCoordSpace}
	V_{3 \pi} (\vec r) &=&\int \frac{d^3q}{(2\pi)^3} V(\vec q \, ) e^{-i \vec{q}\cdot \vec{r}}\nonumber \\
	&=&\tilde V_C (r) + \vec \tau_1 \cdot \vec \tau_2 \tilde W_C (r)  + \vec \sigma_1 \cdot \vec \sigma_2 \big[  \tilde V_S (r) + \vec \tau_1 \cdot \vec \tau_2 \tilde W_S (r) \big] + S_{12} (\hat r )  \big[  \tilde V_T (r) + \vec \tau_1 \cdot \vec \tau_2 \tilde W_T (r) \big]\,, 
	\eeqa
	where $\hat{r} = \vec{r}/|\vec{r}\, |$ and $S_{12}$ is the tensor operator given by
	\begin{align}
		S_{12}(\hat{r}) = 3 \left(\vec{\sigma}_1 \cdot \hat{r}\right)\left(\vec{\sigma}_2 \cdot \hat{r}\right)- \vec{\sigma}_1 \cdot \vec{\sigma}_2\,.
	\end{align}
        The large-distance behavior of the $3\pi$-exchange is dominated by the discontinuity of the momentum-space potentials across the left-hand cut and can be calculated from the spectral functions using
        \beqa
        \label{Disp_CoordSpace}
        V_C (r) &=& - \frac{1}{2 \pi^2 r} \int_{3 M_\pi}^\infty d\mu \mu e^{- \mu r} \, \rho_C (\mu), \nonumber \\
        V_T (r) &=& \frac{1}{6 \pi^2 r^3} \int_{3 M_\pi}^\infty d\mu \mu e^{- \mu r} \, (3 + 3 \mu r + \mu^2 r^2) \, \rho_T (\mu), \nonumber \\       
       V_S (r) &=& \frac{1}{6 \pi^2 r} \int_{3 M_\pi}^\infty d\mu \mu e^{- \mu r} \, \big( \mu^2 \rho_T (\mu ) - 3 \rho_S  (\mu) \big),
        \eeqa
and the analogous expressions for the isovector potentials. 
        	Notice that Eq.~(\ref{Disp_CoordSpace}) also involve admixtures of the short-range contributions, which result from the large-$\mu$ integration region. The chiral expansion of the spectral functions is not expected to converge for values of $\mu$ comparable to the breakdown scale $\Lambda_b$ \cite{Epelbaum:2003gr, Epelbaum:2003xx}. Accordingly, it may be advantageous to remove the corresponding short-range admixtures by using an appropriate regulator as done in Refs.~\cite{Reinert:2017usi}.
	
	\subsection{Calculation of the spectral functions using the Cutkosky cutting rules}
	\label{Subsec:Frame1}
	
	To calculate the spectral functions for the $3\pi$-exchange potential entering Eqs.~(\ref{VMomSpace}) starting from the two-loop expressions in terms of the HB nucleon propagators,  we follow the approach by Kaiser \cite{Kaiser:1999ff,Kaiser:1999jg,Kaiser:2001dm} that makes use of the Cutkosky cutting rules \cite{Cutkosky:1960sp}.
	\begin{figure}[t!]
		\centering
		\includegraphics[width=0.35\linewidth]{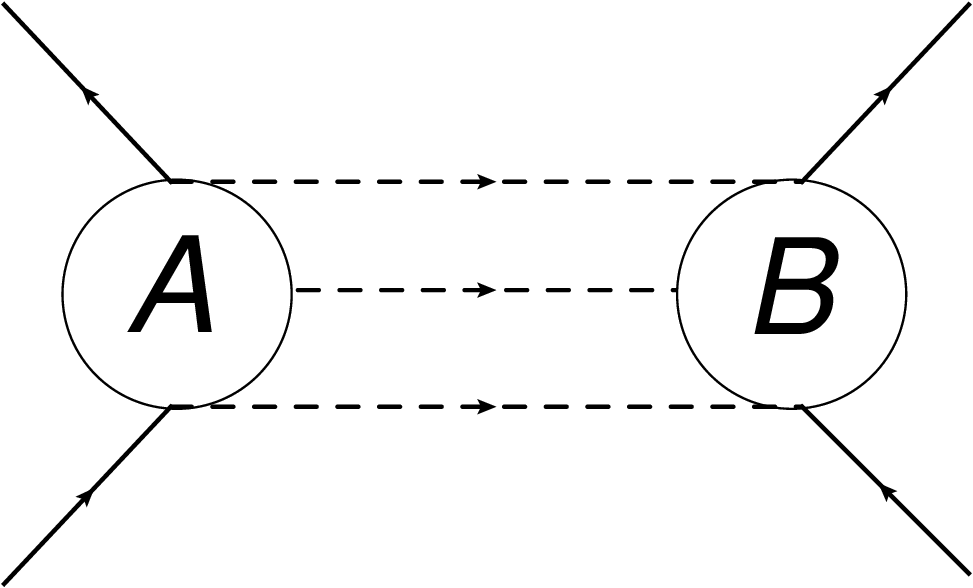}
		\caption{Schematic picture of a $N \bar{N} \rightarrow 3\pi \rightarrow N \bar{N}$ reaction.}
		\label{fig:3pephasespace}
	\end{figure}
	Specifically, the $NN \to NN$ scattering amplitude is considered for time-like momentum transfer $q\cdot q = \mu^2 \ge 9 M_{\pi}^2$, which is technically achieved by performing an analytic continuation to complex-valued momenta $\vec{q}\;  = i \mu\,\vec{e}$, where $\vec{e}$ is some real unity vector. Applying the Cutkosky cutting rules yields the imaginary part of the amplitude as an integral of the squared $\bar{N} N \rightarrow 3 \pi$ on-shell transition amplitudes over the Lorentz-invariant three-pion phase space, see Fig.~\ref{fig:3pephasespace}
	\begin{align}
		\label{Cutkosky}
		\text{Im} \left[ \int \frac{d^4l_1}{(2\pi)^4}\frac{d^4l_2}{(2\pi)^4}\frac{d^4l_3}{(2\pi)^4}(2\pi)^4\delta(l_1+l_2+l_3-q)A\frac{1}{l_1^2-M_{\pi}^2}\frac{1}{l_2^2-M_{\pi}^2}\frac{1}{l_3^2-M_{\pi}^2}B\right] = \frac{1}{2} \int d \Gamma_3 AB\,,
	\end{align}
	where $l_i$ are pion momenta while $A$ and $B$ denote the corresponding transition amplitudes.
	
	Following Ref.~\cite{Kaiser:1999jg}, the phase-space integral on the right-hand side of Eq.~(\ref{Cutkosky}) is most conveniently carried out in the three-pion center-of-mass frame. Specifically, we choose the nucleon incoming and outgoing four-momenta to be 
	\begin{align}
		p_1 =\begin{pmatrix}
			\mu/2\\
			p\; \vec{v}
		\end{pmatrix}, \qquad 	p_1' =\begin{pmatrix}
			-\mu/2\\
			p\; \vec{v}
		\end{pmatrix}
	\end{align}
	and the three pions four vectors are
	\begin{align}
		l_1 =\begin{pmatrix}
			\omega_{1}\\
			\vec{l}_1
		\end{pmatrix}, \qquad 	l_2 =\begin{pmatrix}
			\omega_{2}\\
			\vec{l}_2
		\end{pmatrix},\qquad 	l_3 =\begin{pmatrix}
			\mu - \omega_{1}-\omega_{2}\\
			-\vec{l}_1-\vec{l}_2
		\end{pmatrix}\;.
	\end{align}
	Here, $\vec{v}$ is a unit vector and $p = \sqrt{\mu^2/4-m^2} = i m +
	\mathcal{O} (m^{-1})$, where $m$ is the nucleon mass. 
	For the nucleon propagators entering the expressions for the amplitudes $A$ and $B$, it follows:
	\begin{align}
		\frac{2m}{\left(p_1-l_i\right)^2 - m^2} = \frac{1}{i \vec{v} \cdot \vec{l}_i-\epsilon}, \qquad	\frac{2m}{\left(p_1'+l_i\right)^2 - m^2} = \frac{-1}{i \vec{v} \cdot \vec{l}_i+ \epsilon}\,,
	\end{align}
	with $\epsilon = 0^+$. 
	
	After setting the notation, we move on with simplifying the phase-space integral. Once we applied the Cutkosky rules, we can perform the integration over the time-component in order to obtain the Lorentz-invariant phase space. Thus, our starting expression reads: 
	\beqa 	
	\int d \Gamma_3  AB &=& \int \frac{d^3l_1}{(2\pi)^3 2\omega_1} \frac{d^3l_2}{(2\pi)^32\omega_2} \frac{d^3l_3}{(2\pi)^32\omega_3}(2\pi)^4\delta^{(3)}\big(\vec{l_3}+(\vec{l_1}+\vec{l_2})\big)\delta\big(\mu-(\omega_1+\omega_2+\omega_3)\big) AB\,,\nn
	&=&\int \frac{d^3l_1}{(2\pi)^3 2\omega_1} \frac{d^3l_2}{(2\pi)^32\omega_2} \frac{2\pi}{2\omega_3}\delta\big(\mu-(\omega_1+\omega_2+\omega_3)\big) AB\Big|_{\vec{l}_3=-\vec{l}_1-\vec{l}_2}\,,\label{phase:space:integral:general}
	\eeqa
	where $\omega_i = \sqrt{\vec l_i{}^2 + M_\pi^2}$. 
	To simplify Eq.~(\ref{phase:space:integral:general}) we first
	express $\vec l_1$  and $\vec l_2$ in spherical coordinates and choose the $z$-axis to align in the direction of the nucleon three-momenta, thus $\hat{v} = (0,0,1)$.
	In this frame, the unit vectors of $\vec{l}_1$ and $\vec{l}_2$ can be written as 
	\begin{align}
		\hat{l}_1 = \begin{pmatrix}
			\sqrt{1-x^2} \cos\phi_1\\
			\sqrt{1-x^2} \sin\phi_1\\
			x
		\end{pmatrix}, \quad \hat{l}_2 = \begin{pmatrix}
			\sqrt{1-y^2} \cos\phi_2\\
			\sqrt{1-y^2} \sin\phi_2\\
			y
		\end{pmatrix}, \quad \hat{v} = \begin{pmatrix}
			0\\
			0\\
			1
		\end{pmatrix}
	\end{align}
	with 
	\begin{align}
		\label{KinVar}
		x := \hat{l}_1\cdot \hat{v} = \cos \theta_1,\quad y := \hat{l}_2\cdot \hat{v}= \cos \theta_2, \quad \tilde z := \hat{l}_1 \cdot \hat{l}_2 = x y + \sqrt{1-x^2}\sqrt{1-y^2}\cos(\phi_1 - \phi_2)\,.
	\end{align}
	The integrand of Eq.~(\ref{phase:space:integral:general}) is a scalar function, so it depends on the absolute values $|\vec{l}_1|, \, |\vec{l}_2 |$ and the angles $x,y, \tilde z$. Accordingly, it depends on $\phi_1$ and $\phi_2$ only through $\cos(\phi_1-\phi_2)$. Due to periodicity of cosine, the integral over $\phi_2$ over the whole period does not depend on $\phi_1$:
	\beqa
	\frac{\partial}{\partial \phi_1}\int_0^{2\pi}d\phi_2 f\big(\cos(\phi_1-\phi_2)\big)&=&-\int_0^{2\pi}d\phi_2 \frac{\partial}{\partial \phi_2} f\big(\cos(\phi_1-\phi_2)\big)\,=\,-\Big[f\big(\cos(\phi_1-\phi_2)\big)\Big]_{\phi_2=0}^{\phi_2=2\pi}\,=\,0\;,
	\eeqa
	where $f$ here is an arbitrary differentiable function. Thus, we can set $\phi_1=0$ in the integrand of Eq.~(\ref{phase:space:integral:general}) without affecting the result, so that the integration over $\phi_1$ becomes trivial and gives a factor of $2\pi$. Due to the $\delta$-function, we can trivially perform the integral over $\phi_2$. The only non-zero contribution to the integral emerges if
	\beqa
	\mu&=&\omega_1+\omega_2+\sqrt{l_1^2+l_2^2+2l_1 l_2 z + M_\pi^2}\;,
	\eeqa
	giving
	\beqa
	z&=&\frac{(\mu-\omega_1-\omega_2)^2-l_1^2-l_2^2-M_\pi^2}{2 l_1 l_2}\;,\label{delta:zero}
	\eeqa
	which fixes $\cos(\phi_2^0)$. Here, we define $\phi_2^0$ as the value of $\phi_2$ for which Eq.~(\ref{delta:zero}) is satisfied. Note, that the value of $\phi_2^0$ is not unique.  The solutions always come in pairs, except for one special case to be discussed below.  The first soltion is between $0$ and $\pi$, while the second one is between $\pi$ and $2\pi$. The only exception is the case of $\cos(\phi_0)=-1$, for which there is only one solution $\phi_2^0=\pi$. If we exclude this point, the value of the integral does not change. Let us denote for the case $\cos(\phi_0)\neq -1$ the solutions of Eq.~(\ref{delta:zero}) by $\phi_2^{0,1}$ and $\phi_2^{0,2}$. To evaluate the integral we rewrite the $\delta$-function to
	\beqa
	\delta\big(\mu-(\omega_1+\omega_2+\omega_3)\big)\Big|_{\vec{l}_3=-\vec{l}_1-\vec{l}_2}&=&\frac{\omega_3|_{\tilde z=z}}{l_1 l_2\sqrt{1-x^2}\sqrt{1-y^2}|\sin(\phi_2^0)|}\big(\delta(\phi_2-\phi_2^{0,1}) + \delta(\phi_2-\phi_2^{0,2}) \big)\;,\label{delta:simplified}
	\eeqa
	where we used
	\beqa
	\delta\big(f(x)\big)&=&\sum_j\frac{\delta(x-x_j)}{|f^\prime(x_j)|}\;,
	\eeqa
	with $f(x_j)=0$. Using the identity $\cos^2(\phi_1-\phi_2) = (\tilde z-xy)^2 (1-x^2)^{-1}  (1-y^2)^{-1}$, we can rewrite Eq.~(\ref{delta:simplified}) to
	\beqa
	\delta\big(\mu-(\omega_1+\omega_2+\omega_3)\big)\Big|_{\vec{l}_3=-\vec{l}_1-\vec{l}_2}&=&\frac{\omega_3|_{\tilde z=z}}{l_1 l_2\sqrt{1-x^2-y^2-z^2+2 x y z}}\big(\delta(\phi_2-\phi_2^{0,1})+\delta(\phi_2-\phi_2^{0,2})\big)\;.
	\eeqa
	Finally, we rewrite the radial integration over $l_1$ and $l_2$ into the integration over $\omega_1$ and $\omega_2$ by using
	\beqa
	l_j dl_j&=&\frac{1}{2}d\omega_j^2\,=\,\omega_j d\omega_j\;,
	\eeqa
	with $j=1,2$. With this we arrive at
	\begin{align}
		\int d\Gamma_3 \, AB(\dots) = \frac{1}{64 \pi ^4}\iint d\omega_1 d\omega_2 \iint\frac{dx dy}{\sqrt{1-x^2-y^2-z^2 + 2xyz}}\, AB(\dots) \,.\label{eq_frame1PS}
	\end{align} 
	
	We are now in the position to specify the integration boundaries.  For the $y$ integration, we can use that $1-x^2-y^2-z^2 + 2xyz >0$, yielding the values
	\beqa
	y_{\textrm{min}} &=& x z - \sqrt{1-x^2-z^2+x^2\;z^2}\,,\nonumber \\
	y_{\textrm{max}} &=& x z + \sqrt{1-x^2-z^2+x^2\;z^2}\,,
	\eeqa
	while for the $x$-integration we still have $x_{\textrm{min}} = -1$ and $x_{\textrm{max}} = 1$. For the integration over the pion energies, we obtain
	\begin{align}
		\label{IntFrame1Temp1}
		\iint \limits_{z^2<1} d\omega_1 d\omega_2 \, f(\omega_1, \, \omega_2) = \int_{M_{\pi}}^{\frac{\mu^2-3M_{\pi}^2}{2\mu}}d \omega_{1}\int_{\omega_{2\textrm{min}}}^{\omega_{2\textrm{max}}} d\omega_{2}\, f(\omega_1, \, \omega_2) \, ,
	\end{align} 
	where
	\begin{align}
		\omega_{2\textrm{min}}& = \frac{1}{2} \left(\mu-\omega _1 -\sqrt{\frac{\left(\omega _1^2-M_{\pi}^2\right) \left(\mu  \left(\mu -2 \omega _1\right)-3 M_{\pi}^2\right)}{\mu  \left(\mu -2 \omega _1\right)+M_{\pi}^2}}\right), \\
		\omega_{2\textrm{max}}& = \frac{1}{2} \left(\mu -\omega _1+\sqrt{\frac{\left(\omega _1^2-M_{\pi}^2\right) \left(\mu  \left(\mu -2 \omega _1\right)-3 M_{\pi}^2\right)}{\mu  \left(\mu -2 \omega _1\right)+M_{\pi}^2}}\right)\,.\label{eq_IntegrationOmega1andOmega2}
	\end{align}
	
	Alternatively, one can use the invariant mass $\omega$ for pions $2$ and $3$, i.e.~$\omega^2 = (l_2+l_3)^2 = \mu^2 - 2 \mu \omega_1 + m_\pi^2$, instead of the energy $\omega_1$. Then, Eq.~(\ref{IntFrame1Temp1}) can be cast into the form
	\begin{align}
		\iint \limits_{z^2<1} d\omega_1 d\omega_2 \, f(\omega_1, \, \omega_2)  = \int_{2M_{\pi}}^{\mu-M_{\pi}}dw \frac{w}{\mu}\int_{\hat{\omega}_{2\textrm{min}}}^{\hat{\omega}_{2\textrm{max}}} d\omega_{2} \, f(\omega_1, \, \omega_2)\,, 
	\end{align} 
	with
	\begin{align}
		\hat{\omega}_{2\textrm{min}} = \frac{w^2+\mu ^2-M_{\pi}^2}{4 \mu } -\frac{\sqrt{\left(w^2-4 M_{\pi}^2\right) \lambda(w^2,M_{\pi}^2,\mu^2)}}{4 \mu  w}\,,\nonumber \\
		\hat{\omega}_{2\textrm{max}} = \frac{w^2+\mu ^2-M_{\pi}^2}{4 \mu } +\frac{\sqrt{\left(w^2-4 M_{\pi}^2\right) \lambda(w^2,M_{\pi}^2,\mu^2)}}{4 \mu  w}\,,
	\end{align}
	where $\lambda(x,y,z) = x^2 + y^2 + z^2  -2 x y  - 2 x z- 2y z$ is the K\"allen function. Here we have introduced the hat-notation in order to distinguish between the two integration boundaries for $\omega_{2}$.

	\subsection{Reduction of tensor integrals}
	
	Having discussed the simplification of the integration over the three-pion phase space, we are now in the position to address the treatment of tensor loop integrals emerging after performing the spin-momentum algebra for the amplitudes $A$ and $B$.  Tensor reduction can be easily performed by utilizing a covariant notation for the nucleon spin operator $S_\mu = \frac{i}{2} \gamma_5 \sigma_{\mu\nu} v^\mu$. Notice that in the nucleon rest frame with $v^\mu = (1, \vec 0)$, one has
	\beq
	\label{Spin}
	S^\mu = \bigg( 0, \, \frac{\vec \sigma}{2} \bigg)\,.
	\eeq
	The treatment of the integrals like
	\begin{align}
		\int d \Gamma_3	l_i \cdot S_1\; l_i \cdot S_2 = S_{1\mu} S_{2\nu} \int d \Gamma_3 l_i^{\mu}l_j^{\mu}
	\end{align}
	can be carried out as follows. We start with the generic tensor integral 
	\begin{align}
		\int d\Gamma_3 l_i^\mu l_j^{\nu} = - X g^{\mu \nu} + Y q^\mu q^\nu + Z v^\mu v^\nu + \dots\label{eq_PSReduction}
	\end{align}
	and obtain the scalar functions $X$ and $Y$ using the projection operator $P^{X}_{\mu \nu}$ on the spin-spin part
	\begin{align}
		\label{PX}
		P^{X}_{\mu \nu} = \frac{1}{2} \left(\frac{q_{\mu} q_{\nu}}{\mu^2}+ v_{\mu} v_{\nu}- g_{\mu \nu}\right)\,,
	\end{align}
	and $P^{Y}_{\mu \nu}$ on the tensor part
	\begin{align}
		P^{Y}_{\mu \nu} = \frac{1}{2 \mu^2} \left(\frac{3 q_{\mu} q_{\nu}}{\mu^2}+ v_{\mu} v_{\nu}- g_{\mu \nu}\right)\,.\label{eq_tensorReduction_Projection_TensorPart}
	\end{align}
	This leads to scalar integrals, where the various scalar products involving the pion momenta $l_i$ can be expressed according to the chosen frame. For example, one finds using $v = (0, i \hat v)$:
	\begin{align}
		l_1\cdot l_1 &= l_2 \cdot l_2 = M_{\pi}^2, \quad 	l_1 \cdot l_2 = \omega_{1} \omega_{2} - |\vec{l}_1| |\vec{l}_2| z, \quad 	q\cdot q = \mu^2, \quad 	l_1 \cdot q = \omega_1\: \mu, \quad 	l_2 \cdot q = \omega_2\: \mu, \quad 	q\cdot v = 0\nonumber \\
		l_1 \cdot v &= - i \vec{l}_1 \cdot \vec{v} = -i |\vec{l}_1| x,\quad l_2 \cdot v = - i \vec{l}_2 \cdot \vec{v} = -i |\vec{l}_2| y\,.
	\end{align}

	\subsection{$3\pi$-exchange potential in coordinate space using the Wick-rotation method}
	\label{sec:Wick}
	
	As already emphasized above, we aim at the derivation of the spectral functions for the $3\pi$-exchange contributions. The dispersive spectral representation provides a convenient interface for obtaining the corresponding potentials both in coordinate and momentum spaces. However, for certain classes of diagrams with a large number of nucleon propagators, the spectral representation obtained as described in sec.~\ref{Subsec:Frame1} turns out to be singular \cite{Kaiser:1999jg}. To be more precise, carrying out the spin-momentum and isospin algebra leads to the expressions proportional to $(1-z^2)^{-3/2}$, which cannot be further integrated. To deal with this issue, we follow the approach suggested by Kaiser in Ref.~\cite{Kaiser:2001dm} and refrain from using the method of sec.~\ref{Subsec:Frame1}. Specifically, we compute the coordinate-space representation of the potential in the rest-frame of the nucleons with $v = (1, \vec 0)$ by first performing a Wick-rotation of the pion loop momenta $l_i$, $i \in \{1, 2\}$, by replacing $l_{i 0} \to i l_{i0}$, evaluating the Fourier-transform of $V_{3 \pi} (q)$ with fixed values of $l_{i 0}$ analytically and, finally, performing the remaining integrations over  $l_{i 0}$. To be more precise, consider
	\begin{align}\label{WickRotationFirstSimp}
		V_{3 \pi}(r) &= \int \frac{d^3q}{(2 \pi)^3} e^{i \vec{q}\cdot \vec{r}} \int \frac{d^4l_1}{(2 \pi)^4}\int \frac{d^4l_2}{(2 \pi)^4} \, f
		&= \int \frac{d^3q}{(2 \pi)^3} e^{i \vec{q}\cdot \vec{r}} \int \frac{d^4l_1}{(2 \pi)^4}\int \frac{d^4l_2}{(2 \pi)^4}  \int \frac{d^3l_3}{(2 \pi)^3} \int \frac{d^3x}{(2 \pi)^3} e^{i \vec{x}\cdot\left(\vec{l}_3-\vec{q}+\vec{l}_1-\vec{l}_2\right)}\, f\,,
	\end{align}
	where $f$ is a function of the momenta $l_1$, $l_2$ and $q$ as well as the pion mass.  Performing the Wick rotation, the pion propagators entering the function $f$ turn to
	\begin{align}
		\frac{1}{l_i^2-M_{\pi}^2+ i \epsilon} = 	\frac{1}{l_{i0}^2-\vec{l_i}{}^2-M^2_{\pi}+ i \epsilon}\; \; \rightarrow \; \;  \frac{-1}{\vec{l_i}{}^2+M^2_{i}}\,,
	\end{align}
	where we have  introduced a short-hand notation $M^2_{i}  \equiv l_{i 0}^2+ M_{\pi}^2$. Similarly, for the nucleon propagators one obtains 
	\begin{align}
		\frac{1}{v \cdot l_{i}+i \epsilon} = 	\frac{1}{ l_{i 0}+i \epsilon} \; \; \rightarrow \; \; \frac{-i}{ l_{i 0}+ \epsilon}\,.
	\end{align}
	The singular spectral representation typically appears for the classes of diagrams with factorizable one-pion exchanges, i.e., where the dependence of the function $f$ on $\omega_1$, $\omega_2$ and $\omega_3$ has the form of either $1/(\omega_i^2 \omega_j^2 \omega_k^2)$ or $1/(\omega_i^2 \omega_j^4 \omega_k^2)$.\footnote{The same kind of factorization occurs for the ring-type three-nucleon force contributions, see Ref.~\cite{Bernard:2007sp}.} Then, the integrals over $\vec l_1$ and $\vec l_2$ can be easily performed using
	\begin{align}
		\int  \frac{d^3l_i}{(2\pi)^3} \frac{ e^{i \vec{l}_i \cdot \vec{x}}}{\vec{l}_i{}^{2}+M_i^2}=  \frac{e^{-M_i x}}{4 \pi x},	\qquad \int  \frac{d^3l_i}{(2\pi)^3} \frac{ e^{i \vec{l}_i \cdot \vec{x}}}{\big(\vec{l}_i{}^{2}+M_i^2\big)^2}=  \frac{e^{-M_i x}}{8 \pi M_i}\,.\label{eq_FT_PionProp_WickRo}
	\end{align}
	By introducing the new variables
	\begin{align}
		\label{alphabetagamma}
		\alpha = \sqrt{1+\zeta_1^2}, \qquad \beta =  \sqrt{1+\zeta_2^2}, \qquad 	\gamma = \sqrt{1+\left(\zeta_1-\zeta_2\right)^2}\,,
	\end{align}
	we can express $M_i$ as:
	\begin{align}
		M_i = M_{\pi} \sqrt{1+ \frac{l_{i0}^2}{M_{\pi}^2}} \equiv  M_{\pi} \sqrt{1+ \zeta_i^2}\,,
	\end{align}
	so that the integration over $i_{i0}$ to be performed at the very last step is written in terms of the integral over dimensionless variables  $\zeta_i$ using $dl_{0,i} = d \zeta_i  M_{\pi}$. Notice that the heavy-baryon propagators entering the function $f$ have the form 
	\begin{align}
		\frac{1}{ l_{i 0}+ \epsilon} \; \; \rightarrow \; \; \frac{1}{\zeta_{i} M_{\pi}}\,,
	\end{align}
	leading to principal value integrals once $\epsilon$ is set to zero. Finally, we notice that the integration over $x$ and $q$ is
	trivial given that 
	\begin{align}
		\int \frac{d^3x}{\left(2\pi\right)^3} f\left(\vec{x}\right)	\int \frac{d^3q}{\left(2\pi\right)^3} e^{i \vec{q}\cdot \left(\vec{x}-\vec{r}\right)} = \int \frac{d^3x}{\left( 2\pi\right)^3}  f\left(\vec{x}\right)	\delta \left(\vec{x}-\vec{r}\right) = f\left(\vec{r}\, \right)\,.	
	\end{align}
	Using the above-mentioned technique, the final expressions for the coordinate-space potentials can be given in terms of (principal-value) integrals over $\zeta_1$ and $\zeta_2$. Last but not least, we further emphasize that positive powers of loop momenta that may appear in the function $f$ can be easily expressed in terms of gradients with respect to the corresponding relative distances, which can be evaluated at the very end (see also Ref.~\cite{Bernard:2007sp} for a similar treatment in the case of the three-nucleon force).

	\section{Chiral $3 \pi$-exchange at order N$^3$LO}
\setcounter{equation}{0}
        \def\theequation{\arabic{section}.\arabic{equation}}
        \label{SecIII}	
	
	\subsection{Class-I diagrams}
	
	We start with the class-I diagrams shown in the upper row of Fig.~\ref{fig:3pi_N3LO}. These diagrams do not involve reducible-like topologies, so that their contributions to the NN potential can be conveniently computed by calculating the corresponding Feynman diagrams. The corresponding expressions can be obtained using the Feynman rule
	\begin{align}
		\label{FeynRuleTemp1}
		\frac{g_\mathrm{A}}{4 F_{\pi}^3}\vec{\sigma}\cdot\left\{\tau^a\delta^{bc}\left[4 \alpha \vec{q}_1+ \left(4 \alpha-1\right)\left(\vec{q}_2+ \vec{q}_3\right)\right] + two\; cycl.\; perm.\right\}\,,
	\end{align}
	for the $\pi\pi\pi NN$ vertex and 
	\begin{align}
		\frac{i}{F_{\pi}^2}\left\{\delta^{ab} \delta^{cd}\left[\left(q_1+q_2\right)^2-M_{\pi}^2 + 2 \alpha\left(4M_{\pi}^2-q_1^2-q_2^2-q_3^2-q_4^2\right)\right] + two\; cycl.\; perm.\right\}\,,
	\end{align}
	for the $4 \pi$-vertex. Here, $F_\pi$ is the pion decay constant\footnote{Strictly speaking, the pion decay constant in the effective chiral Lagrangian should be taken at the chiral limit. However, in the calculation of the $3\pi$-exchange potential at orders N$^3$LO and N$^4$LO we are interested in here, the renormalization of the parameters in the lowest-order Lagrangians $\mathcal{L}_\pi^{(2)}$ and $\mathcal{L}_{\pi N}^{(1)}$ is trivial and for this reason we replace, from the very beginning, the various chiral-limit quantities with their physical values.}, the superscripts denote the pion isospin indices while $q_i$ refer to the outgoing pion momenta. Notice that 
	Eq.~(\ref{FeynRuleTemp1}) is given in the nucleon rest frame, but it can be easily generalized to arbitrary frame by expressing the Pauli matrices $\vec \sigma$  in terms of the covariant spin operator using Eq.~(\ref{Spin}). Further, $\alpha$ is an arbitrary parameter that reflects a freedom in the choice of interpolating fields for pions, i.e., the ambiguity in parametrizing the SU(2)-matrix $U (\vec \pi )$, which  transforms under chiral rotations as  $U \to R U L^\dagger$ and is used in the construction of the effective Lagrangian, in terms of the pion fields:
	\beq
	U = 1 + \frac{i}{F_\pi} \vec \tau \cdot \vec \pi \bigg(1 - \alpha \frac{\pi^2}{F^2_\pi} \bigg) - \frac{\pi^2}{2 F_\pi^2} \bigg[ 1 + \bigg( \frac{1}{4} - 2 \alpha \bigg) \frac{\pi^2}{F_\pi^2} \bigg] + \mathcal{O} (\vec \pi\,  ^5)\,.
	\eeq
	Clearly, observable quantities and nuclear potentials must be independent of the arbitrary parameter $\alpha$. An explicit verification of $\alpha$-independence thus provides an important consistency check of the calculation. As pointed out in Ref.~\cite{Kaiser:1999ff}, the contribution of the sunset diagram alone (the first diagram in Fig.~\ref{fig:3pi_N3LO}) is $\alpha$-dependent. In contrast, when considering the contribution of all class-I  diagrams, the $\alpha$-dependence drops out.   
	
	Combining the expressions from all class-I diagrams together and projecting the tensor integral on the spin-spin part using Eq.~(\ref{PX}), we obtain the intermediate expression for the corresponding spectral function 
	\begin{align}
		\textrm{Im}\; W^{\rm I}_S(i \mu)= \frac{-g_{\textrm{A}}^2}{6 \left(32 \pi^2 F_\pi^3\right)^2 } \iint\limits_{z^2<1} d\omega_{1} d \omega_{2} \int_{-1}^{1}dx \int_{y_{\textrm{min}}(x,z)}^{y_{\textrm{max}}(x,z)}  dy\, \frac{\vec{l}_1^{\;2}\left(-1+x^2\right)+\vec{l}_2^{\;2}\left(-1+y^2\right)+ |\vec{l}_1| |\vec{l}_2| \left( x y -z\right)}{\sqrt{1-x^2-y^2-z^2 + 2xyz}}\,.
	\end{align}
	Notice that this result is, as expected, independent of $\alpha$. After performing the angular integrations, changing the variable $\omega_1$ to $\omega$ as discussed in sec.~\ref{Subsec:Frame1} and integrating over $\omega_2$, we obtain the final result in the form
	\begin{align}
		\label{Class1SpinSpinFinal}
		\textrm{Im}\; W^{\rm I}_S(i \mu) =\frac{g_A^2}{6 \mu^4\left(16\pi F_\pi^2 \right)^3} \int_{2M_{\pi}}^{\mu-M_{\pi}} dw
		\sqrt{\left(w^2-4M_{\pi}^2\right)}\sqrt{\lambda^3(w^2, M_{\pi}^2,\mu^2)}\,,
	\end{align}
	which agrees with Eq.~(12) of  Ref.~\cite{Kaiser:1999ff}.
	
	Similarly, using the projection operator on the tensor part $P_{\mu \nu}^Y$ given in Eq.~(\ref{eq_tensorReduction_Projection_TensorPart}), we obtain for the corresponding isovector tensor spectral function the expression 
	\begin{align}
		\textrm{Im}\; W^{\rm I}_T(i \mu) =&\frac{g_A^2 \left(M_{\pi}^2-\mu^2\right)^{-2}}{6\left(32 \pi^2F_\pi^3 \mu\right)^2}\iint\limits_{z^2<1} d \omega_{1} d \omega_{2}  \int_{-1}^{1}dx \int_{y_{\textrm{min}}(x,z)}^{y_{\textrm{max}}(x,z)} dy \frac{1}{\sqrt{1-x^2-y^2-z^2+2x y z}}\left[\vec{l}_1^{\;2} \left(24 \mu ^2 \vec{l}_2^{\;2} \left(z^2+1\right) \right. \right. \nonumber \\
		&\left. \left. +2 \mu  M_{\pi}^2 \left(\mu  \left(x^2+13\right)+4 \omega _2\right)-\left(M_{\pi}^4 \left(x^2-3\right)\right)-\mu ^3 \left(\mu  \left(x^2-19\right)+24 \omega _2\right)\right)+|\vec{l}_1||\vec{l}_2| \left(2 \mu  M_{\pi}^2 \left(\mu  (x y-17 z) \right. \right. \right. \nonumber \\
		&\left.\left. \left. -4 \left(\omega _1+\omega _2\right) z\right)+M_{\pi}^4 (z-x y)-\left(\mu ^4 (x y+7 z)\right)+24 \mu ^2 z \left(\mu  \omega _2+\omega _1 \left(\mu -2 \omega _2\right)\right)\right)+\vec{l}_2^{\,2} \left(2 \mu  M_{\pi}^2 \left(\mu  \left(y^2+13\right)\right. \right. \right. \nonumber \\
		&\left.\left. \left. +4 \omega _1\right)-\left(M_{\pi}^4 \left(y^2-3\right)\right)-\mu ^3 \left(\mu  \left(y^2-19\right)+24 \omega _1\right)\right)+2 \left(\mu ^4 \left(\omega _1 \left(13 \omega _2-7 \mu \right)+\mu  \left(2 \mu -7 \omega _2\right)\right) \right. \right. \nonumber \\
		&\left. \left. +2 \mu ^2 M_{\pi}^2 \left(\omega _1 \left(7 \omega _2-13 \mu \right)+\mu  \left(12 \mu -13 \omega _2\right)\right)+M_{\pi}^4 \left(\omega _1 \left(5 \mu +\omega _2\right)+\mu  \left(23 \mu +5 \omega _2\right)\right)+2 M_{\pi}^6\right)\right]\,.
	\end{align}
	Using the same steps as for the spin-spin part described above, the expression simplifies to 
	\begin{align}
		\label{ClassITemp1} 
		\textrm{Im}\; W^{\rm I}_T(i \mu) =&\frac{2g_A^2 \left(M_{\pi}^2-\mu^2\right)^{-2}}{27 \left(16\pi F_\pi^2 \mu^2 \right)^3} \int_{2M_{\pi}}^{\mu-M_{\pi}} dw\sqrt{\left(w^2-4M_{\pi}^2\right)}\sqrt{\lambda(w^2, M_{\pi}^2,\mu^2)}\left[3 M_{\pi}^4 \left(w^2-M_{\pi}^2\right){}^2 \right.\nonumber \\
		&\left. +3 \mu ^2 M_{\pi}^2 \left(3 M_{\pi}^4+4 w^4-8 M_{\pi}^2 w^2\right)+\mu ^4 \left(37 M_{\pi}^4+21 w^4-42 M_{\pi}^2 w^2\right)-5 \mu ^6 M_{\pi}^2-2 \mu^8\right]\,.
	\end{align}
	This expression differs from the one quoted in Ref.~\cite{Kaiser:1999ff}, but the difference is actually caused by the contribution to the integrand that is antisymmetric with respect to re-labelling the pions $1$ and $2$, which leads to a vanishing result for the spectral function, e.g.: 
	\begin{align}
		\iint\limits_{z^2<1} d \omega_{1} d \omega_{2}f(M_{\pi}, \mu) \left( \omega_{1}-\omega_{2}\right) = \int_{2M_{\pi}}^{\mu -M_{\pi}} dw 	\sqrt{\left(w^2-4M_{\pi}^2\right)}\sqrt{\lambda(w^2, M_{\pi}^2,\mu^2)}f(M_{\pi}, \mu) \left(3M_{\pi}^2-3w^2+\mu^2\right) = 0\,,\label{eq_diff_w1w2_phase_space_int_zero}
	\end{align} 
	where $f(M_{\pi}, \mu)$ is an arbitrary function depending on $M_{\pi}$ and $\mu$. It is then easy to bring Eq.~(\ref{ClassITemp1}) to the form given in Ref.~\cite{Kaiser:1999ff}:
	\begin{align}
		\textrm{Im}\; W^{\rm I}_T(i \mu)=&\frac{g_A^2 \left(M_{\pi}^2-\mu^2\right)^{-2}}{9 \left(16\pi F_\pi^2 \mu^2\right)^3} \int_{2M_{\pi}}^{\mu-M_{\pi}} dw
		\sqrt{\left(w^2-4M_{\pi}^2\right)}\sqrt{\lambda(w^2, M_{\pi}^2,\mu^2)}\nonumber \\
		&\times \left[3w^4 \left(7 \mu ^4+4 \mu ^2 M_{\pi}^2+M_{\pi}^4\right)-2 \mu^8-19 \mu^6 M_{\pi}^2-13 \mu ^4 M_{\pi}^4-17 \mu ^2 M_{\pi}^6-3 M_{\pi}^8\right]\,.
		\label{Class1_WT_mu}                                           
	\end{align}
	Notice that the pre-factor of $\left(M_{\pi}^2-\mu^2\right)^{-2}$ in the spectral function $\textrm{Im}\; W^{\rm I}_T(i \mu)$ is stemming from the pion double-pole contribution of the last diagram in the first raw of Fig.~\ref{fig:3pi_N3LO}.

	\subsection{Class-II diagrams}
	
	The treatment of the class-II contributions is analogous to the previously considered case of class-I diagrams and can, again, be carried out in terms of Feynman graphs. After performing the spin and isospin algebra one encounters the contributions to the isovector spin-spin and tensor parts, which can be obtained using the corresponding projection operators.  For the spin-spin spectral function, we find 
	\begin{align}
		\textrm{Im}\; W^{\rm II}_S(i \mu) =&\frac{2g_A^2}{\left( 16\pi^2  F_\pi^3\right)^2}\iint d \omega_{1}d \omega_{2} \int_{-1}^{1}dx \int_{y_{\textrm{min}}(x,z)}^{y_{\textrm{max}}(x,z)}dy\, \frac{x}{x^2+\epsilon^2} \frac{\left(|\vec{l}_1|x + 2 |\vec{l}_2| y\right)\left(|\vec{l}_1|\left(x^2-1\right)+2 |\vec{l}_2|\left(x y -z\right)\right)}{\sqrt{1-x^2-y^2-z^2+2 x y z}}\,.\label{eq_ImWsClassII_4dim_Rep}
	\end{align}
	Here, we encounter a principal value integral with respect to $x$, which originates from the nucleon propagator. After performing the angular integration, we are left with
	\begin{align}
		\textrm{Im}\; W^{\rm II}_S(i \mu) =-\frac{g_A^2}{3\left( 8\pi  F_\pi^2\right)^3}\iint\limits_{z^2<1} d \omega_{1}d \omega_{2}\left[\vec{l}_1^{\;2}+4|\vec{l}_1||\vec{l}_2| z+2\vec{l}_2^{\;2} \left(3 z^2-1\right)\right]\,.\label{eq_ImWsClassII_2dim_Rep}
	\end{align}
	Using the definition of $z$ in Eq.~(\ref{delta:zero}) and expressing the three momenta $\vec{l}_i$ in terms of energies $\omega_{i}$, one can perform the integration over $\omega_2$: 
	\begin{align}
		\textrm{Im}\; W^{\rm II}_S(i \mu)=&\frac{-g_A^2}{3 \mu^4 \left(16\pi F_\pi^2 \right)^3} \int_{2M_{\pi}}^{\mu-M_{\pi}} dw
		\sqrt{\left(w^2-4M_{\pi}^2\right)}\sqrt{\lambda(w^2, M_{\pi}^2,\mu^2)} \left(M_{\pi}^2-w^2-13\mu^2\right)\left(3 M_{\pi}^2+\mu ^2-3w^2\right)\,.
	\end{align}
	Finally, using Eq.~(\ref{eq_diff_w1w2_phase_space_int_zero}), only one term in the first brackets yields a non-vanishing contribution, and we end up with the expression given in Ref.~\cite{Kaiser:1999ff}:
	\begin{align}
		\label{Class2_WS_mu}
		\textrm{Im}\; W^{\rm II}_S(i \mu)=&\frac{g_A^2}{3 \mu^4 \left(16\pi F_\pi^2 \right)^3} \int_{2M_{\pi}}^{\mu-M_{\pi}} dw
		\sqrt{\left(w^2-4M_{\pi}^2\right)}\sqrt{\lambda(w^2, M_{\pi}^2,\mu^2)} w^2\left(3 M_{\pi}^2+\mu ^2-3w^2\right)\,.
	\end{align}
	
	For the tensor part, we find the intermediate result of 
	\begin{align}
		\textrm{Im}\; W^{\rm II}_T(i \mu) =&\frac{2g_A^2}{\left( 16\pi^2  F_\pi^3\right)^2}\iint\limits_{z^2<1} d \omega_{1}d \omega_{2} \int_{-1}^{1}dx \int_{y_{\textrm{min}}(x,z)}^{y_{\textrm{max}}(x,z)} dy \bigg\{\frac{x}{x^2+\epsilon^2} \frac{|\vec{l}_1| x + 2 |\vec{l}_2|y}{\sqrt{1-x^2-y^2-z^2+2 x y z}}  \nonumber \\
		& \times \left[2 |\vec{l}_1 ||\vec{l}_2| \left(M_{\pi}^2 (z-x y)+\mu  \left(\mu  x y-\mu  z-4 \omega _1 z\right)\right)+ |\vec{l}_1| \left(\mu  \left(\mu  \left(x^2-3\right)+8 \omega _2\right)-M_{\pi}^2 \left(x^2-3\right) \right) \right.\nonumber \\
		& \left.2 \left(-\mu ^2 \omega _1 \left(\mu -2 \omega _2\right)+M_{\pi}^2 \left(\omega _1 \left(\mu +2 \omega _2\right)-\mu  \left(\mu -4 \omega _2\right)\right)+M_{\pi}^4\right)\right]\bigg\}\,.\label{eq_ImWtClassII_4dim_Rep}
	\end{align}
	Performing the angular integrations and the integration over $\omega_2$, we find 
	\begin{align}
		\label{Class2_WT_mu}
		\textrm{Im}\; W^{\rm II}_T(i \mu) =&\frac{2g_A^2\left(\mu^2-M_{\pi}^2\right)^{-1}}{9 \mu^6 \left(16\pi F_\pi^2 \right)^3} \int_{2M_{\pi}}^{\mu-M_{\pi}} dw
		\sqrt{\left(w^2-4M_{\pi}^2\right)^3}\sqrt{\lambda(w^2, M_{\pi}^2,\mu^2)} \nonumber \\
		&\times w^{-2}\left[2 \left(M_{\pi}^3-M_{\pi} w^2\right)^2+\left(3 M_{\pi}^4-9 M_{\pi}^2 w^2+4 w^4\right) \mu ^2 +\mu ^4 w^2-5 \mu ^6\right]\,.
	\end{align}
	This result is identical to the expression given in Ref.~\cite{Kaiser:1999ff}:
	\begin{align}
		\textrm{Im}\; W^{\rm II}_T(i \mu)=&\frac{g_A^2}{6 \mu^6 \left(8 \pi F_\pi^2 \right)^3} \int_{2M_{\pi}}^{\mu-M_{\pi}} dw
		\sqrt{\left(w^2-4M_{\pi}^2\right)^3}\sqrt{\lambda(w^2, M_{\pi}^2,\mu^2)} \nonumber \\
		&\times\left[w^4M_{\pi}^2\left(\mu ^2-M_{\pi}^2\right)^{-1} -2 \mu ^2 w^2 +M_{\pi}^2 \left(\mu ^2+M_{\pi}^2\right) \left(3 \mu ^2+M_{\pi}^2\right)w^{-2}\right]\,.
	\end{align}

	\subsection{Class-III diagrams}
	
	We continue with the purely-irreducible class-III diagrams, which contain two heavy-baryon propagators as displayed in Fig.~\ref{fig:3pi_N3LO}. After performing the spin and isospin algebra, we only find nonvanishing contributions to the isovector spin-spin and tensor parts. For the spin-spin potential, we obtain after the tensor-reduction
	\begin{align}
		\textrm{Im}\; W^{\rm III}_S(i \mu) =&\frac{-g_A^2}{2 \left(32\pi^2  F_\pi^3\right)^2}\iint\limits_{z^2<1} d \omega_{1}d \omega_{2} \int_{-1}^{1}dx \int_{y_{\textrm{min}}(x,z)}^{y_{\textrm{max}}(x,z)}dy \, \frac{x y  \left(x y -z\right)}{\left(x^2+\epsilon^2\right)\left(y^2+\epsilon^2\right)}  \, \frac{2\vec{l}_1^{\;2} x^2 + 5 |\vec{l}_1| |\vec{l}_2| x y + 2 \vec{l}_2^{\,2}y}{\sqrt{1-x^2-y^2-z^2+2 x y z}}\,,
	\end{align}
	which leads to the following compact result after performing the angular integrations:
	\begin{align}
		\textrm{Im}\; W^{\rm III}_S(i \mu) =\frac{g_A^2}{3 \left(8\pi  F_\pi^2\right)^3}\iint\limits_{z^2<1} d \omega_{1}d \omega_{2}\left[5|\vec{l}_1||\vec{l}_2| z+\vec{l}_2^{\;2} \left(-2+6z^2\right)\right]\,.\label{eq_ClassIII_spectralFunction_2dim}
	\end{align}
	To obtain this result we made use of the identity
	\begin{align}
		- \int_{-1}^{1} dx \int_{y_{\textrm{min}}(x,z)}^{y_{\textrm{max}}(x,z)} dy \frac{1}{\sqrt{1-x^2-y^2-z^2+2 x y z}}\frac{x^2}{\left(i x - \epsilon\right)\left(i y + \epsilon\right)}  = 2 \pi z \,,\label{eq_PV_2PiZ}
	\end{align}
	which is derived in \cref{Appendix_Principle Value Integrals}. Performing the integration over $\omega_2$ we finally obtain  
	the expression in terms of the integral over the invariant mass $w$,
	\begin{align}
		\textrm{Im}\; W^{\rm III}_S(i \mu) =&\frac{-g_A^2}{18 \left(16 \mu^2 \pi F_\pi^2 \right)^3} \int_{2M_{\pi}}^{\mu-M_{\pi}} dw
		\sqrt{\left(w^2-4M_{\pi}^2\right)}\sqrt{\lambda^3(w^2, M_{\pi}^2,\mu^2)}\left(7w^2+8M_{\pi}^2\right)w^{-2} \,.
	\end{align}
	To compare,  the result given in Ref.~\cite{Kaiser:1999ff} has the form 
	\begin{align}
		\label{ClassIII_temp1}
		\textrm{Im}\; W^{\rm III}_S(i \mu)=&\frac{g_A^2}{18 \mu^4 \left(16\pi F_\pi^2 \right)^3} \int_{2M_{\pi}}^{\mu-M_{\pi}} dw
		\sqrt{\left(w^2-4M_{\pi}^2\right)}\sqrt{\lambda(w^2, M_{\pi}^2,\mu^2)}\left[9w^4 -5\mu^4 + 30\mu^2 M_{\pi}^2-9 M_{\pi}^4\right]\,.
	\end{align}
	To explicitly demonstrate that both results are identical we can proceed as follows: Starting by \cref{eq_ClassIII_spectralFunction_2dim}, we make use of the definition of $z$ and relable, whenever possible, $\vec{l}_2$ $(\omega_{2})$ to $\vec{l}_1$ $(\omega_{1})$. This yields:
	\begin{align}
		\label{ClassIII_temp2} 
		\textrm{Im}\; W^{\rm III}_S(i \mu)=&\frac{g_A^2}{18 \mu^4 \left(6\pi F_\pi^2 \right)^3} \int_{2M_{\pi}}^{\mu-M_{\pi}} dw
		\sqrt{\left(w^2-4M_{\pi}^2\right)}\sqrt{\lambda(w^2, M_{\pi}^2,\mu^2)}\nonumber \\
		& \times\left[3 \left(M_{\pi}^2-w^2\right)^2+ 4 \left(-19 M_{\pi}^2+22w^2\right)\mu^2-31\mu^4\right]\;.
	\end{align}
	Then, we notice that the difference between Eqs.~(\ref{ClassIII_temp1}) and (\ref{ClassIII_temp2}) is given by 
	\begin{align}
		\frac{g_A^2}{9 \mu^4 \left(16\pi F_\pi^2 \right)^3} \int_{2M_{\pi}}^{\mu-M_{\pi}} dw
		\sqrt{\left(w^2-4M_{\pi}^2\right)}\sqrt{\lambda(w^2, M_{\pi}^2,\mu^2)}\left(3M_{\pi}^2 - 44 \mu^2\right)\left(3M_{\pi}^2-3w^2+\mu^2\right)\;,
	\end{align}
	which is zero according to Eq.~(\ref{eq_diff_w1w2_phase_space_int_zero}). 
	
	For the isovector tensor part, we obtain after performing the tensor reduction 
	\begin{align}
		\textrm{Im}\; W^{\rm III}_T(i \mu) =&-\frac{g_A^2}{2 \left(32 \mu \pi^2  F_\pi^3\right)^2}\iint\limits_{z^2<1} d \omega_{1}d \omega_{2} \int_{-1}^{1}dx \int_{y_{\textrm{min}}(x,z)}^{y_{\textrm{max}}(x,z)} dy \frac{x y }{ \left(x^2+\epsilon^2\right)\left(y^2+\epsilon^2\right)}\frac{\left(2 |\vec{l}_1| x + |\vec{l}_2| y\right)\left( |\vec{l}_1| x +2 |\vec{l}_2| y\right)}{\sqrt{1-x^2-y^2-z^2+2 x y z}} \nonumber \\
		&\times\left[\left(x y -z\right)-2 \frac{\omega_{1} \omega_{2}}{	|\vec{l}_1| |\vec{l}_2| }\right]\;.
	\end{align}
	After the angular integration, the result simplifies to 
	\begin{align}
		\textrm{Im}\; W^{\rm III}_T(i \mu) =\frac{g_A^2}{3 \mu^2 \left( 8 \pi  F_\pi^2\right)^3}\iint\limits_{z^2<1} d \omega_{1}d \omega_{2}\left[15 \omega_{1} \omega_{2} + 5 |\vec{l}_1| |\vec{l}_2| z + 12 \frac{|\vec{l}_2|}{|\vec{l}_1|} \omega_{1} \omega_{2} z + \vec{l}_2^{\;2}\left(-2+6 z^2\right)\right]\;.
	\end{align}
	Note that in analogy to the spin part, we made use of Eq.~(\ref{eq_PV_2PiZ}). In terms of the invariant mass, we finally find for the isovector tensor part of class-III diagrams:
	\begin{align}
		\label{Class3_WT_mu}
		\textrm{Im}\; W^{\rm III}_T(i \mu) =&\frac{g_A^2}{9  \left(16  \mu^2 \pi F_\pi^2 \right)^3} \int_{2M_{\pi}}^{\mu-M_{\pi}} dw
		\sqrt{\left(w^2-4M_{\pi}^2\right)}\sqrt{\lambda(w^2, M_{\pi}^2,\mu^2)}w^{-2} \nonumber \\
		&\times\left[\left(7w^2+8M_{\pi}^2\right)\left(\mu^4 + \mu^2\left(w^2 + M_{\pi}^2\right)-2\left(w^2-M_{\pi}^2\right)^2\right)\right]\,.
	\end{align}
	For completeness, we give the function presented in Ref.~\cite{Kaiser:1999ff}:
	\begin{align}
		\textrm{Im}\; W^{\rm III}_T(i \mu)=&\frac{2g_A^2}{9 \mu^6 \left(18\pi F_\pi^2 \right)^3} \int_{2M_{\pi}}^{\mu-M_{\pi}} dw
		\sqrt{\left(w^2-4M_{\pi}^2\right)}\sqrt{\lambda(w^2, M_{\pi}^2,\mu^2)} \nonumber \\
		&\times\left[ 4 \mu^4 + 9 \mu^2 M_{\pi}^2+ 9 M_{\pi}^4-3w^4 + 6 M_{\pi}^2\left(\mu^4-M_{\pi}^4\right)w^{-2}\right]\,.
	\end{align}
	We have verified that both results are identical.

	\subsection{Class-IV diagrams}
	
	The last type of contributions proportional to $g_A^2$ stems from the class-IV diagrams. Here, we encounter for the first time the appearance of  reducible-like topologies, see class-VI diagrams (1) and (2) in Fig.~\ref{fig:3pi_N3LO}.  To calculate 
	the expressions for the class-IV $3\pi$-exchange potential we use the MUT and calculate the NN momentum-space matrix elements of the operator $V_{\rm Classes- III, \, IV}^{\rm MUT}$ given in Eq.~(\ref{MUTClass34}), keeping only those terms which correspond to the class-IV diagrams.
	The matrix elements of the Weinberg-Tomozawa vertices $H^2_{22}$ are given by
	\begin{align}
		\bra{\pi_{a}(\vec{q}_1) N} H^2_{22}\ket{\pi_{b}(\vec{q}_2)N}& = -\left(\omega_{\vec{q}_1}+\omega_{\vec{q}_2}\right) v_{ab}, \nonumber \\
		\bra{\pi_{a}(\vec{q}_1)\pi_{b}(\vec{q}_2) N} H^2_{22}\ket{N} &= -\left(\omega_{\vec{q}_1}-\omega_{\vec{q}_2}\right) v_{ab}, \nonumber \\
		\bra{N} H^2_{22}\ket{\pi_{a}(\vec{q}_1)\pi_{b}(\vec{q}_2) N} &= \left( \omega_{\vec{q}_1}-\omega_{\vec{q}_2} \right) v_{ab}\,,\label{eq_WT_Vertices}
	\end{align}
	where $a$ and $b$ are the isospin quantum numbers and
	\beq
	v_{ab} = \frac{i}{8 F_\pi^2} \epsilon_{abd} \tau_c \frac{1}{\sqrt{w_{\vec q_1} w_{\vec q_2}}}\,.
	\eeq
	For the $\pi N$ vertex proportional to $g_A$, we have (in the nucleon rest frame) 
	\beq
	\label{MEgA}
	\langle N | H_{21}^1 | \pi_a (\vec q ) N \rangle = i \frac{g_A}{2 F_\pi} \frac{1}{\sqrt{2 w_{\vec q}}} \vec \sigma \cdot \vec q \, \tau_a\,.
	\eeq
	Notice that when applying the canonical formalism to the lowest-order effective pion-nucleon Lagrangian, the time derivative appearing in the Weinberg-Tomozawa vertex generates an  additional contact interaction $H^4_{42}$ in the Hamiltonian, which has four nucleon and two pion fields, see Refs.~\cite{Weinberg:1992yk,Epelbaum:2005bjv,Epelbaum:2007us} for more details. The corresponding matrix elements read 
	\beq
	\langle NN | H_{42}^4 | NN \pi_a (\vec q_1) \pi_b (\vec q_2) \rangle = \frac{1}{16 \sqrt{4 w_{\vec q_1} w_{\vec q_2}} F_\pi^4} \big( 2 \,  \vec \tau_1 \cdot \vec \tau_2 \, \delta^{ab} - \tau_1^a \tau_2^b - \tau_1^b \tau_2^1 \big)\,.
	\eeq
	Having specified all the relevant vertices, it is straightforward to calculate the energy denominators for various class-IV diagrams. When calculating the matrix elements of the terms in Eq.~(\ref{MUTClass34}), one has to sum over all possible sequences of vertices. As already pointed out above, we only keep those contributions whose vertex sequences correspond to the class-IV diagrams. For every diagram, one has the same sequence of vertices at every nucleon line, which ensures that all  contributions generate the same spin-momentum-isospin operators, which will be treated in a separate step. For examples of calculations using the MUT the reader is referred to Refs.~\cite{Epelbaum:2007us,Krebs:2016rqz}. 
	
	We start with the two reducible (box) diagrams and calculate the energy denominators (${\rm ED}$) for e.g.~the graph (2) in Fig.~\ref{fig:3pi_N3LO} using the MUT
	\beq
	\label{Class4PlanarBox}
	{\rm ED}^{\rm IV, \, MUT}_{(2)} = \frac{\omega _1^2+4 \omega _2 \omega _1+4 \omega _2^2}{4 \omega _1^3 \omega _2 \left(\omega _1+\omega _2\right) \left(\omega _1+\omega _2+\omega _3\right)}-\frac{1}{4 \omega _1 \omega _2 \left(\omega _1+\omega _2\right) \omega _3}
	+\frac{1}{\omega _1^2 \left(\omega _2+\omega _3\right){}^2}\,.
	\eeq
	Here, the momentum (energy) of the pion attached to the two $g_A$-vertices is denoted with $\vec l_1$ ($\omega_1$).
	Notice that apart from the actual energy denominators, this expression also includes the factor of $1/(\omega_1 \omega_2 \omega_3)$ stemming from the mode expansion of the pion field entering the vertices, see Eqs.~(\ref{eq_WT_Vertices}), (\ref{MEgA}). 
	Alternatively, following the strategy of Ref.~\cite{Kaiser:1999ff}, we can compute the energy denominators from the S-matrix. This is achieved by writing the expression for the corresponding Feynman diagram and performing the integration over the $0$-th components of the loop momenta:
	\begin{eqnarray}
		{\rm ED}^{\rm IV, \, SMM}_{(2)} &=&\iint \frac{dl_{10}}{2\pi}\frac{dl_{20}}{2\pi} \frac{4 m^2 \left(2l_{20}+l_{10}\right)^2}{\big[\left(p_1-l_1\right)^2-m^2+i \epsilon\big]\big[\left(p_2+l_1\right)^2-m^2+i \epsilon\big]}
		\, \frac{1}{l_1^2-M_{\pi}^2+i \epsilon}\, \frac{1}{l_2^2-M_{\pi}^2+i \epsilon}
		\nonumber \\
		&& {} \times \frac{1}{\left(q-l_1-l_2\right)^2-M_{\pi}^2+i \epsilon}\nonumber \\
		&=& \frac{2m}{[p^2-(\vec{p}-\vec{l}_1)^2 + i \epsilon] \omega _1^2 \left(\omega _2+\omega _3\right)}+\frac{\omega _1^2+4 \omega _2 \omega _1+4 \omega _2^2}{4 \omega _1^3 \omega _2 \left(\omega _1+\omega _2\right) \left(\omega _1+\omega _2+\omega _3\right)}-\frac{1}{4 \omega _1 \omega _2 \left(\omega _1+\omega _2\right) \omega _3}\nonumber \\
		&&{}+\frac{1}{\omega _1^2 \left(\omega _2+\omega _3\right){}^2} + \mathcal{O}\left(m^{-1}\right).\label{eq_PlanarBox3Pi_Relat_expansion}
	\end{eqnarray}
	Clearly, the first term in the last equality corresponds to the iteration of the one and two-pion exchange NN potentials in the Lippmann-Schwinger equation. On the other hand, the second, third and  fourth terms give rise to the irreducible $3\pi$-exchange potential. These terms turn out to be identical to the ones obtained using the MUT, so that both approaches yield the same result for this particular type of diagrams.\footnote{This result is not really unexpected and is in line with the similar findings for the planar two-pion exchange potential \cite{Kaiser:1997mw,Epelbaum:1998ka}.} For the sake of completeness, we give the expression for the energy denominators for the crossed-box diagram (the last class-IV diagram in Fig.~\ref{fig:3pi_N3LO}):
	\begin{align}
		{\rm ED}^{\rm IV}_{(3)} =-\frac{\omega _1^2+4 \omega _1 \omega _2+4 \omega _2^2}{4 \omega _1^3 \omega _2 \left(\omega _1+\omega _2\right) \left(\omega _1+\omega _2+\omega _3\right)}+\frac{1}{4 \omega _1 \omega _2 \left(\omega _1+\omega _2\right) \omega _3}-\frac{1}{\omega _1^2 \left(\omega _2+\omega _3\right){}^2}\,.
		\label{Class4CrossBox}
	\end{align}
	Clearly, this expression coincides with the corresponding result obtained using the S-matrix method. Notice further that the energy denominators for the planar and cross-box topologies in Eqs.~(\ref{Class4PlanarBox}) and (\ref{Class4CrossBox}) are identical up to an overall sign, as one may expect based on the results for the leading two-pion exchange potential \cite{Kaiser:1997mw,Epelbaum:1998ka}. 
	
	Having verified that both the MUT and SMM lead to the same result for the class-IV contributions, we can compute the resulting potentials using the same method as for the previous cases based on the Feynman diagrams. The results for the energy denominators for the planar and cross-box contributions suggest that the irreducible part of the planar box diagrams can be extracted by switching the sign of $i \epsilon$ in one of the heavy-baryon nucleon propagators entering the expression for the planar box Feynman diagram. To see this consider a trivial algebraic example \cite{Kaiser:1999jg}:
	\begin{align}
		\int_{-1}^{1}dx \frac{1}{\left(x + i \epsilon\right)\left(x - i \epsilon\right)} = \frac{\pi}{\epsilon} -2 + \mathcal{O}(\epsilon^2), \quad \quad
		\int_{-1}^{1}dx \frac{1}{\left(x + i \epsilon\right)^2} = -2 + \mathcal{O}(\epsilon^2)\,.
	\end{align}
	Here, the variable $x$ plays the role of the $0$-th component of the loop momentum $l_0$. The divergent $\pi/\epsilon$ term in the first integral reflects the infrared pinch singularity of the two-nucleon bubble diagram in the static limit $m \to \infty$, which gets regularized by the nucleon mass when keeping the kinetic-energy term in the nucleon propagator \cite{Weinberg:1991um}. This term thus behaves as $\sim m$ and represents an iterative contribution to the NN scattering amplitude. Switching the sign of $i \epsilon$ in one of the heavy-baryon propagators results in removing this iterative contribution, yielding the irreducible part of the corresponding diagram. See Ref.~\cite{Zhu:2004vw} for a related discussion. With this in mind, we can directly obtain the class-IV $3\pi$-exchange potential from the corresponding Feynman diagrams. After performing the spin-isospin algebra, we again find only the isovector spin-spin and tensor potentials. Starting with the spin-spin part, we find after applying the projection operator: 
	\begin{align}
		\textrm{Im}\; W^{\rm IV}_S(i \mu) =&-\frac{g_A^2}{ \left(32\pi^2F_\pi^3 \right)^2} \iint\limits_{z^2<1} d \omega_{1} d \omega_{2}  \int_{-1}^{1}dx \int_{y_{\textrm{min}}(x,z)}^{y_{\textrm{max}}(x,z)}dy \frac{\left(x^2-\epsilon^2\right)}{\left(x^2+\epsilon^2\right)^2} \frac{\left(-1+x^2\right)\left(|\vec{l}_1|x + 2 |\vec{l}_2|y\right)^2}{\sqrt{1-x^2-y^2-z^2+2 x y z}}\;.
	\end{align}
	Here and in what follows, the limit $\epsilon \to 0^+$ is understood. 
	After performing the angular integrations (see appendix \ref{Appendix_Principle Value Integrals} for details) and the integration over $\omega_{2}$, we are left with an integral over the invariant mass $w$:
	\begin{align}
		\textrm{Im}\; W^{\rm IV}_S(i \mu) =&\frac{g_A^2}{18 \mu^4 \left(16\pi F_\pi^2 \right)^3} \int_{2M_{\pi}}^{\mu-M_{\pi}} \frac{dw}{w^2}
		\sqrt{\left(w^2-4M_{\pi}^2\right)^3}\sqrt{\lambda(w^2, M_{\pi}^2,\mu^2)} \left[\left(M_{\pi}^2-w^2\right)^2 -2 \mu ^2 \left(M_{\pi}^2+7 w^2\right)+\mu ^4\right] \,.
	\end{align}
	Using the same steps as done for the class-III diagrams, we can cast this expression to the form given in Ref.~\cite{Kaiser:1999ff}: 
	\begin{align}
		\textrm{Im}\; W^{\rm IV}_S(i \mu)=&\frac{g_A^2}{9 \mu^4 \left(16\pi F_\pi^2 \right)^3} \int_{2M_{\pi}}^{\mu-M_{\pi}} dw
		\sqrt{\left(w^2-4M_{\pi}^2\right)}\sqrt{\lambda(w^2, M_{\pi}^2,\mu^2)} \, \left[9w^4-5\mu ^4+ 30 \mu^2 M_{\pi}^2-9 M_{\pi}^4\right]\,.\label{eq_IM_Ws_ClassIV_NK_dw}
	\end{align}
	
	Next, we turn to the isovector tensor part. Applying the corresponding projection-operator, we obtain the four-dimensional representation:
	\begin{align}
		\textrm{Im}\; W^{\rm IV}_T(i \mu) =&\frac{-g_A^2}{ \left(32\mu \pi^2F_\pi^3 \right)^2} \iint\limits_{z^2<1} d \omega_{1} d \omega_{2}  \int_{-1}^{1}dx \int_{y_{\textrm{min}}(x,z)}^{y_{\textrm{max}}(x,z)} dy\frac{\left(x^2-\epsilon^2\right)}{\left(x^2+\epsilon^2\right)^2} \frac{\left(-2M_{\pi}+\vec{l}_1^{\;2}\left(-3+x^2\right)\right)\left(|\vec{l}_1|x + 2 |\vec{l}_2|y\right)^2}{\sqrt{1-x^2-y^2-z^2+2 x y z}}\,.
	\end{align}
	After performing the angular integration, we are left with the two-dimensional integral representation over the energies $\omega_{1,2}$:
	\begin{align}
		\textrm{Im}\; W^{\rm IV}_T(i \mu) =&\frac{g_A^2}{3 \left(8\pi F_\pi^2 \right)^3} \iint\limits_{z^2<1} d \omega_{1} d \omega_{2} \left[4 l_1^2 + 16 l_1 l_2 z + 12 \frac{l_2}{l_1} M_{\pi}^2 z + 12 \frac{l_2^2}{l_1^2} M_{\pi}^2 \left(2z^2 -1\right) + 3 M_{\pi}^2 + 4 l_2^2 \left( 9z^2 -5\right)\right]\,.
	\end{align}
	Again, we can reduce it to a one-dimensional integral over the invariant mass $w$:
	\begin{eqnarray}
		\label{Class4_WT_mu}  
		\textrm{Im}\; W^{\rm IV}_T(i \mu) &=&\frac{g_A^2}{18 \mu^4 \left(8\pi F_\pi^2 \right)^3} \int_{2M_{\pi}}^{\mu-M_{\pi}} dw
		\frac{\big(w^2-4M_{\pi}^2\big)^{\frac{3}{2}}}{\sqrt{\lambda (w^2, M_{\pi}^2,\mu^2)}} \Big[\left(M_{\pi}^2-w^2\right)^4-\left(M_{\pi}^2-w^2\right)^2 \left(M_{\pi}^2+10 w^2\right)\mu ^2 \nonumber \\
		&&{} -2 w^2 \left(M_{\pi}^2-9 w^2\right) \mu ^4-\left(M_{\pi}^2+10 w^2\right)\mu ^6 +\mu ^8\Big]\,.
	\end{eqnarray}
	For comparison, the result from \cite{Kaiser:1999ff} has the form:
	\begin{align}
		\textrm{Im}\; W^{\rm IV}_T(i \mu)=&\frac{g_A^2}{6 \mu^6 \left(8\pi F_\pi^2 \right)^3} \int_{2M_{\pi}}^{\mu-M_{\pi}} dw
		\sqrt{\left(w^2-4M_{\pi}^2\right)}\sqrt{\lambda(w^2, M_{\pi}^2,\mu^2)} \bigg[w^4+7\mu^2M_{\pi}^2  \nonumber \\
		&  -\frac{2}{3} \mu^4 -M_{\pi}^2\left(\mu^2+M_{\pi}^2\right)^2 w^{-2}+4 M_{\pi}^2 \mu^4\left(4M_{\pi}^2-w^2\right)\lambda^{-1}(w^2, M_{\pi}^2,\mu^2)\bigg]\,.
	\end{align}
	We have verified that these to expressions are identical.

	\subsection{Class-V diagrams}
	
	The class-V contributions emerge solely from irreducible topologies. Thus, they can be computed from the corresponding Feynman diagrams and give rise to the isovector spin-spin and tensor interactions. After the tensor reduction, we obtain for the spin-spin part:
	\begin{align}
		\textrm{Im}\; W^{\rm V}_S(i \mu) =&\frac{g_\mathrm{A}^4}{2 \pi \left(8 \pi F_\pi^2 \right)^3} \iint \limits_{z^2<1}d \omega_1 d\omega_2 \int_{-1}^{1}dx \int_{y_{\textrm{min}}(x,z)}^{y_{\textrm{max}}(x,z)} dy \frac{1}{\sqrt{1-x^2-y^2-z^2+2 x y z}}  \frac{1}{ l_1 l_2 (i x-\epsilon) (i y+\epsilon)} \nonumber \\
		&\times \left\{l_1 l_2 (x y-z) \left(l_2^2 y^2-\mu  \left(\omega _1+\omega _2\right)+2 M_{\pi}^2\right)+l_2^2 \left(y^2-1\right) \left(M_{\pi}^2-\mu  \omega _1\right)+l_1^3 l_2 x^2 (x y-z) \right. \nonumber \\ 
		&\left.+l_1^2 \left[l_2^2 \left(x^2 \left(2 y^2-1\right)-y^2\right)+\left(x^2-1\right) \left(M_{\pi}^2-\mu  \omega _2\right)\right]\right\}\,.
	\end{align}
	Here, we used the notation with $l_i \equiv  |\vec{l}_i|$. For the angular integration, we again encounter some non-trivial integrals which read
	\begin{align}
		\int_{-1}^{1} dx \int_{y_{\textrm{min}}(x,z)}^{y_{\textrm{max}}(x,z)}dy \frac{1}{\sqrt{1-x^2-y^2-z^2+2 x y z}}\frac{-1}{\left(i x - \epsilon\right)\left(i y + \epsilon\right)} = \frac{2 \pi \arccos(-z)}{\sqrt{1-z^2}}
	\end{align}
	and 
	\begin{align}
		- 	\int_{-1}^{1} dx \int_{y_{\textrm{min}}(x,z)}^{y_{\textrm{max}}(x,z)}dy\frac{1}{\sqrt{1-x^2-y^2-z^2+2 x y z}}\frac{x^2}{\left(i x - \epsilon\right)\left(i y + \epsilon\right)}  = 2 \pi z \;.
	\end{align}
	Their derivation is explained in detail in  appendix \ref{Appendix_Principle Value Integrals}.  After performing the angular integrals, we find for the spectral function of the isovector spin-spin part 
	\begin{align}
		\label{Class5_WS_mu}
		\textrm{Im}\; W^{\rm V}_S(i \mu) =& \frac{2g_\mathrm{A}^4}{3 \left(8 \pi  F_\pi^2\right)^3 } \iint \limits_{z^2<1} d \omega_1 d\omega_2 \left\{\left(1+ \frac{l_2 z}{l_1}\right) \left[-l_1^2 + 3 \left(l_1 l_2 z + \mu \omega_{1} - 3M_{\pi}^2\right)\right]\right. \nonumber \\
		&\left.+ 3\left(M_{\pi}^2- \mu \omega_{1}\right)\left(z + \frac{l_2}{l_1}\right)\frac{\arccos(-z)}{\sqrt{1-z^2}}\right\}\,.
	\end{align} 
	This expression is identical to the one given in Ref.~\cite{Kaiser:1999jg}. 
	
	We continue with the isovector tensor part. The starting point is again the four-dimensional integral representation after performing the tensor reduction,
	\begin{align}
		\textrm{Im}\; W^{\rm V}_T(i \mu) =&\frac{-g_\mathrm{A}^4\left(\mu^2-M_{\pi}^2\right)^{-1}}{4 \pi \mu^2 \left(8 \pi F_\pi^2 \right)^3} \iint \limits_{z^2<1}d \omega_1 d\omega_2  \int_{-1}^{1}dx \int_{y_{\textrm{min}}(x,z)}^{y_{\textrm{max}}(x,z)} dy \frac{1}{\sqrt{1-x^2-y^2-z^2+2 x y z}}  \frac{1}{ l_1 l_2 (i x-\epsilon) (i y+\epsilon)} \nonumber \\
		& \times \left[M_{\pi}^4 \left(7 \mu ^2-3 l_1^2 x^2-4 l_1 l_2 x y-3 l_2^2 y^2-11 \mu  \left(\omega _1+\omega _2\right)+6 \left(\omega _1+\omega _2\right){}^2\right) \right. \nonumber \\
		& \left. +M_{\pi}^2 \left(\mu  \left(-20 \mu ^2 \left(\omega _1+\omega _2\right) +5 \mu ^3+4 \mu  \left(4 \omega _1^2+11 \omega _2 \omega _1+4 \omega _2^2\right)-12 \omega _1 \omega _2 \left(\omega _1+\omega _2\right)\right) \right.\right. \nonumber \\
		&\left. \left. +2 l_1^2 x^2 \left(2 \mu ^2-2 l_2^2 y^2-\mu  \left(\omega _1+3 \omega _2\right)+3 \omega _2 \left(\omega _1+\omega _2\right)\right)-2 l_2 l_1^3 x^3 y\right. \right. \nonumber \\
		& \left. \left. -2 l_2 l_1 x y \left(l_2^2 y^2+\mu  \left(\mu -3 \left(\omega _1+\omega _2\right)\right)\right)+2 l_2^2 y^2 \left(2 \mu ^2-\mu  \left(3 \omega _1+\omega _2\right)\right. \right. \right. \nonumber \\
		& \left. \left. \left. +3 \omega _1 \left(\omega _1+\omega _2\right)\right)\right)+\mu ^2 \left(l_1^2 x^2 \left(-\mu ^2+4 l_2^2 y^2+2 \mu  \left(\omega _1-4 \omega _2\right)+10 \omega _2 \left(\omega _1+\omega _2\right)\right)\right.\right. \nonumber \\
		& \left. \left.  +2 l_2 l_1^3 x^3 y+2 l_2 l_1 x y \left(l_2^2 y^2+\mu  \left(5 \left(\omega _1+\omega _2\right)-4 \mu \right)\right)+l_2^2 y^2 \left(-\mu ^2+2 \mu  \left(\omega _2-4 \omega _1\right) \right. \right. \right.  \nonumber \\
		&\left. \left. \left.  +10 \omega _1 \left(\omega _1+\omega _2\right)\right)+\mu  \left(\mu -2 \omega _1\right) \left(\mu -2 \omega _2\right) \left(4 \mu -5 \omega _1-5 \omega _2\right)\right)-2 M_{\pi}^6\right]\,.
	\end{align}
	The angular integration can be performed in the same way as for the spin-spin part and we obtain, after simplification, our final result
	\begin{align}
		\textrm{Im}\; W^{\rm V}_T(i \mu) =& \frac{1}{\mu^2} \textrm{Im}\; W^{\rm V}_S(i \mu) - \frac{g_\mathrm{A}^4  \left(\mu ^2-M_{\pi}^2\right)^{-1}}{\mu ^2  \left(8 \pi  F_\pi^2\right)^3} \iint \limits_{z^2<1} d \omega_1 d \omega_2\left[ \left(6 \mu^2 + 2 M_{\pi}^2\right)\left(\omega_1+\omega_2\right) \right. \nonumber \\ 
		&\left. -\mu \left(4 \mu^2 + 3 M_{\pi}^2\right)\right] \left[ \left(\left(\mu^2 + M_{\pi}^2\right)\left(2 \omega_1 -\frac{\mu}{2}\right)- 2 \mu \omega_1 \omega_2\right) \frac{\arccos(-z)}{l_1 l_2 \sqrt{1-z^2}} + \mu + 2 z \omega_1 \frac{l_2}{l_1}\right]\label{eq_WT_linearCombi_FinalRes}\,.
	\end{align}
	This expression agrees with the one obtained by Kaiser in Ref.~\cite{Kaiser:1999jg} up to the sign in front of the second term on the right-hand side, which
	is not correct  \cite{NorbertPrivComm}.

	\subsection{Class-VI diagrams}
	
	Class-VI diagrams diagrams involve reducible-like topologies, and the resulting potentials are thus expected to be scheme-dependent. Using the MUT, we start with calculating  the energy denominators for the two irreducible class-VI diagrams by evaluating the matrix elements of the operators in Eqs.~(\ref{3pioperatorclass2}), (\ref{MUTClass67add}) for the corresponding topology:
	\begin{align}
		\label{EDClass6irreducible}
		{\rm ED}^{\rm VI}_{(4)} =- {\rm ED}^{\rm VI}_{(3)}&=- \frac{\omega _1^2+\omega _2 \omega _1+\omega _2^2}{4 \omega _1^3 \omega _2^3 \left(\omega _1+\omega _2\right) \omega _3}-\frac{1}{4 \omega _1^3 \omega _2^2 \left(\omega _1+\omega _3\right)}+\frac{1}{4 \omega _1^2 \omega _2^3 \left(\omega _2+\omega _3\right)} \nonumber \\
		& \quad +\frac{1}{2 \omega _1^2 \omega _2^2 \left(\omega _2+\omega _3\right){}^2}+\frac{\omega _1+2 \omega _2}{4 \omega _1^3 \omega _2^2 \left(\omega _1+\omega _2\right) \left(\omega _1+\omega _2+\omega _3\right)}\,.
	\end{align}
        For the class-VI diagrams, we label the pion momenta at the second nucleon from bottom to top as $\{l_3, l_2, l_1\}$,  $\{l_1, l_2, l_3\}$,  $\{l_2, l_3, l_1\}$ and  $\{l_1, l_3, l_2\}$ for diagrams (1)--(4), in order.  
	For the irreducible diagrams (1) and (2), we expect the same result to emerge in the S-matrix method. We have verified that this is indeed the case by considering the corresponding Feynman diagrams and performing the integrations over the zeroth components of the loop momenta $l_1^0$ and $l_2^0$.
        
	For the two reducible-like diagrams, the MUT leads to the expression 
	\begin{align}
		\label{EDClass5Reducible}
		{\rm ED}^{\rm VI, \, MUT}_{(2)} =-{\rm ED}^{\rm VI, \, MUT}_{(1)}& = - {\rm ED}^{\rm VI}_{(4)} + \frac{1}{2 \omega _1^2 \omega _2^2 \omega _3^2}\,.
	\end{align}
	This result differs from the one obtained in the SMM by the last term,
	\begin{align}
		{\rm ED}^{\rm VI, \, SMM}_{(2)} =-{\rm ED}^{\rm VI, \, SMM}_{(1)}& = - {\rm ED}^{\rm VI}_{(4)} \,,
	\end{align}
	which reflects the scheme dependence of the reducible-like contributions to the potential. 
	Thus, the two approaches yield  different results for the class-VI $3\pi$-exchange potential. 
	
	Before proceeding with the calculation of the class-VI potential using the method of unitary transformation, it is instructive to rederive the results of Ref.~\cite{Kaiser:1999jg} using the SMM. 
	Here, we proceed as follows: We flip the sign of one of the $\epsilon$ in the nucleon propagator for each of the two reducible diagrams in such a way that the nucleon-propagator structure coincides with that of the irreducible diagrams of class VI.
This procedure removes the iterative contributions in the SMM.
        Afterwards, we group together the contributions of diagrams (1) and (3) as well as (2) and (4) by adding one reducible and one irreducible diagram with the same propagator structure, perform the tensor reduction and solve the angular integrals. Proceeding this way, we observe, in line with the findings of Ref.~\cite{Kaiser:1999jg}, that the class-VI diagrams only generate the isovector spin-spin and tensor potentials when calculated using the SMM. For the spin-spin part, we find the four-dimensional representation:
	\begin{align}
		\textrm{Im}\; W^{{\rm VI, \, SMM}}_S(i \mu) =& \frac{-g_\mathrm{A}^4}{2 \pi \left(8 \pi F_\pi^2 \right)^3} \iint \limits_{z^2<1}d \omega_1 d\omega_2 \int_{-1}^{1}dx \int_{y_{\textrm{min}}(x,z)}^{y_{\textrm{max}}(x,z)} dy \frac{i}{\sqrt{1-x^2-y^2-z^2+2 x y z}} \frac{\left(l_1 x+2 l_2 y\right)}{ l_1^2 l_2 (i x+\epsilon )^2 (i y-\epsilon)} \nonumber \\
		&\times	 \left[l_1^2 l_2 x^2 (x y-z)+l_2 \left(M_{\pi}^2-\mu  \omega _1\right) (x y-z)+l_1 \left(x^2-1\right) \left(l_2^2 y^2-\mu  \omega _2+M_{\pi}^2\right)\right].
	\end{align}
	Performing the same steps as for the  class-V contributions, we find,  after some simplifications, the following double-integral representation:
	\begin{align}
		\label{SSMM}
		\textrm{Im}\; W^{\rm VI, \, SMM}_S(i \mu) =& \frac{2g_\mathrm{A}^4}{ \left(8 \pi  F_\pi^2\right)^3 } \iint \limits_{z^2<1} d \omega_1 d\omega_2 \left\{\left(1+ \frac{l_2 z}{l_1}\right) \left[3l_1^2 + 5 \left(M_{\pi}^2-l_1 l_2 z +
		- \mu \omega_{1} \right)\right]\right. \nonumber \\
		&\left.- \left(M_{\pi}^2- \mu \omega_{1}\right)\left(z + \frac{l_2}{l_1}\right)\frac{\arccos(-z)}{\sqrt{1-z^2}}\right\}\,,
	\end{align} 
	which agrees with the expression given in Ref.~\cite{Kaiser:1999jg}. For the isovector tensor part,  we find
	\begin{eqnarray}
		\textrm{Im}\; W^{\rm VI, \, SMM}_T(i \mu) &=& \frac{1}{\mu^2}\textrm{Im}\; W^{\rm VI, \, SMM}_S(i \mu)\nonumber \\
		&+&\frac{2 g_\mathrm{A}^4}{ \pi \left(8 \pi F_\pi^2 \right)^3\mu^2} \iint \limits_{z^2<1}d \omega_1 d\omega_2  \int_{-1}^{1}dx \int_{y_{\textrm{min}}(x,z)}^{y_{\textrm{max}}(x,z)} dy \frac{i}{\sqrt{1-x^2-y^2-z^2+2 x y z}}	\frac{\left(l_1 x+2 l_2 y\right)}{l_1^2 l_2 ( i x+ \epsilon)^2 (y-i \epsilon )} \nonumber \\ 
		&\times& \left[\omega _1 \left(l_1^2 x^2 \omega _2+\mu  l_1 l_2 (x y-z)+\omega _1 \left(l_2^2 y^2-\mu  \omega _2\right)+M_{\pi}^2 \left(\omega _1+\omega _2\right)\right)\right]\,.
	\end{eqnarray}
	Performing the angular integrations in the usual way, we finally obtain: 
	\begin{eqnarray}
		\label{TSMM}  
		\textrm{Im}\; W^{\rm VI, \, SMM}_T(i \mu) &=& \frac{1}{\mu^2}\textrm{Im}\; W^{\rm VI, \, SMM}_S(i \mu)\nonumber \\
		&-&\frac{4 g_\mathrm{A}^4}{ \left(8 \pi F_\pi^2 \right)^3\mu^2} \iint \limits_{z^2<1}d \omega_1 d\omega_2 \frac{\omega_{1}}{l_1^2} \bigg\{l_1^2 \left(\mu +2 \omega _2\right)+l_1 l_2 z \left(4 \mu +\omega _1+\omega _2\right)-2 M_{\pi}^2 \left(\omega _1+\omega _2\right)+2 \mu  \omega _1 \omega _2 \nonumber \\
		&+&  2 l_2^2 \omega _1 \left(2 z^2-1\right) + \frac{l_1}{l_2}\left[\left(\omega_{1} \omega_{2} + l_1 l_2 z\right)\mu -M_{\pi}^2\left(\omega_{1}+\omega_{2}\right)\right]\frac{\arccos(-z)}{\sqrt{1-z^2}}\bigg\}\,. 
	\end{eqnarray}
	This expression also agrees with the one given in Ref.~\cite{Kaiser:1999jg}.
	
	Having discussed the potentials defined within the SMM, we now turn to the results in the MUT. We first emphasize that the additional term in Eq.~(\ref{EDClass5Reducible}) does not only affect the corresponding expressions for the isovector spin-spin and tensor potentials, but also leads to the emergence of new structures.  To see this we observe that the class-VI potential can  be schematically written in a general form
	\begin{align*}
		O_{\rm spin} \cdot \Big[ O_{\rm mom}^1\big(
		{\rm ED}_{\rm reducible} + {\rm ED}_{\rm irreducible} 
		\big)
		+
		O_{\rm mom}^2\big(
		{\rm ED}_{\rm reducible} - {\rm ED}_{\rm irreducible} 
		\big)
		\, \vec{\tau}_1 \cdot \vec{\tau}_2\Big],
	\end{align*}
	where $O_{\rm spin}$ and $O_{\rm mom}^{1,2}$ are spin- and momentum-dependent operators, respectively. In the SMM, one has 
	$ {\rm ED}_{\rm reducible} = - {\rm ED}_{\rm irreducible} $ leading to a vanishing isoscalar part of the potential. On the other hand, in the MUT one has
	$ {\rm ED}_{\rm reducible} \neq - {\rm ED}_{\rm irreducible} $ so that the resulting potential also has a nonvanishing isoscalar component.  
	
	To derive the class-VI contributions using the MUT, we first observe that the energy denominators for the reducible diagram (1) in Eq.~(\ref{EDClass5Reducible}) can be obtain by performing the $l_1^0$, $l_2^0$ integrations of the following expression written in terms of the Feynman propagators:
	\begin{align}
		{\rm ED}^{\rm VI, \, MUT}_{(1)}=-\iint \frac{d^0l_1}{(2\pi)}\frac{d^0l_2}{(2\pi)} \frac{2 v\cdot l_2+ v \cdot l_1}{\left(v \cdot l_1+ i \epsilon\right)^2\left(v \cdot l_2+ i \epsilon\right)\left(l_{01}^2-\omega_1^2+ i \epsilon\right)\left(l_{02}^2-\omega_2^2+ i \epsilon\right)\big(\left(l_{01}+l_{02}\right)^2-\omega_3^2+ i \epsilon\big)},\label{matching_4dim_C6_reducibleDiagrams}
	\end{align}
	with $\omega_{1,2} = \sqrt{\vec{l}_{1,2}^{\, 2}+ M_{\pi}^2}$ and  $\omega_{3} = \sqrt{(\vec{q}-\vec{l}_1-\vec{l}_2)^2+ M_{\pi}^2}$.
	It is important to keep in mind the assignment of the heavy-baryon propagators to the nucleons 1 and 2, namely
	\begin{align*}
		\frac{1}{\left(v\cdot l_1 + i \epsilon\right)_{\rm nucleon \, 1}\; \left(v\cdot l_1 + i \epsilon\right) _{\rm nucleon \, 2}\; \left(v\cdot l_2+ i \epsilon\right) _{\rm nucleon \, 2}}\,.
	\end{align*}
	With this in mind, we follow the usual steps to calculate the contributions of the two reducible diagrams (those of the irreducible diagrams are identical to the ones obtained in the SMM). When performing the angular integration for the reducible diagrams, we encounter the integral
	\begin{align*}
		\int_{-1}^{1} dx \int_{y_{\textrm{min}}(x,z)}^{y_{\textrm{max}}(x,z)} dy \frac{1}{\sqrt{1-x^2-y^2-z^2 + 2 x y z}}	\frac{i}{\left(x - i \epsilon\right)\left(i y + \epsilon\right)}\,,
	\end{align*}
	which can be calculated as follows. First, we multiply and divide each propagator with its complex conjugated expression leading to 
	\begin{align*}
		\int_{-1}^{1} dx \int_{y_{\textrm{min}}(x,z)}^{y_{\textrm{max}}(x,z)} dy \frac{1}{\sqrt{1-x^2-y^2-z^2 + 2 x y z}}\left[\mathrm{P} \frac{1}{x} \mathrm{P}\frac{1}{y} - \pi^2 \delta(x) \delta(y) + i \pi \left( \mathrm{P}\frac{1}{x} \delta(y)+\mathrm{P}\frac{1}{y} \delta(x)\right)\right]\,.
	\end{align*}
	Then, using the results from appendix \ref{Appendix_Principle Value Integrals}, we know that only the first two terms give non-vanishing contributions, for which we finally obtain 
	\begin{align}
		&\int_{-1}^{1} dx \int_{y_{\textrm{min}}(x,z)}^{y_{\textrm{max}}(x,z)} dy \frac{1}{\sqrt{1-x^2-y^2-z^2 + 2 x y z}}	\frac{i}{\left(x - i \epsilon\right)\left(i y + \epsilon\right)} =\frac{2 \pi \arcsin(z)}{\sqrt{1-z^2}}- \frac{\pi^2}{\sqrt{1-z^2}} =- \frac{2 \pi \arccos(z)}{\sqrt{1-z^2}}\;.
	\end{align}
	Then, putting the contributions of all four diagrams together, we find the following results for the isoscalar spin-spin and tensor potentials
	\begin{eqnarray}
		\label{VSMUT} 
		\textrm{Im}\; V^{\rm VI, \, MUT}_S(i \mu) &=&  \frac{3 g_\mathrm{A}^4}{\left(32 \pi F_\pi^3\right)^2}  \iint \limits_{z^2<1}  d \omega_{1} d \omega_{2}\left(\frac{l_2}{l_1}+z\right)\left[\frac{\mu \omega_{1}-M_{\pi}^2}{\sqrt{1-z^2}}\right]\nonumber \\
		&=&{} - \frac{g_\mathrm{A}^4 \left( \mu- 3 M_{\pi}\right)^2}{ 70 \pi \left(32 F_\pi^3\right)^2} \left[2M_{\pi}^2 - 12 \mu M_{\pi}- 2\mu^2 + 15 \frac{M_{\pi}^3}{\mu}+ 2 \frac{M_{\pi}^4}{\mu^2}+ 3 \frac{M_{\pi}^5}{\mu^3}\right], \\
		\textrm{Im}\; V^{\rm VI, \, MUT}_T(i \mu)& =&  \frac{1}{\mu^2}\textrm{Im}\; V^{\rm VI}_S+ \frac{6 g_\mathrm{A}^4}{\left(32 \pi \mu F_\pi^3\right)^2}\iint\limits_{z^2<1} d\omega_{1} d\omega_{2} \frac{\omega_{1}}{l_1 l_2 \sqrt{1-z^2}} \left\{ \left(\omega_{1} \omega_{2} + l_1 l_2 z\right) \mu - M_{\pi}^2\left(\omega_{1}+\omega_{2}\right)\right\} \nonumber \\
		&=& {}\frac{g_\mathrm{A}^4 \left( \mu- 3 M_{\pi}\right)}{ 70 \pi \left(32 \mu F_\pi^3\right)^2} \left[ 6\mu^3 + 18 \mu^2 M_{\pi} -44  \mu M_{\pi}^2-27 M_{\pi}^3- 81 \frac{M_{\pi}^4}{\mu}+ 9 \frac{M_{\pi}^5}{\mu^2}+27  \frac{M_{\pi}^6}{\mu^3}\right].
		\label{VTMUT}                                              
	\end{eqnarray}
	
	For the the isovector contributions, we can express the spectral functions obtained using the MUT as
	\begin{eqnarray}
		\label{Class6_WS_mu_MUT}  
		\textrm{Im}\; W^{\rm VI, \, MUT}_S(i \mu) & =&  \textrm{Im}\; W^{\rm VI, \, SMM}_S(i \mu) - \frac{2}{3}	\textrm{Im}\; V^{\rm VI, \, MUT}_S(i \mu) \,, \\
		\textrm{Im}\; W^{\rm VI, \, MUT}_T(i \mu) & =&  \textrm{Im}\; W^{{\rm VI, \, SMM}}_T(i \mu) - \frac{2}{3}	\textrm{Im}\; V^{\rm VI, \, MUT}_T(i \mu) \,,
		\label{Class6_WT_mu_MUT}                                                    
	\end{eqnarray}
	where the expressions for the spectral functions on the right-hand side of the above equations are given in Eqs.~(\ref{SSMM}), (\ref{TSMM}), (\ref{VSMUT}) and (\ref{VTMUT}). 
	
	\subsection{Class-VII diagrams}
	
	To derive the class-VII contributions, we can again make use of the Feynman-diagram technique since the corresponding topologies are purely irreducible. Here, it is advantageous to label the diagrams in such a way that one obtains the delta-distribution when adding the diagrams.	
	 Performing the spin-isospin-momentum algebra, we find that the diagrams only lead to nonvanishing contributions for the isoscalar spin-spin and tensor structures. For the spin-spin part, we obtain 
	\begin{align}
		\textrm{Im}\; V^{\rm VII}_S(i \mu)  &= 	\frac{3 i g_\mathrm{A}^4 }{ 4 \pi^2\left(32 \pi F_\pi^3\right)^2} \iint \limits_{z^2<1}d\omega_1 d\omega_2 \int_{-1}^{1}dx \int_{y_{\textrm{min}}(x,z)}^{y_{\textrm{max}}(x,z)} dy\frac{1}{\sqrt{1-x^2-y^2-z^2+2 x y z}} \frac{\epsilon}{ \left(y^2+\epsilon ^2\right) }\nonumber \\
		& \quad \times \frac{\left(2 l_1 x+l_2 y\right)}{l_1 l_2 (x-i \epsilon )\left(l_1 x+l_2 y-i \epsilon \right)} \left[\mu ^3 \left(4 \omega _1+3 \omega _2\right)-\mu ^4+M_{\pi}^2 \left(-\mu ^2 \right. \right.          \nonumber \\
		& \quad \left. \left. +2 l_1^2 x^2+2 l_1 l_2 x y+3 l_2^2 y^2+\mu  \omega _2-2 \omega _2^2\right) -2 l_1^2 x^2 \left(l_2^2 y^2+\omega _2^2\right) \right. \nonumber \\
		& \quad \left.-2 l_1 l_2 x y \left(l_2^2 y^2+\mu  \left(3 \omega _2-2 \mu \right)+\omega _1 \left(4 \mu -2 \omega _2\right)\right)\right. \nonumber \\
		& \quad \left. +\mu ^2 \left(l_2^2 y^2-2 \left(2 \omega _1^2+4 \omega _2 \omega _1+\omega _2^2\right)\right)-2 l_2^2 \mu  y^2 \left(2 \omega _1+\omega _2\right)\right. \nonumber \\
		& \quad \left.+2 l_2^2 y^2 \omega _1 \omega _2+4 \mu  \omega _1 \omega _2 \left(\omega _1+\omega _2\right)+2 M_{\pi}^4\right].
	\end{align}
	We immediately see that we have a form of the $\delta$-distribution with respect to $y$ as mentioned above. We can thus trivially perform this integration and obtain the following expression:
	\begin{align}
		\textrm{Im}\; V^{\rm VII}_S(i \mu)  &= 	\frac{-3 i g_\mathrm{A}^4 }{ 2 \pi \left(32 \pi F_\pi^3\right)^2} \iint \limits_{z^2<1}d\omega_1 d\omega_2 \int_{-1}^{1} dx \frac{ \Theta \big(\sqrt{\left(x^2-1\right) \left(z^2-1\right)}-x z\big) }{l_1 l_2 \sqrt{1-x^2-z^2} (x-i \epsilon )}\nonumber \\
		& \quad \times \Theta \big(x z+\sqrt{\left(x^2-1\right) \left(z^2-1\right)}\big) \left[M_{\pi}^2 \left(\mu ^2-2 l_1^2 x^2-\mu  \omega _2+2 \omega _2^2\right)+2 l_1^2 x^2 \omega _2^2 \right. \nonumber \\
		& \quad \left.+\mu  \left(\mu -2 \omega _1\right) \left(\mu -\omega _2\right) \left(\mu -2 \omega _1-2 \omega _2\right)-2 M_{\pi}^4\right]\;.
	\end{align}
	The $x$-integration can be performed in the usual way by multiplying and dividing with the complex conjugate of $\left(x-i \epsilon\right)$ and making use of the principal value relations and the delta-distribution identities as discussed in appendix \ref{Appendix_Principle Value Integrals}. Notice that due to the $\Theta$-function, the integration boundaries of $x$ change from $[-1, \, 1]$  to $[-\sqrt{1-z^2}, \, \sqrt{1-z^2}]$. The principal value integration yields zero, whereas the delta-distribution part gives
	\begin{eqnarray}
		\label{Class7_VS_mu}
		\textrm{Im}\; V^{\rm VII}_S(i \mu)  &=& \frac{3 g_\mathrm{A}^4}{2 \left(32 \pi F_\pi^3\right)^2} \iint \limits_{z^2<1} \frac{d\omega_1 d\omega_2}{ l_1 l_2 \sqrt{1-z^2}}\left[\mu  \left(\mu -2 \omega _1\right) \left(\mu -\omega _2\right) \left(\mu -2 \omega _1-2 \omega _2\right) +M_{\pi}^2 \left(\mu ^2-\mu  \omega _2+2 \omega _2^2\right)-2 M_{\pi}^4\right] \nonumber \\
		&=& \frac{g_\mathrm{A}^4 \left( \mu- 3 M_{\pi}\right)^2}{ 35 \pi \left(32 F_\pi^3\right)^2} \left[2M_{\pi}^2 - 12 \mu M_{\pi}- 2\mu^2 + 15 \frac{M_{\pi}^3}{\mu}+ 2 \frac{M_{\pi}^4}{\mu^2}+ 3 \frac{M_{\pi}^5}{\mu^3}\right].
	\end{eqnarray}
	
	For the tensor part, we follow the same steps and obtain after the tensor reduction
	\begin{align}
		\textrm{Im}\; V^{\rm VII}_T(i \mu)  &= 	\frac{-3 i g_\mathrm{A}^4 }{ 4 \pi^2\left(32 \pi \mu F_\pi^3\right)^2} \iint \limits_{z^2<1}d\omega_1 d\omega_2  \int_{-1}^{1}dx \int_{y_{\textrm{min}}(x,z)}^{y_{\textrm{max}}(x,z)} dy\frac{1}{\sqrt{1-x^2-y^2-z^2+2 x y z}} \frac{\epsilon}{ \left(y^2+\epsilon ^2\right) }\nonumber \\
		& \quad \times \frac{\left(2 l_1 x+l_2 y\right)}{l_1 l_2 (x-i \epsilon )\left(l_1 x+l_2 y-i \epsilon \right)} \left[-\mu ^3 \left(4 \omega _1+5 \omega _2\right)+\mu ^4 \right. \nonumber \\
		&\left. \quad  +M_{\pi}^2 \left(\mu ^2-2 l_1^2 x^2-2 l_1 l_2 x y-3 l_2^2 y^2-3 \mu  \omega _2+6 \omega _2^2\right)+2 l_1^2 x^2 \left(l_2^2 y^2+3 \omega _2^2\right)\right. \nonumber \\
		& \left. \quad +2 l_1 l_2 x y \left(l_2^2 y^2+\mu  \left(5 \omega _2-2 \mu \right)+\omega _1 \left(4 \mu -6 \omega _2\right)\right)+\mu ^2 \left(-y^2l_2^2+4 \omega _1^2+6 \omega _2^2 \right. \right. \nonumber \\
		& \left. \left. \quad +16 \omega _1 \omega _2\right)+2 \mu  \left(l_2^2 y^2 \left(2 \omega _1+\omega _2\right)-6 \omega _1 \omega _2 \left(\omega _1+\omega _2\right)\right)-6 l_2^2 y^2 \omega _1 \omega _2-2 M_{\pi}^4\right].
	\end{align}
	The angular integration can be performed in the same way as for the spin-spin part, leading to    
	\begin{eqnarray}
		\label{Class7_VT_mu} 
		\textrm{Im}\; V^{\rm VII}_T(i \mu)  &=& \frac{3 g_\mathrm{A}^4}{2 \left(32 \pi F_\pi^3\right)^2} \iint \limits_{z^2<1} d\omega_1 d\omega_2 \frac{1}{ l_1 l_2 \sqrt{1-z^2} \mu^2}\left[\mu  \left(\mu -2 \omega _1\right) \left(\mu -2 \omega _1-2 \omega _2\right) \left(\mu -3 \omega _2\right) \right. \nonumber \\
		&&{} \left. \quad +M_{\pi}^2 \left(\mu ^2-3 \mu  \omega _2+6 \omega _2^2\right)-2 M_{\pi}^4\right]\nonumber \\
		&=& \frac{g_\mathrm{A}^4 \left( \mu- 3 M_{\pi}\right)}{ 35 \pi \left(32 \mu F_\pi^3\right)^2} \left[ \mu^3 + 3 \mu^2 M_{\pi} + 2 \mu M_{\pi}^2+ 6 M_{\pi}^3+ 18 \frac{M_{\pi}^4}{\mu}- 9 \frac{M_{\pi}^5}{\mu^2}-27  \frac{M_{\pi}^6}{\mu^3}\right].
	\end{eqnarray}    
	Our results for the class-VII contributions in Eqs.~(\ref{Class7_VS_mu}) and (\ref{Class7_VT_mu}) are identical to those of Ref.~\cite{Kaiser:1999jg}. 
	
	\subsection{Class-VIII diagrams}
	
	Class-VIII contributions are generated from one reducible-like and one irreducible diagram. By calculating the matrix elements of the operators in Eqs.~(\ref{3pioperator_gA^6}) and (\ref{ga6corterm}) corresponding to the two class-VIII diagrams, we find the following expressions for the energy denominators:
	\begin{eqnarray}
		\label{EDClass8irreducible}
		{\rm ED}^{\rm VIII}_{(2)} &=&\frac{\omega _1^4+\omega _2 \omega _1^3+\omega _2^2 \omega _1^2+\omega _2^3 \omega _1+\omega _2^4}{4 \omega _1^5 \omega _2^5 \left(\omega _1+\omega _2\right) \omega _3}-\frac{1}{4 \omega _1^5 \omega _2^2 \left(\omega _1+\omega _3\right)}-\frac{1}{4 \omega _1^4 \omega _2^2 \left(\omega _1+\omega _3\right){}^2}\nonumber \\
		&&{} -\frac{1}{4 \omega _1^2 \omega _2^5 \left(\omega _2+\omega _3\right)}-\frac{1}{4 \omega _1^2 \omega _2^4 \left(\omega _2+\omega _3\right){}^2}-\frac{1}{4 \omega _1^3 \omega _2^3 \left(\omega _1+\omega _2\right) \left(\omega _1+\omega _2+\omega _3\right)},\nonumber \\
		{\rm ED}^{\rm VIII, \, MUT}_{(1)} &=&{\rm ED}^{\rm VIII}_{(2)}  + \frac{1}{\omega _1^2 \omega _2^2 \omega _3^4}\,,
	\end{eqnarray}
	where the pions at nucleon $1$ are labeled as $1$, $3$ and $2$ from bottom to top. 
	When using the SMM, one  of course obtains the same expression for the irreducible diagram (2) after performing the integration over $l_1^0$, $l_2^0$. However, in this approach, the energy denominators for the reducible-like diagram (1) come out identically to those of the irreducible graph (2) \cite{Kaiser:1999jg},
	\beq
	\label{EDClass6reducibleSMM}
	{\rm ED}^{\rm VIII, \, SMM}_{(1)} ={\rm ED}^{\rm VIII}_{(2)}  \,,
	\eeq
	while in the MUT, the results differ by the last term in Eq.~(\ref{EDClass8irreducible}). This suggests that both methods lead to different potentials for this class of contributions.  
	
	The expression in the second equality of Eq.~(\ref{EDClass8irreducible}) can be cast into the form 
	\begin{align}
		{\rm ED}^{\rm VIII, \, MUT}_{(1)}  =& \int \frac{dl_{01}}{\left(2 \pi\right)}\int \frac{dl_{02}}{\left(2 \pi\right)}
		\frac{1}{(-l_{01}+ i \epsilon)^2} \, \frac{1}{(-l_{02}+ i \epsilon)^2} \, \frac{1}{l_{01}^2-\omega_{1}^2+ i \epsilon} \,
		\frac{1}{l_{02}^2-\omega_{2}^2+ i \epsilon} \frac{1}{\left(l_{01}+l_{02}\right)^2-\omega_{3}^2+ i \epsilon}\,,\label{eq_matching_Ladder_Diagram_HB}
	\end{align}
	which allows one to treat the reducible-like diagram (1) on the same footing as the irreducible graph (2) by applying our standard methods. Performing the tensor reduction, we find that the class-VIII diagrams generate nonvanishing isoscalar and isovector contributions of the central, spin-spin and tensor types. For the central potentials we can use the spectral representation as done for all previously considered cases. After performing the tensor reduction, the spectral functions of the central potentials are found to be
	\begin{eqnarray}
		\textrm{Im} \; V_{C, \, \rm (1)}^{\rm VIII, \, MUT}(i \mu) &=& \frac{3 g_{A}^6 \mu ^2}{\left(16 F_\pi^2\right)^3 \pi^4} \iint \limits_{z^2<1}  d \omega_1 d \omega_2  \int_{-1}^{1}dx \int_{y_{\textrm{min}}(x,z)}^{y_{\textrm{max}}(x,z)} dy  \frac{ \left(1-x^2-y^2-z^2+2 x y z\right)}{\sqrt{1-x^2-y^2-z^2+ 2x y z}(x+i \epsilon )^2 (\epsilon -i y)^2} \nonumber \\
		&=&- \frac{6 g_{A}^6 \mu ^2}{\left(16 \pi F_\pi^2\right)^3 } \iint \limits_{z^2<1}  d \omega_1 d \omega_2 \left(1-\frac{z \arccos(z)}{\sqrt{1-z^2}}\right) ,\nonumber \\
		\textrm{Im} \; 	W_{C,  \, \rm (1)}^{\rm VIII, \, MUT}(i \mu) &=& - \frac{7}{6}  \, \textrm{Im} \; V_{C, \, \rm (1)}^{\rm VIII, \, MUT}(i \mu) \,,
	\end{eqnarray}       
	while diagram (2) yields
	\begin{eqnarray}
		\textrm{Im} \; V_{C, \, \rm (2)}^{\rm VIII}(i \mu) &=&\frac{3 g_{A}^6 \mu ^2}{\left(16 F_\pi^2\right)^3 \pi^4} \iint \limits_{z^2<1}  d \omega_1 d \omega_2 \int_{-1}^{1}dx \int_{y_{\textrm{min}}(x,z)}^{y_{\textrm{max}}(x,z)} dy \frac{ \left(1-x^2-y^2-z^2+2 x y z\right)}{\sqrt{1-x^2-y^2-z^2+ 2x y z}(x+i \epsilon )^2 (i y+\epsilon)^2} \nonumber \\
		&=& {}- \frac{6 g_{A}^6 \mu ^2}{\left(16 \pi F_\pi^2\right)^3 }\iint \limits_{z^2<1}  d \omega_1 d \omega_2 \left(1+\frac{z \arccos(-z)}{\sqrt{1-z^2}}\right), \label{CrossBoxDiagram3PE_eq_SF_V_C}\\
		\textrm{Im} \; 	W_{C,  \, \rm (2)}^{\rm VIII}(i \mu)&=&\frac{7}{6}  \, \textrm{Im} \; V_{C, \, \rm (2)}^{\rm VIII}(i \mu) \,.
		\label{CrossBoxDiagram3PE_eq_SF_W_C}
	\end{eqnarray}
	Combining the two results together we obtain our final expressions for the class-VIII central contributions in the MUT:
	\begin{eqnarray}
		\textrm{Im} \; V_{C}^{\rm VIII, \, MUT}(i \mu) &=&-\frac{3 g_{A}^6 \mu ^2}{2 \left(8 \pi F_\pi^2\right)^3} \iint \limits_{z^2<1}  d \omega_1 d \omega_2  \left(1+ \frac{z}{\sqrt{1-z^2}}\left[\arccos(-z)-\frac{\pi}{2}\right]\right), \label{eq_Im_Vc_ClassVIII_MUT}  \\
		\textrm{Im} \; W_{C}^{\rm VIII, \, MUT}(i \mu) &=&-\frac{7 g_{A}^6 \mu ^2}{ \pi^2 \left(16 F_\pi^2\right)^3} \iint \limits_{z^2<1}  d \omega_1 d \omega_2  \frac{z}{\sqrt{1-z^2}} \nonumber \\
		&=&
		\frac{7 g_{A}^6 (\mu -3 M_{\pi})^2 }{60 \pi \left( 4 F_\pi\right)^6 }\left(4 \mu ^2+9 \mu M_{\pi}-2 M_{\pi}^2- \frac{3 M_{\pi}^3}{\mu}\right).
		\label{eq_Im_Wc_ClassVIII_MUT}                                          
	\end{eqnarray}
	This has to be compared with the results of Ref.~\cite{Kaiser:1999jg} obtained using the SMM:
	\begin{eqnarray}
		\label{Class8VCSMM}
		\textrm{Im} \; V_{C}^{{\rm VIII, \, SMM}}(i \mu) &=&-\frac{3 g_{A}^6 \mu ^2}{2 \left(8 \pi F_\pi^2\right)^3} \iint \limits_{z^2<1}  d \omega_1 d \omega_2  \left(1+ \frac{z}{\sqrt{1-z^2}}\arccos(-z)\right), \\	\label{Class9Temp2}
		\textrm{Im} \; W_{C}^{{\rm VIII, \, SMM}}(i \mu) &=& 0\,.
	\end{eqnarray}                                                        
	Notice that the difference between the two results originates from the last term in Eq.~(\ref{EDClass8irreducible}), whose contributions to the spectral functions is analytically calculable:
	\begin{eqnarray}
		\label{Class9Temp3}
		\textrm{Im} \; V_{C}^{\rm VIII, \, MUT}(i \mu)  -  \textrm{Im} \; V_{C}^{{\rm VIII, \, SMM}}(i \mu) &=&-\frac{6}{7}  \, \textrm{Im} \; W_{C}^{\rm VIII, \, MUT}(i \mu) \,.
	\end{eqnarray}                                                                                                 
	
	We now turn to the isovector spin-spin and tensor contributions of the class-VIII diagrams. As already pointed out above and discussed in Ref.~\cite{Kaiser:1999jg}, the approach we pursued so far leads for this particular type of contributions to singular expressions for the spectral functions in terms of the double integrals over $w_1$ and $w_2$. We therefore follow an alternative path and derive the corresponding coordinate-space potentials using the Wick-rotation method  introduced in sec.~\ref{sec:Wick}, which does not rely on the spectral representation. 
	For the sake of compactness, we introduce a short-hand notation with
	\begin{align}
		\langle\langle \dots\rangle \rangle = \int \frac{d^3q}{(2 \pi)^3} e^{i \vec{q}\cdot \vec{r}} \int \frac{d^3l_1}{(2 \pi)^3}\int \frac{d^3l_2}{(2 \pi)^3}  \int \frac{d^3l_3}{(2 \pi)^3} \int \frac{d^3x}{(2 \pi)^3} e^{i \vec{x}\cdot\left(\vec{l}_3-\vec{q}+\vec{l}_1-\vec{l}_2\right)}\, (\dots )\,.
		\label{NotationTemp1}
	\end{align}
	Before proceeding with the calculation of the spin-spin and tensor contributions, it is instructive to reproduce the already known results that could be obtained using the spectral representation. The isoscalar central coordinate-space potential can be written as
	\begin{eqnarray}
		V^{\rm VIII}_C(r)&=&- \frac{3 g_A^6}{32 F_\pi^6} 	\big\langle \big\langle  \big(
		{\rm ED}^{\rm VIII}_{(1)} + {\rm ED}^{\rm VIII}_{(2)} \big) \nonumber  \\
		&& {} \times\big[\big(\vec{l}_1 \cdot \vec{l}_3\big)^2 \vec{l}_2 \cdot \vec{l}_2 -2  \vec{l}_1 \cdot \vec{l}_2  \vec{l}_1 \cdot \vec{l}_3 \vec{l}_2 \cdot \vec{l}_3 +  \vec{l}_1 \cdot \vec{l}_1 \big( \vec{l}_2 \cdot \vec{l}_3\big)^2 + \big( \vec{l}_1 \cdot \vec{l}_2\big)^2  \vec{l}_3 \cdot \vec{l}_3- \vec{l}_1 \cdot \vec{l}_1  \vec{l}_2 \cdot \vec{l}_2  \vec{l}_3 \cdot \vec{l}_3\big]\big\rangle \big\rangle \,,
	\end{eqnarray}       
	with the expressions for the energy denominators in terms of the heavy-baryon propagators given in Eq.~(\ref{eq_matching_Ladder_Diagram_HB}) and
		\begin{eqnarray}
			{\rm ED}^{\rm VIII}_{(2)} & =&  \int \frac{d l_{01}}{\left(2 \pi \right)}\frac{dl_{02}}{\left(2 \pi \right)} \frac{1}{\left(-l_{01}+ i \epsilon\right)^2\left(-l_{02}+ i \epsilon\right)^2\left(l_{01}^2-\omega_{1}^2+ i \epsilon\right)\left(l_{02}^2-\omega_{2}^2+ i \epsilon\right) \big( (l_{02}-l_{01})^2-\omega_{3}^2+ i \epsilon \big) } .
		\end{eqnarray}
		Notice that the energy denominators of the irreducible class-VIII and class-IX diagrams fulfill the following relation (see the next section for the explicit expressions for the class-IX energy denominators)
		\beq
		\label{Class9Temp1}
		{\rm ED}^{\rm VIII}_{(2)}  +  {\rm ED}^{\rm IX}_{(1)} + {\rm ED}^{\rm IX}_{(2)} = \frac{\omega_{1}^2+\omega_{2}^2}{2 \omega_{1}^4 \omega_{2}^4\omega_{3}^2}\,,
		\eeq
		where for the class-IX diagrams the assignment of the pion labels corresponds to $1$, $3$ and $2$ from bottom to top at the nucleon $1$.
		The expressions for the combined energy denominators in terms of the pion energies for all four class-IX diagrams are given at the beginning of the next section. To reproduce the results of the SMM for the central potentials, we
		use Eq.~(\ref{EDClass6reducibleSMM}) and express $ {\rm ED}^{\rm VIII}_{(2)}$ in terms of
		\beqa
		{\rm ED}^{\rm IX}_{(1)} &=& \int \frac{d l_{01}}{\left(2 \pi \right)}\frac{dl_{02}}{\left(2 \pi \right)} \frac{1}{\left(l_{01}- i \epsilon\right)^2\left(-l_{02}+ i \epsilon\right)\left(l_{02}-l_{01}+ i \epsilon\right)\left(l_{01}^2-\omega_{1}^2+ i \epsilon\right) \left(l_{02}^2-\omega_{2}^2+ i \epsilon\right)\big( (l_{02}-l_{01})^2-\omega_{3}^2+ i \epsilon\big)} \,, \nonumber \\
		{\rm ED}^{\rm IX}_{(2)} &=& \int \frac{d l_{01}}{\left(2 \pi \right)}\frac{dl_{02}}{\left(2 \pi \right)} \frac{1}{\left(-l_{01}+ i \epsilon\right)\left(l_{02}- i \epsilon\right)^2\left(l_{01}-l_{02}+ i \epsilon\right)\left(l_{01}^2-\omega_{1}^2+ i \epsilon\right) \left(l_{02}^2-\omega_{2}^2+ i \epsilon\right)\big( (l_{02}-l_{01})^2-\omega_{3}^2+ i \epsilon\big) } \,,
		\nonumber
		\eeqa
		using Eq.~(\ref{Class9Temp1}) to arrive at 
		\begin{align}
			V_{C}^{\rm VIII, \, SMM}(r) &= \frac{6 g_{A}^6 }{ \left(16 \pi F_\pi^2\right)^3 r^7}\bigg\{-e^{-3x}\left(1 + x\right)\left(4 + 5x+ 3x^2\right) \nonumber \\
			& \quad+ \dashint_{-\infty}^{\infty}  \frac{d \zeta_1 d \zeta_2 }{\pi^2 \zeta_1 \zeta_2} \frac{\partial^2}{\partial \zeta_1\partial \zeta_2}e^{-x\left(\alpha+\beta + \gamma \right)} \left[x \left(\alpha ^2 (\beta +\gamma )+\alpha  \left(\beta ^2+6 \beta  \gamma +\gamma ^2\right)+\beta  \gamma  (\beta +\gamma )\right) \nonumber \right. \\
			& \left. \quad+ \alpha ^2+6 \alpha  (\beta +\gamma )+\beta ^2+6 \beta  \gamma +\gamma ^2+\alpha  \beta  \gamma  x^2 (\alpha +\beta +\gamma )+6 x^{-1} (\alpha +\beta +\gamma )+6 x^{-2}\right] \bigg\}\label{eq_Vc_ClassVIII_WR_SMM},
		\end{align}
		where we have introduced the dimensionless variable $x = M_{\pi} r$, while the dimensionless parameters $\alpha$, $\beta$ and $\gamma$ are specified in Eq.~(\ref{alphabetagamma}). Further, $\dashint$ denotes the principal value integrals. 
		Notice that the first (analytical) term on the right-hand side of the above equation stems from the last term in Eq.~(\ref{Class9Temp1}).
		In order to calculate the double integral in Eq.~(\ref{eq_Vc_ClassVIII_WR_SMM}), we make use of the relation
		\begin{align}
			\dashint_{-\infty}^{\infty} d \zeta f(\zeta)\zeta^{-1} = \int_{0}^{\infty} d\zeta \left[f(\zeta)-f(-\zeta)\right]\zeta^{-1}
		\end{align}
		given in Ref.~\cite{Kaiser:2001dm}.
		We have then verified numerically that the potential in Eq.~(\ref{eq_Vc_ClassVIII_WR_SMM}) agrees with the one obtained using the spectral function specified in Eq.~(\ref{Class8VCSMM}).  
		The result for the isoscalar central potential using the MUT can be obtained analogously by using Eq.~(\ref{EDClass8irreducible}) for ${\rm ED}^{\rm VIII, \, MUT}_{(1)}$. Notice that the last term in  Eq.~(\ref{EDClass8irreducible}) cancels against the term on the right-hand side of Eq.~(\ref{Class9Temp1}), leading to
		\begin{align}
			V_{C}^{\rm VIII, \, MUT}(r) &= \frac{6 g_{A}^6 }{ \left(16 \pi F_\pi^2\right)^3 r^7} \dashint_{-\infty}^{\infty}  \frac{d \zeta_1 d \zeta_2 }{\pi^2 \zeta_1 \zeta_2} \frac{\partial^2}{\partial \zeta_1\partial \zeta_2}e^{-x\left(\alpha+\beta + \gamma \right)} \left[x \left(\alpha ^2 (\beta +\gamma )+\alpha  \left(\beta ^2+6 \beta  \gamma +\gamma ^2\right) \nonumber \right.\right. \\
			& \left. \left. \quad+\beta  \gamma  (\beta +\gamma )\right)+ \alpha ^2+6 \alpha  (\beta +\gamma )+\beta ^2+6 \beta  \gamma +\gamma ^2+\alpha  \beta  \gamma  x^2 (\alpha +\beta +\gamma )\right. \nonumber \\
			& \left. \quad+6 x^{-1} (\alpha +\beta +\gamma )+6 x^{-2}\right].\label{eq_Vc_ClassVIII_WR_with_Correc}
		\end{align}
		We further emphasize that the difference between the two results,
		\beq
		V_{C}^{\rm VIII, \, MUT}(r)  - V_{C}^{\rm VIII, \, SMM}(r) = \frac{6 g_{A}^6 }{ \left(16 \pi F_\pi^2\right)^3} \, \frac{e^{-3x}}{r^7}\left(1 + x\right)\left(4 + 5x+ 3x^2\right)\,,
		\eeq
		agrees exactly with the Fourier transform of the expression in Eq.~(\ref{Class9Temp3}).
		
		Next, consider the isovector central contribution. It is easy to see that starting from the representation 
		\begin{eqnarray}
			W^{\rm VIII}_C(r)&=& \frac{7 g_A^6}{64 F_\pi^6}
			\big\langle \big\langle  \big(
			{\rm ED}^{\rm VIII}_{(1)} - {\rm ED}^{\rm VIII}_{(2)} \big)  \nonumber \\
			&& {} \times \big[\big(\vec{l}_1 \cdot \vec{l}_3\big)^2 \vec{l}_2 \cdot \vec{l}_2 -2  \vec{l}_1 \cdot \vec{l}_2  \vec{l}_1 \cdot \vec{l}_3 \vec{l}_2 \cdot \vec{l}_3 +  \vec{l}_1 \cdot \vec{l}_1 \big( \vec{l}_2 \cdot \vec{l}_3\big)^2 + \big( \vec{l}_1 \cdot \vec{l}_2\big)^2  \vec{l}_3 \cdot \vec{l}_3- \vec{l}_1 \cdot \vec{l}_1  \vec{l}_2 \cdot \vec{l}_2  \vec{l}_3 \cdot \vec{l}_3\big]\big\rangle \big\rangle\,,
		\end{eqnarray}
		one obtains
		\beqa
		\label{Class8_Wc_r_MUT}
		W_{C}^{\rm VIII, \, MUT}(r) &=&- \frac{7 g_{A}^6 }{ \left( 16 \pi F_\pi^2\right)^3}\frac{e^{-3 x}}{r^7}\left(1 + x\right)\left(4 + 5 x+ 3x^2\right), \\
		W_{C}^{\rm VIII, \, SMM}(r) &=& 0\,,\nonumber
		\eeqa
		in agreement with the previously obtained results in Eqs.~(\ref{eq_Im_Vc_ClassVIII_MUT}) and (\ref{Class8VCSMM}). Similarly, the isoscalar spin-spin and tensor potentials can be obtained from 
		\begin{eqnarray}
			V^{\rm VIII}_{S, T}(r)& =& - \frac{3 g_A^6}{32 F_\pi^6}	\big\langle \big\langle  \big(
			{\rm ED}^{\rm VIII}_{(1)} - {\rm ED}^{\rm VIII}_{(2)} \big)  \nonumber \\
			&& {} \times\big(\vec{l}_1\cdot \vec{\sigma}_1 \vec{l}_2\cdot \vec{l}_3 + \vec{l}_1 \cdot \vec{l}_3 \vec{l}_2\cdot \vec{\sigma}_1- \vec{l}_1 \cdot \vec{l}_2 \vec{l}_3\cdot \vec{\sigma}_1\big) \big(\vec{l}_1\cdot \vec{\sigma}_2 \vec{l}_2 \cdot \vec{l}_3+ \vec{l}_1 \cdot \vec{l}_3 \vec{l}_2\cdot \vec{\sigma}_2 - \vec{l}_1 \cdot \vec{l}_2 \vec{l}_3 \cdot \vec{\sigma}_2\big) \big\rangle \big\rangle\,,
		\end{eqnarray}
		which leads to 
		\beqa
		\label{Class8_VS_r_MUT}
		V_{S}^{\rm VIII, \, MUT}(r) &=& \frac{g_A^6 e^{- 3 x}}{\left(16 \pi F_\pi^2\right)^3 r^7} \left(8 +18 x +16x^2 +6x^3 +2x^4 -x^5\right),  \\
		\label{Class8_VT_r_MUT}
		V_{T}^{\rm VIII, \, MUT}(r) &=& - \frac{g_A^6 e^{- 3 x}}{\left(16 \pi F_\pi^2\right)^3 r^7}\left(1+ x \right) \left(16+26 x+18x^2 +6x^3 +x^4\right),  \\
		V_{S}^{\rm VIII, \, SMM}(r) &=&V_{T}^{\rm VIII, \, SMM}(r) = 0\,,
		\eeqa        
		upon inserting the expressions for the energy denominators.
                The expressions for the potentials $V_{S}^{\rm VIII, \, MUT}(r)$ and $V_{T}^{\rm VIII, \, MUT}(r)$ correspond to the spectral functions
\beqa
\label{VS8MUTSpF}
\textrm{Im}\;V_{S}^{\rm VIII, \, MUT}(i \mu) &=& -\frac{2 g_A^6}{7\pi \left(32 F_{\pi}^2\right)^3}\bigg[8 \mu^4 -21 \mu^3 M_{\pi} - 56 \mu^2 M_{\pi}^2  +147 \mu M_{\pi}^3 + 112 M_{\pi}^4- 399\frac{M_{\pi}^5}{\mu} +81\frac{M_{\pi}^7}{\mu^3}\bigg], \\
 \label{VT8MUTSpF}               
			\textrm{Im}\;V_{T}^{\rm VIII, \, MUT}(i \mu) &=& -\frac{2 g_A^6}{35 \pi \left(32 F_{\pi}^2\right)^3}\left[64 \mu^2 -105 \mu M_{\pi} - 560 M_{\pi}^2 + 945\frac{M_{\pi}^3}{\mu}+ 273\frac{M_{\pi}^5}{\mu^3}+ 1215\frac{M_{\pi}^7}{\mu^5 }\right] .
		\eeqa
		It remains to derive the isovector spin-spin and tensor contributions starting from the representation
		\begin{eqnarray}
			W^{\rm VIII}_{S, T}(r)&=&\frac{7g_A^6}{64 F_\pi^6}\big\langle \big\langle  \big(
			{\rm ED}^{\rm VIII}_{(1)} + {\rm ED}^{\rm VIII}_{(2)} \big)  \nonumber \\
			&& {} \times \big(\vec{l}_1 \cdot \vec{\sigma}_1 \vec{l}_2 \cdot \vec{l}_3+\vec{l}_1 \cdot  \vec{l}_3 \vec{l}_2 \cdot \vec{\sigma}_1- \vec{l}_1 \cdot \vec{l}_2 \vec{l}_3 \cdot \vec{\sigma}_1\big) \big( \vec{l}_1 \cdot \vec{\sigma}_2 \vec{l}_2 \cdot \vec{l}_3+\vec{l}_1 \cdot  \vec{l}_3 \vec{l}_2 \cdot \vec{\sigma}_2- \vec{l}_1 \cdot \vec{l}_2 \vec{l}_3 \cdot \vec{\sigma}_2\big) \big\rangle \big\rangle \,.
		\end{eqnarray}
		Following the same steps as before, we obtain 
		\beqa
		W_{S}^{\rm VIII, \, MUT}(r)  &=& -\frac{7g_A^6}{6 \left( 16 \pi F_\pi^2\right)^3 r^7}\dashint_{-\infty}^{\infty}  \frac{d \zeta_1 d \zeta_2 }{\pi^2 \zeta_1 \zeta_2} \frac{\partial^2}{\partial \zeta_1 \partial \zeta_2}e^{-x\left(\alpha+\beta + \gamma \right)} \left[12x^{-2}+12 x^{-1} (\alpha +\beta +\gamma )-\alpha ^2 \beta ^2 \gamma ^2 x^4 \right.  \nonumber \\ 
		&&{}\left. +2 x \left(\alpha ^2 (\beta +\gamma )+\alpha  \left(\beta ^2+6 \beta  \gamma +\gamma ^2\right)+\beta  \gamma  (\beta +\gamma )\right) \right.  \nonumber \\ 
		&&{}\left. +2 \left(\alpha ^2+6 \alpha  (\beta +\gamma )+\beta ^2+6 \beta  \gamma +\gamma ^2\right)+2 \alpha  \beta  \gamma x^2(\alpha+\beta+\gamma)\right],\label{eq_Ws_ClassVIII_MUT} \\
		W_{T}^{\rm VIII, \, MUT}(r)  &=& \frac{7g_A^6}{6 \left( 16 \pi F_\pi^2\right)^3 r^7} \dashint_{-\infty}^{\infty}  \frac{d \zeta_1 d \zeta_2 }{\pi^2 \zeta_1 \zeta_2} \frac{\partial^2}{\partial \zeta_1 \partial \zeta_2}e^{-x\left(\alpha+\beta + \gamma \right)} \left[36x^{-2}+36x^{-1}+9\alpha ^2 \beta ^2 \gamma ^2 x^4 \right. \nonumber \\
		&&{} \left. +x^2 \left(\alpha ^2 (3 \beta +\gamma ) (\beta +3 \gamma )+10 \alpha  \beta  \gamma  (\beta +\gamma )+3 \beta ^2 \gamma^2\right)\right. \nonumber \\
		&&{} \left.  +2 x \left(5 \alpha ^2 (\beta +\gamma )+\alpha  \left(5 \beta ^2+18 \beta  \gamma +5 \gamma ^2\right)+5 \beta  \gamma  (\beta +\gamma )\right)\right. \nonumber \\
		&&{} \left. +2 \left(5 \alpha ^2+18 \alpha  (\beta +\gamma )+5 \beta ^2+18 \beta  \gamma +5 \gamma ^2\right) +3 \alpha  \beta  \gamma  x^3 (\alpha  (\beta +\gamma )+\beta  \gamma )\right].\label{eq_Wt_ClassVIII_MUT}       
		\eeqa
		The corresponding SMM results have the form 
		\beqa
		\label{Class8_WS_r_SMM}
		W_{S}^{\rm VIII, \, SMM}(r)  &=& W_{S}^{\rm VIII, \, MUT}(r) + \frac{7g_A^6}{6 \left( 16 \pi F_\pi^2\right)^3} \, \frac{e^{- 3x}}{r^7}
		\,(8 +18 x +16x^2 +6x^3 +2x^4 -x^5)\,,  \\
		W_{T}^{\rm VIII, \, SMM}(r)  &=& W_{T}^{\rm VIII, \, MUT}(r) - \frac{7g_A^6}{6 \left( 16 \pi F_\pi^2\right)^3}\, \frac{e^{- 3x}}{r^7} \, (1+x) (16+26x+18 x^2+6 x^3+x^4) \,.
		\label{Class8_WT_r_SMM}
		\eeqa  
		While these expressions are less compact than the ones given in Ref.~\cite{Kaiser:2001dm}, we have verified that our SMM results are identical to those given in that paper. Notice further that the differences in the isovector spin-spin and tensor potentials obtained using the SMM and MUT are stemming from the last term in Eq.~(\ref{EDClass8irreducible}) and can be calculated using the spectral function representation.

		\subsection{Class-IX diagrams}
		
		Class-IX diagrams also involve reducible topologies. To calculate the energy denominators for the four diagrams of class IX, we use the following labeling of the pion lines: Starting from the bottom, the first pion at nucleon $1$ is labelled with $l_1$, the second one with $l_3$ and the last one with $l_2$. 
		With this in mind, we obtain the following expressions for the energy denominators by calculating the matrix elements of the operators in Eqs.~(\ref{3pioperator_gA^6}) and (\ref{ga6corterm}) corresponding to the class-IX diagrams:
		\begin{eqnarray}
			\label{EDClass9irreducible}
			{\rm ED}^{\rm IX}_{(1)} &=&\frac{\omega _1^2+\omega _2^2}{4 \omega _1^4 \omega _2^4 \omega _3^2} + \frac{-2 \omega _1^5-4 \omega _2 \omega _1^4-2 \omega _2^2 \omega _1^3+2 \omega _2^4 \omega _1+\omega _2^5}{4 \omega _1^5 \omega _2^5 \left(\omega _1+\omega _2\right){}^2 \omega _3}-\frac{1}{4 \omega _1^5 \omega _2^2 \left(\omega _1+\omega _3\right)}
			+\frac{1}{4 \omega _1^2 \omega _2^4 \left(\omega _2+\omega _3\right){}^2}
			\nonumber\\
			&& {} +\frac{1}{4 \omega _1^3 \omega _2^2 \left(\omega _1+\omega _2\right){}^2 \left(\omega _1+\omega _2+\omega _3\right)}+\frac{1}{2 \omega _1^2 \omega _2^5 \left(\omega _2+\omega _3\right)}\,,
			\nonumber \\
			{\rm ED}^{\rm IX}_{(2)} &=&\frac{\omega _1^2+\omega _2^2}{4 \omega _1^4 \omega _2^4 \omega _3^2}+\frac{\omega _1^5+2 \omega _2 \omega _1^4-2 \omega _2^3 \omega _1^2-4 \omega _2^4 \omega _1-2 \omega _2^5}{4 \omega _1^5 \omega _2^5 \left(\omega _1+\omega _2\right){}^2 \omega _3}+\frac{1}{4 \omega _1^4 \omega _2^2 \left(\omega _1+\omega _3\right){}^2} -\frac{1}{4 \omega _1^2 \omega _2^5 \left(\omega _2+\omega _3\right)}\nonumber\\
			&& {}+\frac{1}{4 \omega _1^2 \omega _2^3 \left(\omega _1+\omega _2\right){}^2 \left(\omega _1+\omega _2+\omega _3\right)}+\frac{1}{2 \omega _1^5 \omega _2^2 \left(\omega _1+\omega _3\right)} \,, \nonumber \\
			{\rm ED}^{\rm IX, \, SMM}_{(3)} &=& -\frac{\omega _1^2+\omega _2^2}{4 \omega _1^4 \omega _2^4 \omega _3^2}+\frac{-2 \omega _1^5-4 \omega _2 \omega _1^4-2 \omega _2^2 \omega _1^3+2 \omega _2^4 \omega _1+\omega _2^5}{4 \omega _1^5 \omega _2^5 \left(\omega _1+\omega _2\right){}^2 \omega _3}+\frac{1}{2 \omega _1^2 \omega _2^5 \left(\omega _2+\omega _3\right)} +\frac{1}{4 \omega _1^2 \omega _2^4 \left(\omega _2+\omega _3\right){}^2}\nonumber \\
			&& {}+\frac{1}{4 \omega _1^3 \omega _2^2 \left(\omega _1+\omega _2\right){}^2 \left(\omega _1+\omega _2+\omega _3\right)}-\frac{1}{4 \omega _1^5 \omega _2^2 \left(\omega _1+\omega _3\right)}\,,\nonumber \\
			{\rm ED}^{\rm IX, \, SMM}_{(4)} &=&-\frac{\omega _1^2+\omega _2^2}{4 \omega _1^4 \omega _2^4 \omega _3^2}+\frac{\omega _1^5+2 \omega _2 \omega _1^4-2 \omega _2^3 \omega _1^2-4 \omega _2^4 \omega _1-2 \omega _2^5}{4 \omega _1^5 \omega _2^5 \left(\omega _1+\omega _2\right){}^2 \omega _3}+\frac{1}{4 \omega _1^4 \omega _2^2 \left(\omega _1+\omega _3\right){}^2} +\frac{1}{2 \omega _1^5 \omega _2^2 \left(\omega _1+\omega _3\right)}\nonumber\\
			&&{} -\frac{1}{4 \omega _1^2 \omega _2^5 \left(\omega _2+\omega _3\right)}+\frac{1}{4 \omega _1^2 \omega _2^3 \left(\omega _1+\omega _2\right){}^2 \left(\omega _1+\omega _2+\omega _3\right)}\,.
		\end{eqnarray}
                While the energy denominators ${\rm ED}^{\rm IX}_{(1)} $ and  ${\rm ED}^{\rm IX}_{(2)} $ corresponding to the two irreducible diagrams are the same in both approaches, those corresponding to the reducible-like diagrams turn out to be different in the MUT:
		\begin{eqnarray}
			\label{eq_ResultEnergyDenomMUTClassIX}
			{\rm ED}^{\rm IX, \, MUT}_{(3)} &=& {\rm ED}^{\rm IX, \, SMM}_{(3)}  - \frac{1}{2 \omega _1^2 \omega _2^2 \omega _3^4}\,,\nonumber \\
			{\rm ED}^{\rm IX, \, MUT}_{(4)} &=& {\rm ED}^{\rm IX, \, SMM}_{(4)}  - \frac{1}{2 \omega _1^2 \omega _2^2 \omega _3^4}\,. 
		\end{eqnarray}  
		Consequently, the SMM and MUT are expected to yield different class-IX potentials. 
		
		The class-IX contributions have been derived using the SMM in Refs.~\cite{Kaiser:1999jg,Kaiser:2001dm}. Specifically, the expressions for the spectral functions of the isoscalar central, spin-spin and tensor potentials as well as of the isovector central potential have been worked out in Ref.~\cite{Kaiser:1999jg}. For the isovector spin-spin and tensor contributions, the method based on the Cutkosky cutting rules described in sec.~\ref{Subsec:Frame1} yields ill-defined results  (similarly to the class-VIII case). In Ref.~\cite{Kaiser:2001dm}, the corresponding $r$-space potentials were obtained in the SMM using the Wick-rotation method. In the following, we rederive the above-mentioned expressions from Ref.~\cite{Kaiser:1999jg,Kaiser:2001dm} and subsequently work out the corresponding results using the MUT. 
		
		The isoscalar and isovector central potentials can be calculated in the usual way by using the Cutkosky cutting rules. Combining the expressions for all four Feynman diagrams, switching the sign in front of $i \epsilon$ in one of the nucleon propagators in diagrams (3) and (4) to eliminate the iterative contributions and projecting the result on the isoscalar central channel we obtain 
		\begin{align}
			\textrm{Im} \; V_{C}^{\rm IX, \, SMM}(i \mu) =&  \frac{3 g_A^6 \mu^2}{2 \pi \left(8 \pi F_\pi^2\right)^3}\iint\limits_{z^2<1} d \omega_{1} d \omega_{2} \int_{-1}^{1}dx \int_{y_{\textrm{min}}(x,z)}^{y_{\textrm{max}}(x,z)} dy\frac{1}{\sqrt{1-x^2-y^2-z^2 + 2 x y z}}\frac{y}{\left(y^2+\epsilon ^2\right)} \nonumber \\
			&\times l_2\frac{\left(1-x^2-y^2-z^2 + 2 x y z\right)}{(x+i \epsilon )^2 (l_1 x+l_2 y+i \epsilon )} \,.
		\end{align}
		The angular integrations are carried out in detail in appendix \ref{subsec_SolvingImVcClassIX}, yielding the following double-integral representation:
		\begin{align}
			\label{Class9Temp1a}
			\textrm{Im} \; V_{C}^{\rm IX, \, SMM}(i \mu) =& \frac{3 g_A^6 \mu^2}{2\left(8 \pi F_\pi^2 \right)^3}\iint \limits_{z^2<1} d \omega_{1} d \omega_{2} \bigg\{1 - \frac{1}{\sqrt{1-z^2}}\left[\pi \left(\frac{l_2}{2l_1}+z\right)-z \arccos(-z)\right]\bigg\}\,.
		\end{align}
		For the isovector central contribution, we obtain
		\begin{eqnarray}
			\label{Class9WCSMM} 
			\textrm{Im} \; W_{C}^{\rm IX, \, SMM}(i \mu) &=&  \frac{2 g_A^6 \mu^2}{\pi \left(16 \pi F_\pi^2\right)^3}\iint \limits_{z^2<1} d \omega_{1} d \omega_{2}  \int_{-1}^{1}dx \int_{y_{\textrm{min}}(x,z)}^{y_{\textrm{max}}(x,z)} dy \frac{1}{\sqrt{1-x^2-y^2-z^2 + 2 x y z}} \frac{\epsilon}{\left(y^2+\epsilon ^2\right)}\nonumber\\
			&\times& l_2 \frac{\left(1-x^2-y^2-z^2 + 2 x y z\right)}{ (x+i \epsilon )^2 (l_1 x+l_2 y+i \epsilon )}\,.
		\end{eqnarray}
		The angular integrations can be solved in a similar way as done for the class-VII contributions, leading to 
		\begin{eqnarray}
			\label{Class9Temp2a}
			\textrm{Im} \; W_{C}^{\rm IX, \, SMM}(i \mu) &=&  - \frac{ g_A^6 \mu^2}{\left(16 F_\pi^2\right)^3 \pi^2} \iint \limits_{z^2<1} d \omega_{1} d \omega_{2} \frac{l_2^2}{\sqrt{1-z^2}}\nonumber \\
			&=&\frac{g_A^6 \left(\mu - 3 M_{\pi}\right)^2}{30 \pi \mu \left(4 F_\pi\right)^6}\left(3M_{\pi}^3+ 2 \mu M_{\pi}^2- 9 \mu^2 M_{\pi} - 4 \mu^3\right)\,.
		\end{eqnarray}
		Both expressions for $\textrm{Im} \; V_{C}^{\rm IX, \, SMM}(i \mu) $ in Eq.~(\ref{Class9Temp1a}) and for $\textrm{Im} \; W_{C}^{\rm IX, \, SMM}(i \mu)$ in Eq.~(\ref{Class9Temp2a}) coincide with the results given in Ref.~\cite{Kaiser:1999jg}. 
		
                The remaining contributions can be most conveniently  derived using the Wick-rotation method. The resulting expressions for the spin-spin and tensor potentials in coordinate space will turn out to be very simple and will be used to extract the corresponding spectral functions.
            For the isovector spin-spin and tensor contributions, we restrict ourselves to the expressions in coordinate space as also done in Ref.~\cite{Kaiser:2001dm}. 

                Following the same strategy as for the class-VIII diagrams, we begin with reproducing the already known results for the central potentials. The isoscalar central term has the form
		\begin{eqnarray}
			\label{class9VC}
			V_C^{\rm IX}(r) &=&-\frac{3g_A^6}{32 F_\pi^6} \big\langle \big\langle  \big(   {\rm ED}^{\rm IX}_{(1)}  +    {\rm ED}^{\rm IX}_{(2)}  +    {\rm ED}^{\rm IX}_{(3)}  +    {\rm ED}^{\rm IX}_{(4)} \big) \big[ \vec{l}_1 \cdot \vec{l}_3  \vec{l}_1 \cdot \vec{l}_3 \vec{l}_2 \cdot \vec{l}_2-2\vec{l}_1 \cdot \vec{l}_2 \vec{l}_1 \cdot \vec{l}_3\vec{l}_2 \cdot \vec{l}_3 \nonumber \\ 
			&&{}  + \vec{l}_1 \cdot \vec{l}_2 \vec{l}_1 \cdot \vec{l}_2 \vec{l}_3 \cdot \vec{l}_3 + \vec{l}_1 \cdot \vec{l}_1 \big( \vec{l}_2 \cdot \vec{l}_3 \vec{l}_2 \cdot \vec{l}_3 - \vec{l}_2 \cdot \vec{l}_2 \vec{l}_3 \cdot \vec{l}_3\big)\big] \big\rangle \big\rangle  \,.
		\end{eqnarray}
		We replace the first two energy denominators using Eq.~(\ref{Class9Temp1}), write the expressions for the energy denominators ${\rm ED}^{\rm IX}_{(3)}$ and ${\rm ED}^{\rm IX}_{(4)}$ in terms of the Feynman propagators with $\vec{l}_3 = \vec{q}- \vec{l}_1 + \vec{l}_2$,
		\begin{align}
		{\rm ED}^{\rm IX, \, SMM}_{(3)} &= \int \frac{d l_{01}}{\left(2 \pi \right)}\frac{dl_{02}}{\left(2 \pi \right)} \frac{-1}{\left(-l_{01}+ i \epsilon\right)^2\left(-l_{02}+ i \epsilon\right)\left(l_{01}-l_{02}+ i \epsilon\right)\left(l_{01}^2-\omega_{1}^2+ i \epsilon\right) \left(l_{02}^2-\omega_{2}^2+ i \epsilon\right)\big( (l_{02}-l_{01})^2-\omega_{3}^2+ i \epsilon\big)}
		\,, \nonumber \\
		{\rm ED}^{\rm IX, \, SMM}_{(4)} &= \int \frac{d l_{01}}{\left(2 \pi \right)}\frac{dl_{02}}{\left(2 \pi \right)} \frac{-1}{\left(-l_{01}+ i \epsilon\right)\left(-l_{02}+ i \epsilon\right)^2\left(l_{01}-l_{02}+ i \epsilon\right)\left(l_{01}^2-\omega_{1}^2+ i \epsilon\right) \left(l_{02}^2-\omega_{2}^2+ i \epsilon\right)\big( (l_{02}-l_{01})^2-\omega_{3}^2+ i \epsilon\big) }\,,
		\nonumber
		\end{align}
		and obtain for the isoscalar central part in the SMM
		\begin{align}
			V_{C}^{\rm IX, \, SMM}(r) &= \frac{6 g_{A}^6 }{ \left(16 \pi F_\pi^2\right)^3}\bigg\{\frac{e^{-3x}}{r^7}\left(1 + x\right)\left(4 + 5x+ 3x^2\right) \nonumber \\
			& \quad- \dashint_{-\infty}^{\infty}  \frac{d \zeta_1 d \zeta_2 }{\pi^2 \zeta_1 \zeta_2} \frac{\partial^2}{\partial \zeta_1\partial \zeta_2}e^{-x\left(\alpha+\beta + \gamma \right)} \left[x \left(\alpha ^2 (\beta +\gamma )+\alpha  \left(\beta ^2+6 \beta  \gamma +\gamma ^2\right)+\beta  \gamma  (\beta +\gamma )\right) \nonumber \right. \\
			& \left. \quad+ \alpha ^2+6 \alpha  (\beta +\gamma )+\beta ^2+6 \beta  \gamma +\gamma ^2+\alpha  \beta  \gamma  x^2 (\alpha +\beta +\gamma )+6 x^{-1} (\alpha +\beta +\gamma )+6 x^{-2}\right] \bigg\}\,.\label{eq_Vc_ClassIX_WR}
		\end{align}
		We have verified numerically that this result is consistent with the one in Eq.~(\ref{Class9Temp1a}) obtained using the spectral-function method.
		
		For the isovector central potential, we start with the expression 
		\begin{eqnarray}
			\label{class9WC}
			W_C^{\rm IX}(r) &=&\frac{g_A^6}{64 F_\pi^6}  \big\langle \big\langle  \big( -  {\rm ED}^{\rm IX}_{(1)}  -    {\rm ED}^{\rm IX}_{(2)}  +    {\rm ED}^{\rm IX}_{(3)}  +    {\rm ED}^{\rm IX}_{(4)} \big) \big[ \vec{l}_1 \cdot \vec{l}_3  \vec{l}_1 \cdot \vec{l}_3 \vec{l}_2 \cdot \vec{l}_2-2\vec{l}_1 \cdot \vec{l}_2 \vec{l}_1 \cdot \vec{l}_3\vec{l}_2 \cdot \vec{l}_3  \nonumber \\ 
			&&{}  + \vec{l}_1 \cdot \vec{l}_2 \vec{l}_1 \cdot \vec{l}_2 \vec{l}_3 \cdot \vec{l}_3 + \vec{l}_1 \cdot \vec{l}_1 \big( \vec{l}_2 \cdot \vec{l}_3 \vec{l}_2 \cdot \vec{l}_3 - \vec{l}_2 \cdot \vec{l}_2 \vec{l}_3 \cdot \vec{l}_3\big)\big] \big\rangle \big\rangle  \,.
		\end{eqnarray}
		Using Eq.~(\ref{EDClass9irreducible}), we see that the sum of the energy denominators in the SMM takes a particularly simple form 
		\begin{align}
			-  {\rm ED}^{\rm IX}_{(1)}  -    {\rm ED}^{\rm IX}_{(2)}  +    {\rm ED}^{\rm IX, \, SMM}_{(3)}  +    {\rm ED}^{\rm IX, \, SMM}_{(4)} = -\frac{\omega_{1}^2+\omega_{2}^2}{2 \omega_{1}^4 \omega_{2}^4 \omega_{3}^2}\,,
		\end{align}
		which leads to a compact analytical expression for the corresponding $r$-space potential
		\begin{align}
			\label{Class9_WC_r_SMM}
			W^{\rm IX, \, SMM}_C(r)= \frac{2g_{A}^6 }{ \left(16 \pi F_\pi^2\right)^3}\frac{e^{-3x}}{r^7}\left(1 + x\right)\left(4 + 5x+ 3x^2\right)\,.
		\end{align}
		Clearly, the same expression is obtained from the spectral function in Eq.~(\ref{Class9Temp2a}). 
		
		We now move on with calculating the isoscalar spin-spin and tensor potentials using the SMM, which are given by  
		\begin{eqnarray}
			\label{class9VST}
			V^{\rm IX}(r) &=& \frac{3 g_{\textrm{A}}^6}{32 F_\pi^6}  \Big\langle \Big\langle  \big(  \vec{l}_1\cdot \vec{\sigma}_1 \vec{l}_2\cdot\vec{l}_3 + \vec{l}_1 \cdot\vec{l}_3 \vec{l}_2 \cdot \vec{\sigma}_1  - \vec{l}_1 \cdot \vec{l}_2 \vec{l}_3 \cdot \vec{\sigma}_1\big)\nonumber \\
			&\times & \Big[\big(  -  {\rm ED}^{\rm IX}_{(1)}  +    {\rm ED}^{\rm IX}_{(2)}  +    {\rm ED}^{\rm IX}_{(3)}  -    {\rm ED}^{\rm IX}_{(4)} \big) \vec{l}_1 \cdot \vec{\sigma}_2 \vec{l}_2 \cdot \vec{l}_3 + \big(   {\rm ED}^{\rm IX}_{(1)}  -    {\rm ED}^{\rm IX}_{(2)}  -    {\rm ED}^{\rm IX}_{(3)}  +    {\rm ED}^{\rm IX}_{(4)}  \big)\vec{l}_1 \cdot \vec{l}_3 \vec{l}_2 \cdot \vec{\sigma}_2 \nonumber \\
			&+& \big( -  {\rm ED}^{\rm IX}_{(1)}  -    {\rm ED}^{\rm IX}_{(2)}  +    {\rm ED}^{\rm IX}_{(3)}  +    {\rm ED}^{\rm IX}_{(4)}  \big) \vec{l}_1 \cdot \vec{l}_2 \vec{l}_3 \cdot \vec{\sigma}_2\Big] \Big\rangle \Big\rangle \,. 
		\end{eqnarray}
		For the SMM, the first two combinations of the energy denominators yield vanishing results, while the last one leads to the simple expression
		\beq
		-  {\rm ED}^{\rm IX}_{(1)}  -    {\rm ED}^{\rm IX}_{(2)}  +    {\rm ED}^{\rm IX, \, SMM}_{(3)}  +    {\rm ED}^{\rm IX, \, SMM}_{(4)}  =  -\frac{\omega_{1}^2+\omega_{2}^2}{\omega_{1}^4 \omega_{2}^4 \omega_{3}^2} \,,
		\eeq
		which allows one to obtain the corresponding $r$-space potentials in a compact analytical form
		\begin{eqnarray}
			\label{Class9VSSMM}
			V_S^{\rm IX, \, SMM}(r) &=&\frac{2 g_\textrm{A}^6}{\left(16 \pi F_\pi^2\right)^3}\frac{e^{-3x}}{r^7}\left(2 + 2 x+x^2\right)\left(2-2x-4x^2-x^3\right), \\
			V_T^{\rm IX, \, SMM}(r) &=&\frac{2 g_\textrm{A}^6}{\left(16 \pi F_\pi^2\right)^3}\frac{e^{-3x}}{r^7}\left(2 + 2 x+x^2\right)\left(2+x-x^2-x^3\right).
			\label{Class9VTSMM}
		\end{eqnarray}
		These expressions coincide with the ones given in Ref.~\cite{Kaiser:2001dm}. Notice further that the spectral functions corresponding to the above equations have the form 
		\begin{eqnarray}
			\label{Class9_VS_mu_SMM}
			\textrm{Im} \; V_{S}^{\rm IX, \, SMM}(i \mu) &=&\frac{g_A^6}{70\pi \left(4 F_{\pi}\right)^6}\Bigg[- 4\mu^4 +105 \mu^3 M_{\pi} - 336 \mu^2 M_{\pi}^2   +420 M_{\pi}^4 +357\frac{M_{\pi}^5}{\mu} -702\frac{M_{\pi}^7}{\mu^3}\Bigg]  \,,  \\
			\textrm{Im} \; V_{T}^{\rm IX, \, SMM}(i \mu) &=& \frac{ g_A^6}{70 \pi \left(4 F_{\pi}\right)^6}\left[16 \mu^2 -105 \mu M_{\pi} +112 M_{\pi}^2 + 987\frac{M_{\pi}^5}{\mu^3}- 2106\frac{M_{\pi}^7}{\mu^5 }\right]   \,.
			\label{Class9_VT_mu_SMM}
		\end{eqnarray}
		
		It remains to derive the isovector spin-spin and tensor contributions by evaluating 
		\begin{eqnarray}
			\label{class9WST}
			W^{\rm IX}(r) &=&-\frac{g_A^6}{64 F_\pi^6}  \Big\langle \Big\langle  \big( \vec{l}_1 \cdot \vec{\sigma}_1 \vec{l}_2 \cdot \vec{l}_3+\vec{l}_2 \cdot \vec{\sigma}_1 \vec{l}_1 \cdot \vec{l}_3- \vec{l}_3 \cdot \vec{\sigma}_1 \vec{l}_1\cdot \vec{l}_2\big) \nonumber \\
			&\times & \Big[ \big(   {\rm ED}^{\rm IX}_{(1)}  -    {\rm ED}^{\rm IX}_{(2)}  +    {\rm ED}^{\rm IX}_{(3)}  -    {\rm ED}^{\rm IX}_{(4)}     \big) \vec{l}_1 \cdot \vec{\sigma}_2 \vec{l}_2 \cdot \vec{l}_3 + \big(  -  {\rm ED}^{\rm IX}_{(1)}  +    {\rm ED}^{\rm IX}_{(2)}  -    {\rm ED}^{\rm IX}_{(3)}  +    {\rm ED}^{\rm IX}_{(4)}   \big) \vec{l}_2 \cdot \vec{\sigma}_2 \vec{l}_1 \cdot \vec{l}_3 \nonumber \\
			&+&  \big(   {\rm ED}^{\rm IX}_{(1)}  +    {\rm ED}^{\rm IX}_{(2)}  +    {\rm ED}^{\rm IX}_{(3)}  +   {\rm ED}^{\rm IX}_{(4)}    \big) \vec{l}_3 \cdot \vec{\sigma}_2 \vec{l}_1 \cdot \vec{l}_2\Big] \Big\rangle \Big\rangle\,.
		\end{eqnarray}
		Inserting the expressions for the energy denominators in the SMM from Eq.~(\ref{EDClass9irreducible}) and writing the last combination of the energy denominators in the form 
		\beq
		\label{Class9Temp3a}
		{\rm ED}^{\rm IX}_{(1)}  +    {\rm ED}^{\rm IX}_{(2)}  +    {\rm ED}^{\rm IX, \, SMM}_{(3)}  +   {\rm ED}^{\rm IX, \, SMM}_{(4)}
		= 2 \big( {\rm ED}^{\rm IX, \, SMM}_{(3)}  +    {\rm ED}^{\rm IX, \, SMM}_{(4)} \big)+ \frac{\omega_1^2+\omega_2^2}{\omega_1^4 \omega_2^4 \omega_3^2}\,,
		\eeq
		we obtain for the tensor part, after some simplifications
		\begin{eqnarray}
			W_T^{\rm IX, \, SMM}(r) &=& \frac{g_A^6 e^{-3 x}}{3 \left( 16 \pi F_\pi^2\right)^3 r^7} \left(2+2 x+x^2\right) \left(2+ x- x^2-x^3\right)  \\
			&+& \frac{g_A^6}{3 (16 \pi F_\pi^2)^3 \pi ^2 r^7} \dashint_{-\infty}^{\infty} \frac{d \zeta_1 d \zeta_2 \, e^{-x (\alpha +\beta +\gamma )} }{\zeta_1\left(\zeta_1-\zeta_2\right)\zeta_2^2} \left[2 \gamma ^2+\alpha ^2 \left(\beta ^2 \gamma ^2 x^4+x^2 \left(\beta ^2+2 \beta  \gamma +3 \gamma ^2\right)+\beta  \gamma  x^3 (\beta +3 \gamma ) \right. \right. \nonumber \\
			&+& \left. \left.  2 x (\beta +\gamma )+2\right)+\alpha  x \left(2 \gamma ^2+\beta ^2 \left(\gamma ^2 x^2+2 \gamma  x+2\right)+2 \beta  \gamma ^2 x\right) +\beta ^2 \left(\gamma ^2 x^2+2 \gamma  x+2\right)+2 \beta  \gamma ^2 x\right].
			\nonumber
		\end{eqnarray}
		Notice that the first (analytical) contribution on the right-hand side of the above equation stems from the last term in Eq.~(\ref{Class9Temp3a}).  The apparent singularity of the integrand at $\zeta_1 = \zeta_2$  can be dealt with by first  writing: 
		\begin{align}
			\frac{1}{\zeta_1\left(\zeta_1-\zeta_2\right)\zeta_2^2} = 	\frac{1}{\left(\zeta_1-\zeta_2\right)\zeta_2^3} - 	\frac{1}{\zeta_1\zeta_2^3}\;.
		\end{align}
		For the first term on the right-hand side, we perform a variable shift $\zeta_1 \rightarrow \zeta_1 + \zeta_2$, which removes the singularity and changes the expressions in the numerator as $\gamma = \sqrt{1+ \left(\zeta_1- \zeta_2\right)^2} \rightarrow \sqrt{1+\zeta_1^2} = \alpha$ and 
		$\alpha = \sqrt{1 + \zeta_1^2} \rightarrow \sqrt{1+ \left(\zeta_1+ \zeta_2\right)^2} = \gamma '$. We then change  $\zeta_2$ to $-\zeta_2$ to replace $\gamma '$ with $\gamma$ and arrive at 
		(symbolically): 
		\begin{align}
			\dashint_{-\infty}^{\infty}  d \zeta_1 d \zeta_2 \frac{1}{\zeta_1\left(\zeta_1-\zeta_2\right)\zeta_2^2} f\left(\zeta_1, \zeta_2\right) \rightarrow - \dashint_{-\infty}^{\infty} d \zeta_1 d \zeta_2  \frac{2}{\zeta_1 \zeta_2^3} f\left(\zeta_1, \zeta_2\right).
		\end{align}
		Further, using partial integration, we have:
		\begin{align}
			- \dashint_{-\infty}^{\infty} d \zeta_1 d \zeta_2  \frac{2}{\zeta_1 \zeta_2^3} f\left(\zeta_1, \zeta_2\right) =  - \dashint_{-\infty}^{\infty}  \frac{d \zeta_1 d \zeta_2 }{\zeta_1 \zeta_2} \frac{\partial^2}{\partial \zeta_2^2} f\left(\zeta_1, \zeta_2\right) \;.
		\end{align}
		After these manipulations, the expression for the tensor potential takes the form 
		\begin{eqnarray}
			\label{Class9Temp4}
			W_{T}^{\rm IX, \, SMM }(r) & =& \frac{g_A^6}{3 \left( 16 \pi F_\pi^2\right)^3 r^7}\bigg\{\ e^{-3 x}\left(2+2 x+x^2\right) \left(2+ x- x^2-x^3\right)   \\
			&-& \dashint_{-\infty}^{\infty}  \frac{d \zeta_1 d \zeta_2 }{\pi^2 \zeta_1 \zeta_2} \frac{\partial^2}{\partial \zeta_2^2}e^{-x\left(\alpha+\beta + \gamma \right)} \Big[2 \gamma ^2+\alpha ^2 \left(\beta ^2 \gamma ^2 x^4+x^2 \left(\beta ^2+2 \beta  \gamma +3 \gamma ^2\right)+\beta  \gamma  x^3 (\beta +3 \gamma ) \right.   \nonumber \\
			&+&  \left.  2 x (\beta +\gamma )+2\right)+\alpha  x \left(2 \gamma ^2+\beta ^2 \left(\gamma ^2 x^2+2 \gamma  x+2\right)+2 \beta  \gamma ^2 x\right)  +\beta ^2 \left(\gamma ^2 x^2+2 \gamma  x+2\right)+2 \beta  \gamma ^2 x\Big] \bigg\}.
			\nonumber
		\end{eqnarray}
		Finally, following the same procedure as for the tensor part, we obtain for the isovector spin-spin contribution 
		\begin{eqnarray}
			\label{Class9Temp5}
			W_{S}^{\rm IX, \, SMM}(r)  &=& \frac{g_A^6}{3 \left( 16 \pi F_\pi^2\right)^3 r^7}\bigg\{\ e^{-3 x}\left(2+2 x+x^2\right) \left(2-2 x-4 x^2-x^3\right)  \nonumber \\
			&-& \dashint_{-\infty}^{\infty}  \frac{d \zeta_1 d \zeta_2 }{\pi^2 \zeta_1 \zeta_2} \frac{\partial^2}{\partial \zeta_2^2}e^{-x\left(\alpha+\beta + \gamma \right)} \Big[\alpha ^2 \beta ^2 \gamma ^2 x^6+2 x^3 \left(\alpha ^2 (\beta +\gamma )+\alpha  \left(7 \beta ^2+6 \beta  \gamma +\gamma ^2\right)    \beta  \gamma  (7 \beta +\gamma )\right) \nonumber \\
			&+& \left. 2 x^2 \left(\alpha ^2+6 \alpha  (\beta +\gamma )+7 \beta ^2+6 \beta  \gamma +\gamma ^2\right) \right.         + 2 \beta  x^4 \left(\alpha ^2 (2 \beta +\gamma )+\alpha  \gamma  (7 \beta +\gamma )+2 \beta  \gamma ^2\right)    \nonumber \\
			&+&  4 \alpha  \beta ^2 \gamma  x^5 (\alpha +\gamma ) +  12 x (\alpha +\beta +\gamma )+12\Big]x^{-2} \bigg\}.
		\end{eqnarray}
		The expressions in Eqs.~(\ref{Class9Temp4}) and (\ref{Class9Temp5}) are identical to those quoted in \cite{Kaiser:2001dm}. 
		
		We are now in the position to derive the class-IX contributions using the MUT. This can be easily done by calculating the contributions stemming from the differences in the energy denominators of the reducible-like topologies specified in Eq.~(\ref{eq_ResultEnergyDenomMUTClassIX}), which lead to simple analytic expressions for the potentials. Using Eqs.~(\ref{class9VC}), (\ref{class9WC}), (\ref{class9VST}) and  (\ref{class9WST}) and keeping only the above-mentioned differences in the energy denominators by setting
		\beqa
		{\rm ED}^{\rm IX}_{(1)} &=& {\rm ED}^{\rm IX}_{(2)}   = 0\,, \nonumber \\
		{\rm ED}^{\rm IX}_{(3)} &=& {\rm ED}^{\rm IX}_{(4)} = - \frac{1}{2 \omega _1^2 \omega _2^2 \omega _3^4} \,, 
		\eeqa
		we obtain for the various correction terms
		\beqa
		\delta V^{\rm IX}_{C}(r) &=& -\frac{6 g_{\textrm{A}}^6}{\left(16 \pi F_\pi^2\right)^3} \frac{e^{-3x}}{r^7} \left(1+x\right)\left(4+5 x + 3x^2\right), \nonumber \\
		\delta V^{\rm IX}_S (r) &=& -\frac{ g_{\textrm{A}}^6}{\left(16 \pi F_\pi^2\right)^3} \frac{e^{-3x}}{r^7} \left(8+18x+20x^2+14x^3+6x^4+x^5\right), \nonumber \\
		\delta V^{\rm IX}_T (r) &=& \frac{ g_{\textrm{A}}^6}{\left(16 \pi F_\pi^2\right)^3} \frac{e^{-3x}}{r^7} \left(4+6x+4x^2-2x^3-3x^4-x^5\right), \nonumber \\
		\delta W^{\rm IX}_C (r) &=& \frac{ g_{\textrm{A}}^6}{\left(16 \pi F_\pi^2\right)^3} \frac{e^{-3x}}{r^7} \left(1+x\right)\left(4+5 x + 3x^2\right), \nonumber \\
		\delta W^{\rm IX}_S (r) &=&\frac{g_A^6}{6 \left( 16 \pi F_\pi^2\right)^3} \frac{e^{-3 x}}{r^7} \left(8+18x+20x^2+14x^3+6x^4+x^5\right), \nonumber \\
		\delta W^{\rm IX}_T (r) &=&\frac{g_A^6}{6 \left( 16 \pi F_\pi^2\right)^3} \frac{e^{-3 x}}{r^7} \left(-4-6 x-4 x^2+2x^3+3x^4+x^5\right).
		\eeqa
		Adding these contributions to the SMM results in Eqs.~(\ref{Class9Temp1a}),    (\ref{Class9VSSMM}),  (\ref{Class9VTSMM}),   (\ref{Class9Temp2a}),   (\ref{Class9Temp4}) and       (\ref{Class9Temp5}), we end up with the following $r$-space potentials in the MUT:
		\beqa
		\label{Class9_VC_r_MUT}
		V_{C}^{\rm IX, \, MUT}(r) &=&- \frac{6 g_{A}^6 }{ \left(16 \pi F_\pi^2\right)^3}\, \dashint_{-\infty}^{\infty}  \frac{d \zeta_1 d \zeta_2 }{\pi^2 \zeta_1 \zeta_2} \frac{\partial^2}{\partial \zeta_1\partial \zeta_2}e^{-x\left(\alpha+\beta + \gamma \right)} \Big[x \left(\alpha ^2 (\beta +\gamma )+\alpha  \left(\beta ^2+6 \beta  \gamma +\gamma ^2\right)+\beta  \gamma  (\beta +\gamma )\right) \nonumber \\
		&+& \alpha ^2+6 \alpha  (\beta +\gamma )+\beta ^2+6 \beta  \gamma +\gamma ^2+\alpha  \beta  \gamma  x^2 (\alpha +\beta +\gamma )+6 x^{-1} (\alpha +\beta +\gamma )+6 x^{-2}\Big] \,, \\
		\label{Class9_VS_r_MUT}
		V^{\rm IX, \, MUT}_{S}(r) &=& -\frac{ g_{\textrm{A}}^6 M_\pi}{\left(16 \pi F_\pi^2\right)^3} \frac{e^{-3x}}{r^6} \left(18+40x+38x^2+18x^3+3x^4\right), \\
		\label{Class9_VT_r_MUT}
		V^{\rm IX, \, MUT}_{T}(r) &=& \frac{ g_{\textrm{A}}^6}{\left(16 \pi F_\pi^2\right)^3} \frac{e^{-3x}}{r^7} \left(12+18x+8x^2-8x^3-9x^4-3x^5\right), \\
		\label{Class9_WC_r_MUT}
		W^{\rm IX, \, MUT}_{C} (r) &=& \frac{3 g_{\textrm{A}}^6}{\left(16 \pi F_\pi^2\right)^3} \frac{e^{-3x}}{r^7} \left(1+x\right)\left(4+5 x + 3x^2\right), \\
		\label{Class9_WS_r_MUT}
		W_{S}^{\rm IX, \, MUT}(r)&=& \frac{g_A^6}{3 \left( 16 \pi F_\pi^2\right)^3 r^7}\bigg\{\frac{e^{-3 x}}{2} \left(16+18 x-10 x^3-6x^4-x^5\right) \nonumber \\
		&-& \dashint_{-\infty}^{\infty}  \frac{d \zeta_1 d \zeta_2 }{\pi^2 \zeta_1 \zeta_2} \frac{\partial^2}{\partial \zeta_2^2}e^{-x\left(\alpha+\beta + \gamma \right)} \Big[\alpha ^2 \beta ^2 \gamma ^2 x^6+2 x^3 \left(\alpha ^2 (\beta +\gamma )+\alpha  \left(7 \beta ^2+6 \beta  \gamma +\gamma ^2\right) + \beta  \gamma  (7 \beta +\gamma )\right)  \nonumber \\
		&+&\left.2 x^2 \left(\alpha ^2+6 \alpha  (\beta +\gamma )+7 \beta ^2+6 \beta  \gamma +\gamma ^2\right) +2 \beta  x^4 \left(\alpha ^2 (2 \beta +\gamma )+\alpha  \gamma  (7 \beta +\gamma )+2 \beta  \gamma ^2\right) \right. \nonumber \\
		&+&4 \alpha  \beta ^2 \gamma  x^5 (\alpha +\gamma ) + 12 x (\alpha +\beta +\gamma )+12\Big]x^{-2} \bigg\},
	 \\
		\label{Class9_WT_r_MUT}
		W_{T}^{\rm IX, \, MUT}(r) & =& \frac{g_A^6}{3 \left( 16 \pi F_\pi^2\right)^3 r^7}\bigg\{\frac{e^{-3 x}}{2} \left(4+6x-4x^3-3x^4-x^5\right) \nonumber \\
		&-& \dashint_{-\infty}^{\infty}  \frac{d \zeta_1 d \zeta_2 }{\pi^2 \zeta_1 \zeta_2} \frac{\partial^2}{\partial \zeta_2^2}e^{-x\left(\alpha+\beta + \gamma \right)} \Big[2 \gamma ^2+\alpha ^2 \left(\beta ^2 \gamma ^2 x^4+x^2 \left(\beta ^2+2 \beta  \gamma +3 \gamma ^2\right)+\beta  \gamma  x^3 (\beta +3 \gamma ) \right.  \\
		&+& \left. 2 x (\beta +\gamma )+2\right)+\alpha  x \left(2 \gamma ^2+\beta ^2 \left(\gamma ^2 x^2+2 \gamma  x+2\right)+2 \beta  \gamma ^2 x\right)  +\beta ^2 \left(\gamma ^2 x^2+2 \gamma  x+2\right)+2 \beta  \gamma ^2 x\Big] \bigg\}.
		\nonumber
		\eeqa
		For the sake of completeness, we also provide the expressions for the corresponding spectral functions (except for the isovector spin-spin and tensor potentials):
		\beqa
                \textrm{Im} \; V_{C}^{\rm IX, \, MUT}(i \mu) &=&- \frac{g_A^6 \left(\mu - 3 M_{\pi}\right)^2}{10 \pi \mu \left(4 F_\pi\right)^6}\left(3M_{\pi}^3+ 2 \mu M_{\pi}^2- 9 \mu^2 M_{\pi} - 4 \mu^3\right) \nonumber \\
                &+&\frac{3 g_A^6 \mu^2}{2\left(8 \pi F_\pi^2 \right)^3}\iint \limits_{z^2<1} d \omega_{1} d \omega_{2} \bigg\{1 - \frac{1}{\sqrt{1-z^2}}\left[\pi \left(\frac{l_2}{2l_1}+z\right)-z \arccos(-z)\right]\bigg\}
                , \label{Class9_VC_mu_MUT} \\
		\textrm{Im} \; V_{S}^{\rm IX, \, MUT}(i \mu) &=&\frac{2 g_A^6}{35\pi \left(32 F_{\pi}^2\right)^3}\Bigg[16 \mu^4 +105 \mu^3 M_{\pi} - 504 \mu^2 M_{\pi}^2 -315 \mu M_{\pi}^3 +840 M_{\pi}^4  \nonumber \\
		&+& 1827\frac{M_{\pi}^5}{\mu} -2673\frac{M_{\pi}^7}{\mu^3}\Bigg], \label{Class9_VS_mu_MUT} \\
		\textrm{Im} \; V_{T}^{\rm IX, \, MUT}(i \mu) &=&  \frac{2 g_A^6}{35 \pi \left(32 F_{\pi}^2\right)^3}\left[48 \mu^2 -315 \mu M_{\pi}  + 448 M_{\pi}^2 - 525\frac{M_{\pi}^3}{\mu}+ 4851\frac{M_{\pi}^5}{\mu^3}- 8019\frac{M_{\pi}^7}{\mu^5 }\right] \,, \label{Class9_VT_mu_MUT} \\
		\textrm{Im} \; W_{C}^{\rm IX, \, MUT}(i \mu) &=& \frac{ g_A^6 \left(\mu - 3 M_{\pi}\right)^2}{20 \pi \mu \left( 4 F_\pi\right)^6}\left(3M_{\pi}^3+ 2 \mu M_{\pi}^2- 9 \mu^2 M_{\pi} - 4 \mu^3\right)\,. \label{Class9_WC_mu_MUT} 
		\eeqa
		\section{Chiral $3 \pi$-exchange at order N$^4$LO}
               \setcounter{equation}{0} 
\def\theequation{\arabic{section}.\arabic{equation}}
                \label{SecIV}
	
		We now turn our attention to the subleading $3\pi$-exchange potential that appears at N$^4$LO in the chiral expansion. Following Ref.~\cite{Kaiser:2001dm}, we divide the corresponding contributions into five classes as visualized in Fig.~\ref{fig:3pi_N4LO}. One observes that class-XI and class-XIII diagrams include reducible-like topologies, so that the corresponding potentials can, in principle, have a different form when using the MUT and SMM.
		However, the reducible-like class-XI diagrams have the same structure as the $2\pi$-exchange box NN diagrams (and also the reducible-like class-IV graphs), for which no scheme-dependence appears in the resulting potentials.
		
		To clarify the situation with the reducible-like class-XIII contributions we follow the same approach as in the N$^3$LO case considered in the previous section and start with the Fock-space expressions for the potentials obtained in the method of unitary transformation, which have the form \cite{Krebs:2012yv}: 
		\beqa
		V_{\rm MUT}^{\rm XIII, XIV}&=& \eta \bigg[   \alpha_9 \bigg(
		H_{21}^{(1)}  \frac{\lambda^1}{E_\pi}  H_{21}^{(1)}  \eta  \
		H_{21}^{(1)}  \frac{\lambda^1}{E_\pi}  H_{21}^{(1)}  \
		\frac{\lambda^2}{E_\pi^2}  H_{22}^{(3)}  
		-   H_{21}^{(1)}  \
		\frac{\lambda^1}{E_\pi}  H_{21}^{(1)}  \eta  H_{22}^{(3)}  \
		\frac{\lambda^2}{E_\pi^2}  H_{21}^{(1)}  \frac{\lambda^1}{E_\pi}  \
		H_{21}^{(1)} \bigg)\nn
		&& {} + 
		\alpha_{10} \bigg(   H_{21}^{(1)}  \frac{\lambda^1}{E_\pi}  \
		H_{21}^{(1)}  \eta  H_{21}^{(1)}  \frac{\lambda^1}{E_\pi^2}  \
		H_{21}^{(1)}  \frac{\lambda^2}{E_\pi}  H_{22}^{(3)}  -   \
		H_{21}^{(1)}  \frac{\lambda^1}{E_\pi}  H_{21}^{(1)}  \eta  \
		H_{22}^{(3)}  \frac{\lambda^2}{E_\pi}  H_{21}^{(1)}  \
		\frac{\lambda^1}{E_\pi^2}  H_{21}^{(1)}  \bigg) \nn
		&& {}+
		\alpha_{11} \bigg( H_{21}^{(1)}  \frac{\lambda^1}{E_\pi}  \
		H_{21}^{(1)}  \eta  H_{21}^{(1)}  \frac{\lambda^1}{E_\pi}  \
		H_{22}^{(3)}  \frac{\lambda^1}{E_\pi^2}  H_{21}^{(1)}   - 
		H_{21}^{(1)}  \frac{\lambda^1}{E_\pi}  H_{21}^{(1)}  \eta  \
		H_{21}^{(1)}  \frac{\lambda^1}{E_\pi^2}  H_{22}^{(3)}  \
		\frac{\lambda^1}{E_\pi}  H_{21}^{(1)}  \bigg) \nn
		&& {}
		+ \frac{1}{2} \bigg( -   H_{21}^{(1)}  \frac{\lambda^1}{E_\pi}  H_{21}^{(1)}  \eta  \
		H_{21}^{(1)}  \frac{\lambda^1}{E_\pi}  H_{21}^{(1)}  \
		\frac{\lambda^2}{E_\pi^2}  H_{22}^{(3)}   -   H_{21}^{(1)}  \
		\frac{\lambda^1}{E_\pi}  H_{21}^{(1)}  \eta  H_{21}^{(1)}  \
		\frac{\lambda^1}{E_\pi}  H_{22}^{(3)}  \frac{\lambda^1}{E_\pi^2}  \
		H_{21}^{(1)}  \nn
		&& {} -   H_{21}^{(1)}  \frac{\lambda^1}{E_\pi}  \
		H_{21}^{(1)}  \eta  H_{21}^{(1)}  \frac{\lambda^1}{E_\pi^2}  \
		H_{21}^{(1)}  \frac{\lambda^2}{E_\pi}  H_{22}^{(3)}   -   \
		H_{21}^{(1)}  \frac{\lambda^1}{E_\pi}  H_{21}^{(1)}  \eta  \
		H_{21}^{(1)}  \frac{\lambda^1}{E_\pi^2}  H_{22}^{(3)}  \
		\frac{\lambda^1}{E_\pi}  H_{21}^{(1)}  \nn
		&&  {} -   H_{21}^{(1)}  \
		\frac{\lambda^1}{E_\pi}  H_{21}^{(1)}  \eta  H_{22}^{(3)}  \
		\frac{\lambda^2}{E_\pi}  H_{21}^{(1)}  \frac{\lambda^1}{E_\pi^2}  \
		H_{21}^{(1)}   -   H_{21}^{(1)}  \frac{\lambda^1}{E_\pi}  \
		H_{21}^{(1)}  \eta  H_{22}^{(3)}  \frac{\lambda^2}{E_\pi^2}  \
		H_{21}^{(1)}  \frac{\lambda^1}{E_\pi}  H_{21}^{(1)}  \nn
		&& {}  + 2   \,
		H_{21}^{(1)}  \frac{\lambda^1}{E_\pi}  H_{21}^{(1)}  \
		\frac{\lambda^2}{E_\pi}  H_{21}^{(1)}  \frac{\lambda^1}{E_\pi}  \
		H_{21}^{(1)}  \frac{\lambda^2}{E_\pi}  H_{22}^{(3)}   + 2   \,
		H_{21}^{(1)}  \frac{\lambda^1}{E_\pi}  H_{21}^{(1)}  \
		\frac{\lambda^2}{E_\pi}  H_{21}^{(1)}  \frac{\lambda^1}{E_\pi}  \
		H_{22}^{(3)}  \frac{\lambda^1}{E_\pi}  H_{21}^{(1)}  \nn
		&& {}  + 2  \,
		H_{21}^{(1)}  \frac{\lambda^1}{E_\pi}  H_{21}^{(1)}  \
		\frac{\lambda^2}{E_\pi}  H_{21}^{(1)}  \frac{\lambda^3}{E_\pi}  \
		H_{21}^{(1)}  \frac{\lambda^2}{E_\pi}  H_{22}^{(3)}   + 2  \,
		H_{21}^{(1)}  \frac{\lambda^1}{E_\pi}  H_{21}^{(1)}  \
		\frac{\lambda^2}{E_\pi}  H_{21}^{(1)}  \frac{\lambda^3}{E_\pi}  \
		H_{22}^{(3)}  \frac{\lambda^1}{E_\pi}  H_{21}^{(1)}  \nn
		&& {}  -  \
		H_{21}^{(1)}  \frac{\lambda^1}{E_\pi}  H_{21}^{(1)}  \
		\frac{\lambda^2}{E_\pi}  H_{22}^{(3)}  \eta  H_{21}^{(1)}  \
		\frac{\lambda^1}{E_\pi^2}  H_{21}^{(1)}   +  H_{21}^{(1)}  \
		\frac{\lambda^1}{E_\pi}  H_{21}^{(1)}  \frac{\lambda^2}{E_\pi}  \
		H_{22}^{(3)}  \frac{\lambda^2}{E_\pi}  H_{21}^{(1)}  \
		\frac{\lambda^1}{E_\pi}  H_{21}^{(1)}  \nn
		&& {} -   H_{21}^{(1)}  \frac{\lambda^1}{E_\pi}  \
		H_{22}^{(3)}  \frac{\lambda^1}{E_\pi}  H_{21}^{(1)}  \eta  \
		H_{21}^{(1)}  \frac{\lambda^1}{E_\pi^2}  H_{21}^{(1)}  -   H_{21}^{(1)}  \
		\frac{\lambda^1}{E_\pi^2}  H_{21}^{(1)}  \eta  H_{21}^{(1)}  \
		\frac{\lambda^1}{E_\pi}  H_{21}^{(1)}  \frac{\lambda^2}{E_\pi}  \
		H_{22}^{(3)}  \bigg) \bigg] \eta \; + \; {\rm h.c.}\,. 
		\eeqa
		In Ref.~\cite{Krebs:2012yv}, it was found that renormalizability of the N$^3$LO three-nucleon force leads to the following constraint on the
		phases $\alpha_9$, $\alpha_{10}$ and $\alpha_{11}$: 
		\beq
		\alpha_{10} = - \alpha_{11} = - \frac{1}{4} (1 - 2 \alpha_9 )\,.
		\eeq
		Moreover, the expressions for the three-nucleon force turned out to be independent of the arbitrary remaining phase $\alpha_9$. 
		We have calculated the matrix elements of these operators corresponding to the reducible-like class-XIII diagrams and verified, that the corresponding energy denominators have the same form as in the SMM. Thus, both the MUT and SMM lead to the same $3\pi$-exchange potentials at N$^4$LO.
		
		\section{Summary and conclusions}
\setcounter{equation}{0}
                \def\theequation{\arabic{section}.\arabic{equation}}
               \label{SecV}
		
               The main results of our work can be summarized as follows:
               \vspace{-0.25cm}
		\begin{itemize}
			\item We have rederived the $3 \pi$-exchange nucleon-nucleon potential using the S-matrix method as done by Kaiser in Refs.~\cite{Kaiser:1999ff,Kaiser:1999jg,Kaiser:2001dm}. We provide a detailed description of the relevant computational techniques such as the Cutkosky cutting rules for deriving the corresponding spectral functions and the Wick-rotation approach for directly calculating the potentials in coordinate space. We also discuss the reduction of tensor integrals and provide details for solving the relevant integrals. Last but not least, we provide important intermediate steps and expressions when deriving the $3 \pi$-exchange potentials in order to facilitate reproducibility of these results. The summary of the obtained expressions for the N$^3$LO  potentials is provided in Table \ref{Tab1}. We have succeeded to verify all results of Refs.~\cite{Kaiser:1999ff,Kaiser:1999jg,Kaiser:2001dm} and have corrected the result of Ref.~\cite{Kaiser:1999jg} for the class-V isovector tensor potential.
			\begin{table*}
				\begin{ruledtabular}
					\begin{tabular*}{\textwidth}{@{\extracolsep{\fill}}lllllll}
						&&& &&&\\[-11pt]
						&$V_C$ &  $V_S$ &  $V_T$ & $W_C$ &  $W_S$ &  $W_T$ \\
						&&& &&&\\[-10pt]     
						\hline \\[-7pt] 
						Class-I     & --- & --- & --- & --- & Eq.~(\ref{Class1SpinSpinFinal}) & Eq.~(\ref{Class1_WT_mu}) \\
						Class-II     & --- & --- & --- & --- & Eq.~(\ref{Class2_WS_mu}) & Eq.~(\ref{Class2_WT_mu})$^{\rm a}$ \\         
						Class-III     & --- & --- & --- & --- & Eq.~(\ref{ClassIII_temp1}) & Eq.~(\ref{Class3_WT_mu})$^{\rm a}$ \\          
						Class-IV     & --- & --- & --- & --- & Eq.~(\ref{eq_IM_Ws_ClassIV_NK_dw}) & Eq.~(\ref{Class4_WT_mu})$^{\rm a}$ \\        
						Class-V     & --- & --- & --- & --- & Eq.~(\ref{Class5_WS_mu})$^{\rm a}$ & Eq.~(\ref{eq_WT_linearCombi_FinalRes})$^{\rm b}$ \\
						Class-VI     & --- & --- & --- & --- & Eq.~(\ref{SSMM}) & Eq.~(\ref{TSMM})$^{\rm a}$ \\             
						Class-VII     & --- & Eq.~(\ref{Class7_VS_mu}) &  Eq.~(\ref{Class7_VT_mu}) & --- & --- & --- \\  
						Class-VIII     & Eqs.~(\ref{Class8VCSMM},\ref{eq_Vc_ClassVIII_WR_SMM}) & --- &  ---& --- & Eq.~(\ref{Class8_WS_r_SMM})$^{\rm a,\, c}$ & Eq.~(\ref{Class8_WT_r_SMM})$^{\rm a,\, c}$ \\
						Class-IX     & Eqs.~(\ref{Class9Temp1a}$^{\rm a}$,\ref{eq_Vc_ClassIX_WR}) & Eqs.~(\ref{Class9_VS_mu_SMM},\ref{Class9VSSMM})  &  Eqs.~(\ref{Class9_VT_mu_SMM},\ref{Class9VTSMM})  & Eqs.~(\ref{Class9Temp2a},\ref{Class9_WC_r_SMM}) & Eq.~(\ref{Class9Temp5})$^{\rm a,\, c}$ & Eq.~(\ref{Class9Temp4})$^{\rm a,\, c}$ \\                
					\end{tabular*}
					{$^a$ The expression has a different form but was verified to be identical to the one from Refs.~\cite{Kaiser:1999ff,Kaiser:1999jg,Kaiser:2001dm}. \hfill}
					\newline
					{$^b$ We have corrected the result of Ref.~\cite{Kaiser:1999jg}. \hfill}
					\newline
					{$^c$ Only coordinate-space results have been calculated. \hfill}
					\vspace{-0.1cm}
					\caption{Summary of the obtained results for the N$^3$LO $3\pi$-exchange potentials using the SMM. Unless explicitly stated otherwise, a single expression quoted refers to the result for the corresponding spectral function. The potentials in momentum and coordinate spaces can be calculated using Eqs.~(\ref{SpectralMomSpace}), (\ref{Disp_CoordSpace}). Whenever two equations are quoted, the second one gives the expression for the potential in coordinate space. }
					\label{Tab1}
                                      \end{ruledtabular}
                                      \vspace{0.35cm}
			\end{table*}
                        \\[-13pt]
			\item The main motivation of this study was a concern that the existing expressions for the $3\pi$-exchange NN potential worked out in Refs.~\cite{Kaiser:1999ff,Kaiser:1999jg,Kaiser:2001dm} are not off-shell consistent with the nuclear forces and current operators derived in Refs.~\cite{Epelbaum:1998ka,Epelbaum:1999dj,Epelbaum:2002gb,Epelbaum:2005fd,Epelbaum:2005bjv,Epelbaum:2007us,Bernard:2007sp,Bernard:2011zr,Krebs:2012yv,Krebs:2013kha,Kolling:2009iq,Kolling:2011mt,Krebs:2016rqz,
				Krebs:2019aka,Krebs:2020plh,Krebs:2020pii}  using the method of unitary transformation. Such scheme-dependence of nuclear interactions is well known for, e.g., the nonstatic $2\pi$-exchange contributions \cite{Friar:1994zz,Friar:1999sj,Bernard:2011zr}, but for the $3\pi$-exchange, it is expected to affect even the static terms. Specifically, scheme-dependent results for the $3\pi$-exchange NN potential can be expected for class-IV, VI, VIII and IX contributions which include reducible-like diagrams. We have used the method of unitary transformation to derive these contributions. While the expressions for the class-IV diagrams turn out to be identical in both the SMM and MUT, we indeed found different results (and even additional non-vanishing potentials) for the  class-VI, VIII and IX diagrams. Our results for the leading $3\pi$-exchange potential, calculated using the method of unitary  transformation, are summarized in Table \ref{Tab2}.  
			\begin{table*}
				\begin{ruledtabular}
					\begin{tabular*}{\textwidth}{@{\extracolsep{\fill}}lllllll}
						&&& &&&\\[-11pt]
						&$V_C$ &  $V_S$ &  $V_T$ & $W_C$ &  $W_S$ &  $W_T$ \\
						&&& &&&\\[-10pt]     
						\hline \\[-7pt] 
						Class-VI     & --- & Eq.~(\ref{VSMUT}) & Eq.~(\ref{VTMUT}) & --- & Eq.~(\ref{Class6_WS_mu_MUT}) & Eq.~(\ref{Class6_WT_mu_MUT}) \\             
						Class-VIII     & Eqs.~(\ref{eq_Im_Vc_ClassVIII_MUT},\ref{eq_Vc_ClassVIII_WR_with_Correc}) & Eqs.~(\ref{VS8MUTSpF},\ref{Class8_VS_r_MUT}) &  Eqs.~(\ref{VT8MUTSpF},\ref{Class8_VT_r_MUT})& Eqs.~(\ref{eq_Im_Wc_ClassVIII_MUT},\ref{Class8_Wc_r_MUT}) & Eq.~(\ref{eq_Ws_ClassVIII_MUT})$^{c}$ & Eq.~(\ref{eq_Wt_ClassVIII_MUT})$^{c}$ \\
                                          Class-IX     &
                                                          Eqs.~(\ref{Class9_VC_mu_MUT},
                                                         \ref{Class9_VC_r_MUT}) & Eqs.~(\ref{Class9_VS_mu_MUT},\ref{Class9_VS_r_MUT})  &  Eqs.~(\ref{Class9_VT_mu_MUT},\ref{Class9_VT_r_MUT})  &                                                                                                                                                                                                                      Eqs.~(\ref{Class9_WC_mu_MUT},\ref{Class9_WC_r_MUT}) & Eq.~(\ref{Class9_WS_r_MUT})$^{c}$ & Eq.~(\ref{Class9_WT_r_MUT})$^{c}$ \\                
					\end{tabular*}
					{$^c$ Only coordinate-space results have been calculated. \hfill}
					\vspace{-0.1cm}
					\caption{Summary of the results for the N$^3$LO $3\pi$-exchange potentials using the MUT. Only those potentials are given, for which the MUT and  SMM lead to different results. For remaining notation see Table \ref{Tab1}. }
					\label{Tab2}
				\end{ruledtabular} 
			\end{table*}
                        \\[-13pt]
			\item
			For the cases where the two approaches yield different results, we have calculated and compared with each other the corresponding potentials in coordinate space. Our results for the class-VI, VIII and IX contributions are visualized in Figs.~\ref{fig:Class6},  \ref{fig:Class8} and \ref{fig:Class9}, respectively. In all considered cases, the differences between the two schemes appear to be large (i.e., comparable in size with the actual potentials), with the results based on the MUT typically having a larger magnitude as compared to the potentials calculated from S-matrix method.   \\[-13pt]
			\item
			We have also considered the subleading $3\pi$-exchange potential at N$^4$LO, which also features reducible-like diagrams in classes XI and XIII. However, in that case, we found the method of unitary transformation to yield the same results as obtained using the S-matrix method. 
                      \end{itemize}
   \vspace{-0.4cm}                   
		\begin{figure}[t!]
			\begin{center}
				\includegraphics[width=\textwidth]{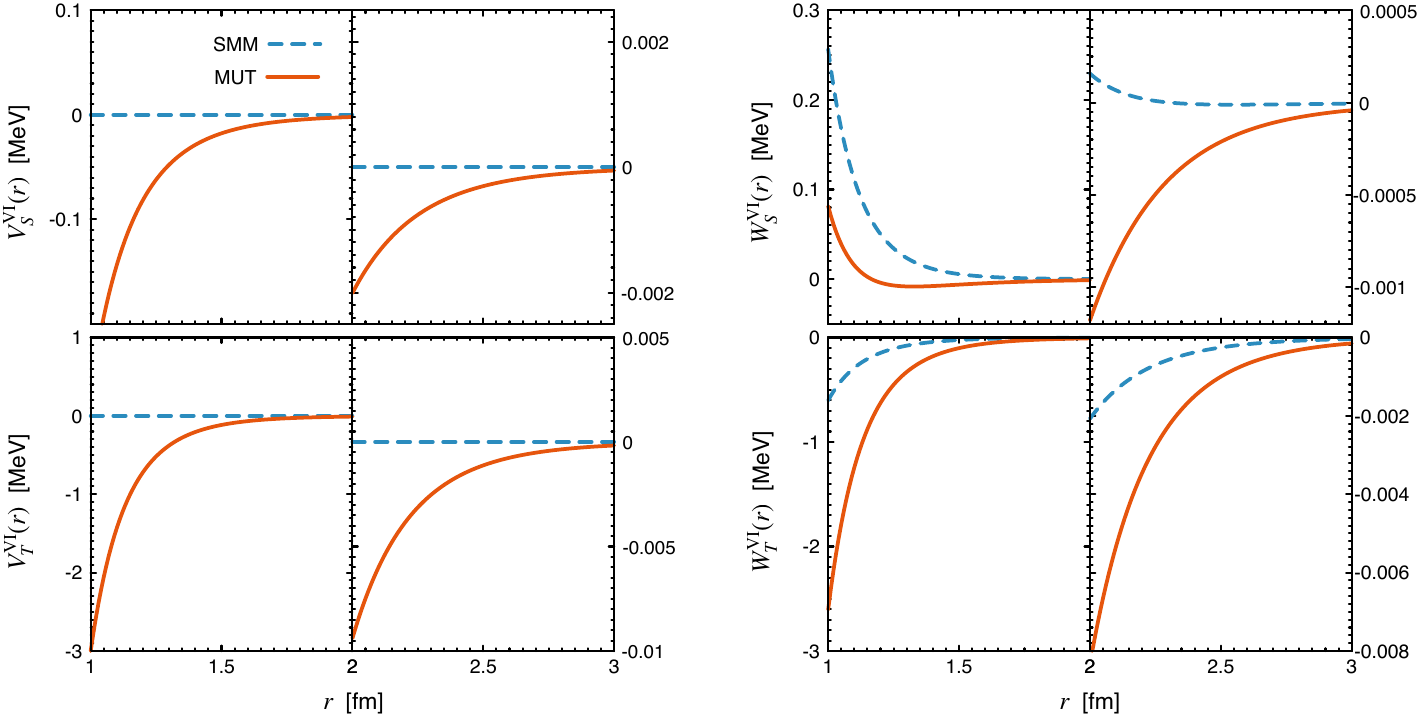}
			\end{center}
			\caption{Class-VI $3\pi$-exchange potentials in coordinate space using the MUT (red solid lines) and the SMM (blue dashed lines). Tie class-VI scalar potentials turn out to be the same in both the SMM and MUT and are not shown.}
			\label{fig:Class6}
		\end{figure}
		\begin{figure}[t!]
			\begin{center}
				\includegraphics[width=\textwidth]{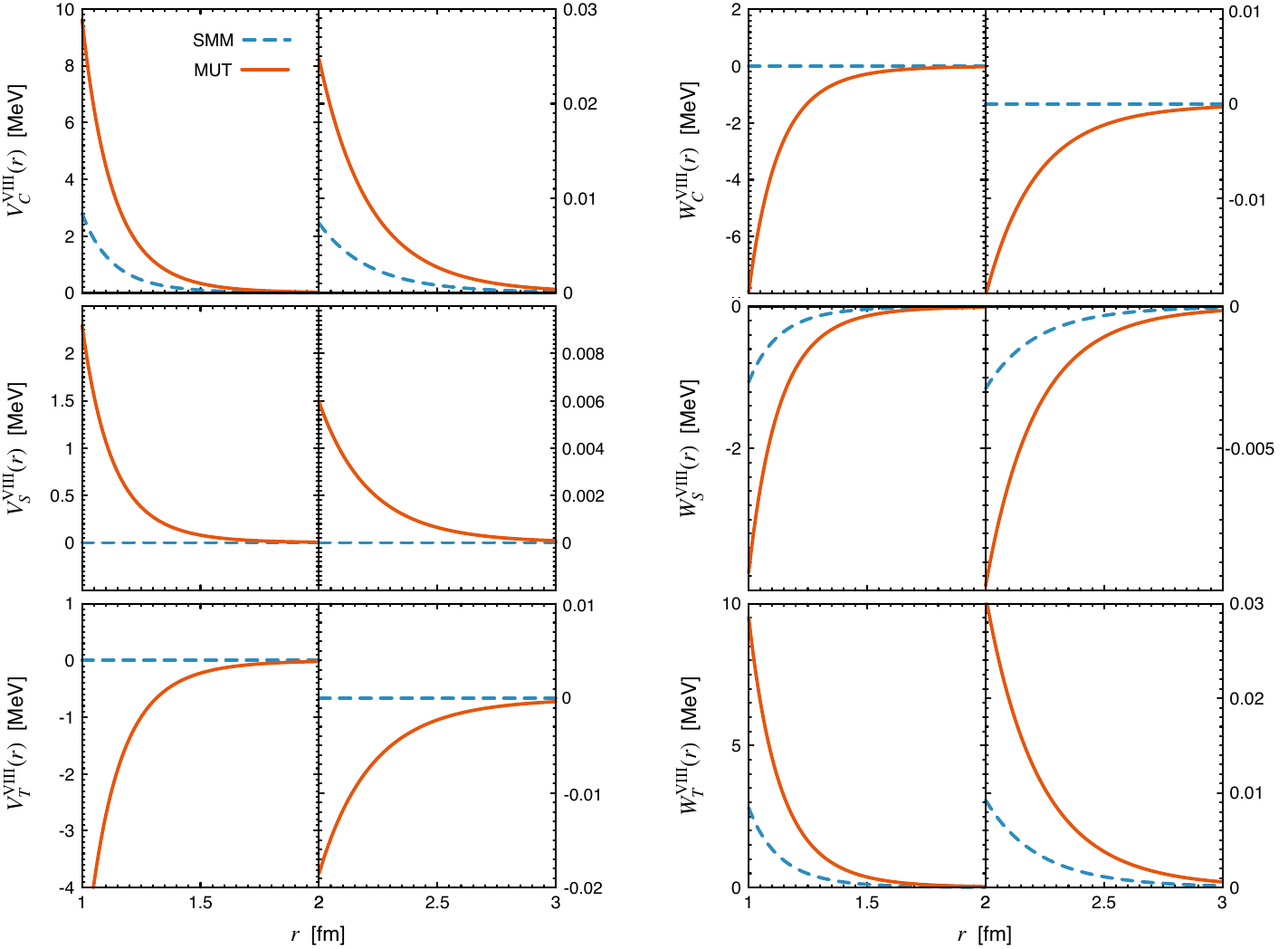}
			\end{center}
			\caption{Class-VIII $3\pi$-exchange potentials in coordinate space using the MUT (red solid lines) and the SMM (blue dashed lines).}
			\label{fig:Class8}
		\end{figure}
		\begin{figure}[t!]
			\begin{center}
				\includegraphics[width=\textwidth]{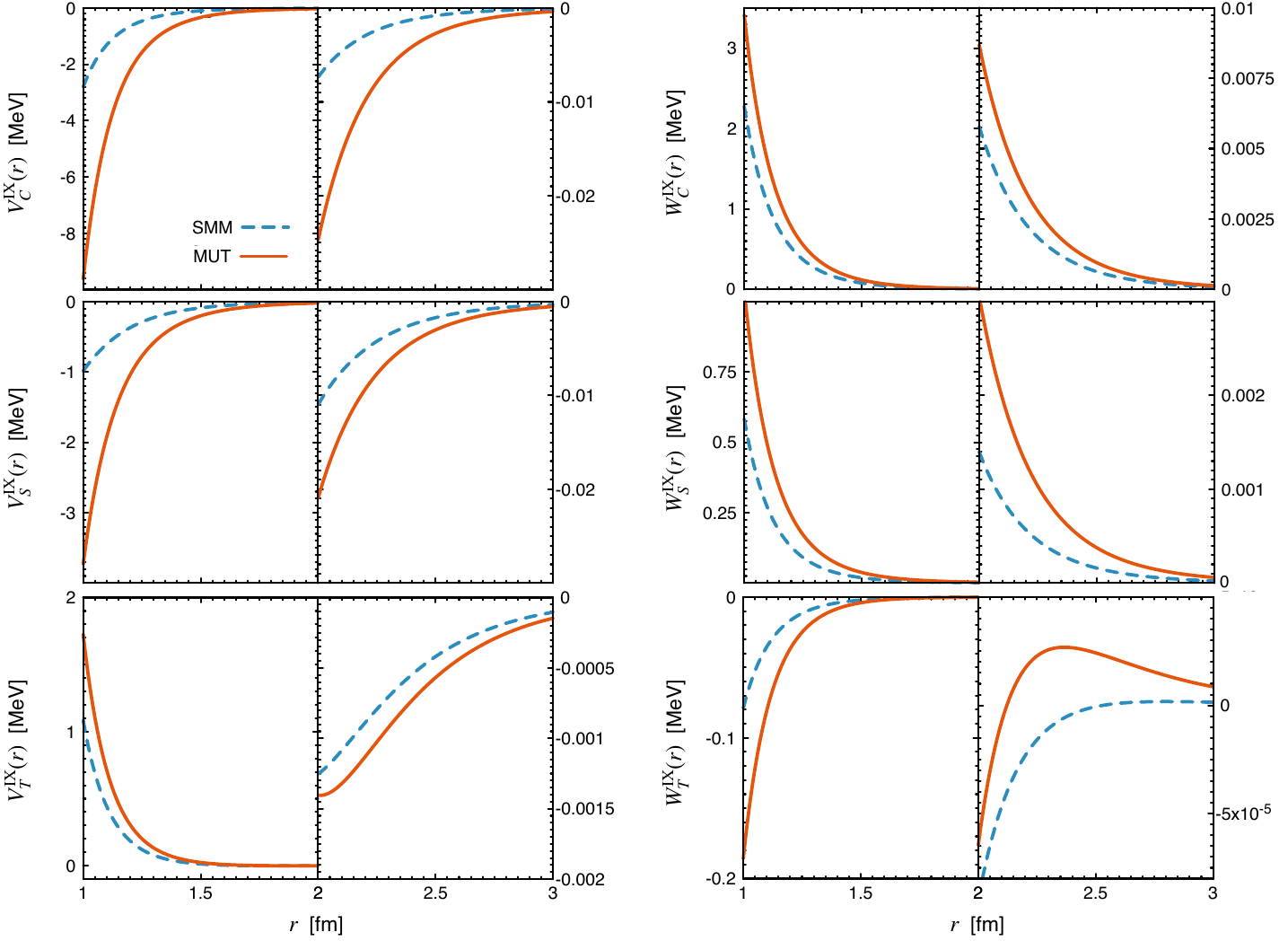}
			\end{center}
			\caption{Class-IX $3\pi$-exchange potentials in coordinate space using the MUT (red solid lines) and the SMM (blue dashed lines). }
			\label{fig:Class9}
		\end{figure}
		The analysis carried out in this work will allow us, in the future, to extend the state-of-the-art high-precision NN chiral potentials of Refs.~\cite{Reinert:2017usi,Epelbaum:2019kcf,Reinert:2020mcu,Epelbaum:2022cyo} by 
		\emph{explicitly} including the chiral $3\pi$-exchange contributions. Clearly, this will require a proper regularization of the obtained potentials by applying a local momentum-space cutoff along the lines of Ref.~\cite{Reinert:2017usi} \footnote{A rigorous symmetry-preserving regularization method for chiral EFT has been recently introduced in Refs.~\cite{Krebs:2023ljo,Krebs:2023gge}. This novel approach allows one to derive consistently regularized three- and more-nucleon forces as well as exchange current operators in harmony with the chiral and gauge symmetries. For two-nucleon potentials considered in this paper, it is, however, not really nesessary to apply this rigorous but computationally more demanding regularization approach \cite{Epelbaum:2019kcf}.}. It is important to emphasize that given the knowledge of the relevant $\pi N$ low-energy constants \cite{Hoferichter:2015tha,Siemens:2016jwj}, the leading and subleading $3\pi$-exchange potentials come out as parameter-free predictions within chiral EFT. For the $2\pi$-exchange, such predictions have already been successfully confronted with the wealth of the available neutron-proton and proton-proton scattering data below pion-production threshold \cite{Reinert:2017usi,Epelbaum:2014efa,Epelbaum:2014sza}, see also a recent work \cite{Yang:2025mhg} for a related discussion in the context of neutron-$\alpha$ scattering. It remains to be seen if one can observe evidence of the chiral $3\pi$-exchange from two-nucleon scattering data, see also Ref.~\cite{Machleidt:2011zz,Entem:2014msa,Entem:2015xwa} for a related discussion. Work along these lines is in progress.

		\section*{Acknowledgments}
		
		We are grateful to Norbert Kaiser for an illuminating email exchange at various stages of the project. 
		This work was supported in part by the by
		DFG and NSFC through funds provided to the Sino-German CRC 110
		``Symmetries and the Emergence of Structure in QCD'' (NSFC Grant
		No. 11621131001, DFG Project-ID 196253076 - TRR 110),
		by ERC  NuclearTheory (grant No. 885150), by MKW NRW
		under the funding code NW21-024-A,  and by the EU Horizon 2020 research and
		innovation programme (STRONG-2020, grant agreement No. 824093).

		\appendix
\renewcommand{\theequation}{\thesection.\arabic{equation}}
                
		\section{Principal value integrals }\label{Appendix_Principle Value Integrals}
		In this appendix, we give an overview of some important principle value integrals and show how to solve them.
		We begin with a simple example: 
		\begin{align*}
			-\int_{-1}^{1}dx	\frac{1}{i x - \epsilon}\,.
		\end{align*}
		Here and in what follows, the limit $\epsilon \to 0^+$ is understood.
		First, we rewrite this expression as
		\begin{align}
			\frac{-1}{i x - \epsilon} = \frac{i x + \epsilon}{x^2 + \epsilon^2} = i \frac{x}{x^2+\epsilon^2} + \frac{\epsilon}{x^2 + \epsilon^2} =  i \mathrm{P} \frac{1}{x}+ \pi \delta(x)\,.
		\end{align}
		In the last step, we used a representation of the delta-distribution as well as the definition of the principal value indicated by the P.
		Thus, it follows for our simple example:
		\begin{align*}
			-\int_{-1}^{1}dx	\frac{1}{i x - \epsilon} = i \underbrace{P \int_{-1}^{1}dx	\frac{1}{x}}_{0}+ \pi \underbrace{\int_{-1}^{1}dx \delta(x)}_{1} = \pi \;.
		\end{align*}
		We continue with a more complicated problem. Consider the following expression:
		\begin{align*}
			-\int_{-1}^{1} dx \int_{y_{\textrm{min}}(x,z)}^{y_{\textrm{max}}(x,z)} dy \frac{1}{\sqrt{1-x^2-y^2-z^2 + 2 x y z}} \frac{1}{\left(i x - \epsilon\right)\left(i y + \epsilon\right)}\;.
		\end{align*}
		The first thing we can do is to multiply the numerator and denominator with the complex conjugate of the denominator. Then we can make use of definitions of the delta distribution as well as of the principal value. We find:
		\begin{align}\label{PV_twoProp}
			\frac{-1}{\left(i x - \epsilon\right)\left(i y + \epsilon\right)}  &= 	\frac{xy + \epsilon^2 + i x \epsilon - i y \epsilon}{\left(x^2 + \epsilon^2\right)\left(y^2 + \epsilon^2\right)} = \mathrm{P} \frac{1}{x} \mathrm{P}\frac{1}{y} + \pi^2 \delta(x) \delta(y) + i \pi \left[ \mathrm{P}\frac{1}{x} \delta(y)-\mathrm{P}\frac{1}{y} \delta(x)\right]\;.
		\end{align}
		The next step is to perform the integration over the distributions. We start with 
		\begin{align}
			\mathrm{P} \int_{-1}^{1} dx\frac{1}{x}   \mathrm{P} \int_{y_{\textrm{min}}(x,z)}^{y_{\textrm{max}}(x,z)} dy \frac{1}{\sqrt{1-x^2-y^2-z^2 + 2 x y z}} \frac{1}{y} \;.
		\end{align}
		Recall that $x,y$ and $z$ are actual angles and thus limited to values between $-1$ and $1$. Further, due to the simplification of the phase space, we have the boundary of $1-x^2-y^2-z^2 + 2 x y z >0$. From this, it follows for the integration area of $y$:
		\begin{align}
			y_{min}(x,z) =	x z - \sqrt{1-x^2-z^2+ x^2 z^2}, \qquad y_{max}(x,z) = x z + \sqrt{1-x^2-z^2+ x^2 z^2}\;.
		\end{align}
		With this at hand, we find:
		\begin{align}
			\mathrm{P} \int_{y_{\textrm{min}}(x,z)}^{y_{\textrm{max}}(x,z)} dy \frac{1}{\sqrt{1-x^2-y^2-z^2 + 2 x y z}} \frac{1}{y} = 0, \quad \textrm{for}\; -1<z<1\; \textrm{and}\; -\sqrt{1-z^2}\le x\le \sqrt{1-z^2}\;.
		\end{align}
		Next in line, we take a closer look at the regions for $-1<x\le \sqrt{1-z^2}$ and $\sqrt{1-z^2}\le x <1$. We begin with the lower region:
		\begin{align}
			\mathrm{P} \int_{y_{\textrm{min}}(x,z)}^{y_{\textrm{max}}(x,z)} dy \frac{1}{\sqrt{1-x^2-y^2-z^2 + 2 x y z}} \frac{1}{y} = - \frac{\pi |z|}{z \sqrt{-1+x^2+z^2}}, \quad \textrm{for}\; -1<z<1\; \textrm{and}\; -1< x\le -\sqrt{1-z^2}\;.
		\end{align}
		The same is done for the upper one:
		\begin{align}
			\mathrm{P} \int_{y_{\textrm{min}}(x,z)}^{y_{\textrm{max}}(x,z)} dy \frac{1}{\sqrt{1-x^2-y^2-z^2 + 2 x y z}} \frac{1}{y} =  \frac{\pi |z|}{z \sqrt{-1+x^2+z^2}}, \quad \textrm{for}\; -1<z<1\; \textrm{and}\; \sqrt{1-z^2} \le  x<1\;.
		\end{align}
		Next, we consider the principal value integral with respect to $x$. From the $y-$integration, we know that we only receive a non-zero value for $-1<x \le -\sqrt{1-z^2}$ and $\sqrt{1-z^2}\le x <1$. Thus, we simply integrate in those regions, that is:
		\begin{align}
			\mathrm{P} \int_{-1}^{-\sqrt{1-z^2}} dx \frac{1}{x} \frac{-\pi | z| }{z \sqrt{-1+x^2+z^2}} +	\mathrm{P} \int_{\sqrt{1-z^2}}^{1} dx \frac{1}{x} \frac{\pi | z| }{z \sqrt{-1+x^2+z^2}}  = \frac{2 \pi  \arcsin(z)}{\sqrt{1-z^2}}, \quad \textrm{for} -1<z<1\;.
		\end{align}
		Combining the results, we find for the principal value part:
		\begin{align}
			\int_{-1}^{1} dx \int_{y_{\textrm{min}}(x,z)}^{y_{\textrm{max}}(x,z)} dy \frac{1}{\sqrt{1-x^2-y^2-z^2 + 2 x y z}}  \mathrm{P} \frac{1}{x} \mathrm{P} \frac{1}{y} = \frac{2\pi \arcsin\left(z\right)}{\sqrt{1-z^2}}\;.
		\end{align}
		We continue with the two delta-distributions. Since the integration is straightforward, we directly give the result:
		\begin{align}
			\pi^2 \int_{-1}^{1} dx \int_{y_{min}}^{y_{max}} dy \frac{1}{\sqrt{1-x^2-y^2-z^2 + 2 x y z}} \delta(x) \delta(y) = \frac{\pi^2}{\sqrt{1-z^2}}\;.
		\end{align}
		Next in line are the mixed terms of the two distributions. Here, both terms yield zero, thus we do not give the derivation. After all the integrations, only two terms survive. They can be simplified to
		\begin{align}
			\frac{2 \pi  \arcsin(z)}{\sqrt{1-z^2}}+	\frac{\pi ^2}{\sqrt{1-z^2}} = \frac{\pi  \left(\pi + 2 \arcsin (z) \right)}{\sqrt{1-z^2}} = \frac{2 \pi \arccos(-z)}{\sqrt{1-z^2}}\;.
		\end{align}
		For completeness, we give the full result:
		\begin{align}
			\int_{-1}^{1} dx \int_{y_{\textrm{min}}(x,z)}^{y_{\textrm{max}}(x,z)} dy \frac{1}{\sqrt{1-x^2-y^2-z^2+2 x y z}}\frac{-1}{\left(i x - \epsilon\right)\left(i y + \epsilon\right)} = \frac{2 \pi \arccos(-z)}{\sqrt{1-z^2}}\;.
		\end{align}
		Next, we derive another important result. Consider the following expression:
		\begin{align*}
			-	\int_{-1}^{1} dx \int_{y_{\textrm{min}}(x,z)}^{y_{\textrm{max}}(x,z)} dy  \frac{1}{\sqrt{1-x^2-y^2-z^2+2 x y z}}\frac{x^2}{\left(i x - \epsilon\right)\left(i y + \epsilon\right)} \;.
		\end{align*}
		We perform the same steps as above and work with the distributions. Since we already know the result for the $y$ integration, we find:
		\begin{align}
			\int_{-1}^{-\sqrt{1-z^2}} dx \; x\frac{-\pi | z| }{z \sqrt{-1+x^2+z^2}}  + \int_{\sqrt{1-z^2}}^{1} dx \; x \frac{\pi | z| }{z \sqrt{-1+x^2+z^2}}  = 2 \pi z, \quad \textrm{for} -1<z<1\;.
		\end{align}
		The remaining integrals do not give a contribution, therefore we can write
		\begin{align}
			- 	\int_{-1}^{1} dx \int_{y_{\textrm{min}}(x,z)}^{y_{\textrm{max}}(x,z)} dy  \frac{1}{\sqrt{1-x^2-y^2-z^2+2 x y z}}\frac{x^2}{\left(i x - \epsilon\right)\left(i y + \epsilon\right)}  = 2 \pi z\;. \label{eq_2piZ_PV}
		\end{align}

		\section{Important integrals for class-VIII contributions}
		\label{subsection_LadderDiagram_IntegralTable}
		
		In this appendix, we collect the expressions for some integrals needed for calculating the class-VIII contributions. The integrals relevant for the crossed-box diagram, i.e., for the
		diagram (2) of class VIII in Fig.~\ref{fig:3pi_N3LO}, read
		\begin{align}
			\int_{-1}^{1} dx   \int_{y_{\text{min}}(x,z)}^{y_{\text{max}}(x,z)} dy  \frac{1}{\sqrt{1-x^2-y^2-z^2+2 x y z}} \frac{x^2}{\left(x + i \epsilon\right)^{2}\left(i y + \epsilon\right)^{2}}&= 2 \pi\,,  \nonumber \\
			\int_{-1}^{1} dx   \int_{y_{\text{min}}(x,z)}^{y_{\text{max}}(x,z)} dy   \frac{1}{\sqrt{1-x^2-y^2-z^2+2 x y z}} \frac{x^4}{\left(x + i \epsilon\right)^{2}\left(i y + \epsilon\right)^{2}}&= 2 \pi \left(1- 2z^2\right)\, , \nonumber \\
			\int_{-1}^{1} dx   \int_{y_{\text{min}}(x,z)}^{y_{\text{max}}(x,z)} dy   \frac{1}{\sqrt{1-x^2-y^2-z^2+2 x y z}}\frac{x y}{\left(x + i \epsilon\right)^{2}\left(i y + \epsilon\right)^{2}}&= -2 \pi \frac{  \arccos(-z)}{\sqrt{1-z^2}}\,, \nonumber \\
			\int_{-1}^{1} dx   \int_{y_{\text{min}}(x,z)}^{y_{\text{max}}(x,z)} dy \frac{1}{\sqrt{1-x^2-y^2-z^2+2 x y z}} \frac{1}{\left(x + i \epsilon\right)^{2}\left(i y + \epsilon\right)^{2}}&= \frac{2 \pi \left(\sqrt{1-z^2}+ z \arccos(-z)\right)}{\left(1-z^2\right)^{3/2}}\, .
			\label{eq_CrossedBoxDiagram_AngularInte_Solutions}
		\end{align} 
		For the reducible-like class-VIII diagram,  i.e., for the diagram (1) in Fig.~\ref{fig:3pi_N3LO}, we give the following integrals: 
		\begin{align}
			\int_{-1}^{1} dx   \int_{y_{\text{min}}(x,z)}^{y_{\text{max}}(x,z)} dy \frac{x^2}{\sqrt{1-x^2-y^2-z^2+2 x y z}} \frac{1}{\left(x + i \epsilon\right)^{2}\left(-i y + \epsilon\right)^{2}}&= 2 \pi \,,\nonumber \\
			\int_{-1}^{1} dx   \int_{y_{\text{min}}(x,z)}^{y_{\text{max}}(x,z)} dy   \frac{x^4}{\sqrt{1-x^2-y^2-z^2+2 x y z}} \frac{1}{\left(x + i \epsilon\right)^{2}\left(-i y + \epsilon\right)^{2}}&= 2 \pi \left(1-2z^2\right) \,,\nonumber\\ 
			\int_{-1}^{1} dx   \int_{y_{\text{min}}(x,z)}^{y_{\text{max}}(x,z)} dy \frac{x y}{\sqrt{1-x^2-y^2-z^2+2 x y z}}\frac{1}{\left(x + i \epsilon\right)^{2}\left(-i y + \epsilon\right)^{2}}&= \frac{2 \pi  \arccos (z)}{\sqrt{1-z^2}}=2 \pi \frac{ \left(\pi- \arccos(-z)\right)}{\sqrt{1-z^2}}\,, \nonumber \\
			\int_{-1}^{1} dx   \int_{y_{\text{min}}(x,z)}^{y_{\text{max}}(x,z)} dy \frac{1}{\sqrt{1-x^2-y^2-z^2+2 x y z}}  \frac{1}{\left(x + i \epsilon\right)^{2}\left(-i y + \epsilon\right)^{2}}&= \frac{2 \pi \left(\sqrt{1-z^2}- \pi z+ z \arccos(-z)\right)}{\left(1-z^2\right)^{3/2}}\,.\label{eq_IntegralSolutionsLadderDiagram}
		\end{align}
		
		\section{Angular integration for the isoscalar central potential of class IX}\label{subsec_SolvingImVcClassIX}
		
		The aim of this appendix is to show an elegant way of solving the angular integrations generated by diagrams of class IX. Our starting point is the following expression:
		\begin{align}
			\textrm{Im} \; V_{C}^{\rm IX, \, SMM}(i \mu) =&  \frac{3 g_A^6 \mu^2}{2 \pi \left(8 \pi F_\pi^2\right)^3}\iint\limits_{z^2<1} d \omega_{1} d \omega_{2} \int_{-1}^{1}dx  \int_{y_{\text{min}}(x,z)}^{y_{\text{max}}(x,z)} dy\frac{1}{\sqrt{1-x^2-y^2-z^2 + 2 x y z}}\nonumber \\
			&\times\frac{y}{\left(y^2+\epsilon ^2\right)} \frac{l_2\left(1-x^2-y^2-z^2 + 2 x y z\right)}{(x+i \epsilon )^2 (l_1 x+l_2 y+i \epsilon )} \,. \label{eq_VcImCIX_Appendix}
		\end{align}
		First, we make use of 
		\begin{align}
			\frac{1}{y \pm i \epsilon} = P \frac{1}{y} \mp i \pi \delta(y) \qquad \Rightarrow \qquad P \frac{1}{y}  = \frac{1}{y \pm i \epsilon}  \pm i \pi \delta(y)\;,
		\end{align}
		in order to replace the principal value distribution in Eq.~(\ref{eq_VcImCIX_Appendix}) by the standard propagator and a $\delta$-distribution. Here, we choose for the propagator $+i \epsilon$ which will become clear in the next step. We have:
		\begin{align}
			\textrm{Im} \; V_{C}^{\rm IX, \, SMM}(i \mu) =&  \frac{3 g_A^6 \mu^2}{2 \pi \left(8 \pi F_\pi^2\right)^3}\iint\limits_{z^2<1} d \omega_{1} d \omega_{2} \int_{-1}^{1}dx  \int_{y_{\text{min}}(x,z)}^{y_{\text{max}}(x,z)} dy\frac{1}{\sqrt{1-x^2-y^2-z^2 + 2 x y z}}\nonumber \\
			&\times\left(\frac{1}{\left(y+ i \epsilon\right)}+ i \pi \delta(y)\right) \frac{l_2\left(1-x^2-y^2-z^2 + 2 x y z\right)}{(x+i \epsilon )^2 (l_1 x+l_2 y+i \epsilon )} \,. 
		\end{align}
		Since the part proportional to the $\delta$-distribution can be trivially solved, we focus on the remaining part, indicated now by $\textrm{Im} \; \hat{V}_{C}^{\rm IX, \, SMM}(i \mu)$. Here, we make use of the permutational symmetry of the three-pion phase space. We begin with splitting the expression into two parts and interchange for one part $x$ $(l_1)$ with $y$ $(l_2)$ and vice versa.
		Afterwards, we add both parts together.  The advantage of this procedure is that we have now reduced the problem to the already solved one: 
		\begin{align}
			\textrm{Im} \; \hat{V}_{C}^{\rm IX, \, SMM}(i \mu) =& \frac{1}{2}\textrm{Im} \; \hat{V}_{C}^{\rm IX, \, SMM}(i \mu)+ \frac{1}{2}\textrm{Im} \; \hat{V}_{C}^{\rm IX, \, SMM}(i \mu)\nonumber \\
			=&  \frac{3 g_A^6 \mu^2}{4 \pi \left(8 \pi F_\pi^2\right)^3}\iint\limits_{z^2<1} d \omega_{1} d \omega_{2} \int_{-1}^{1}dx  \int_{y_{\text{min}}(x,z)}^{y_{\text{max}}(x,z)} dy\frac{1}{\sqrt{1-x^2-y^2-z^2 + 2 x y z}}\nonumber \\
			&\times\frac{1}{\left(y+ i \epsilon\right)} \frac{l_2\left(1-x^2-y^2-z^2 + 2 x y z\right)}{(x+i \epsilon )^2 (l_1 x+l_2 y+i \epsilon )}\nonumber\\
			&+\frac{3 g_A^6 \mu^2}{4 \pi \left(8 \pi F_\pi^2\right)^3}\iint\limits_{z^2<1} d \omega_{1} d \omega_{2} \int_{-1}^{1}dx  \int_{y_{\text{min}}(x,z)}^{y_{\text{max}}(x,z)} dy\frac{1}{\sqrt{1-x^2-y^2-z^2 + 2 x y z}}\nonumber\\
			&\times\frac{1}{\left(x+ i \epsilon\right)} \frac{l_1\left(1-x^2-y^2-z^2 + 2 x y z\right)}{(y+i \epsilon )^2 (l_1 x+l_2 y+i \epsilon )}\nonumber \\
			=&-\frac{3 g_A^6 \mu^2}{4 \pi \left(8 \pi F_\pi^2\right)^3}\iint\limits_{z^2<1} d \omega_{1} d \omega_{2} \int_{-1}^{1}dx  \int_{y_{\text{min}}(x,z)}^{y_{\text{max}}(x,z)} dy\frac{1}{\sqrt{1-x^2-y^2-z^2 + 2 x y z}}\nonumber \\
			&\times\frac{1}{\left(x+ i \epsilon\right)^2} \frac{\left(1-x^2-y^2-z^2 + 2 x y z\right)}{(-i y+ \epsilon )^2 (l_1 x+l_2 y+i \epsilon )} \left(l_1 x + l_2 y + i \left(l_1+l_2\right)\epsilon\right)\nonumber \\
			=&-\frac{3 g_A^6 \mu^2}{4 \pi \left(8 \pi F_\pi^2\right)^3}\iint\limits_{z^2<1} d \omega_{1} d \omega_{2} \int_{-1}^{1}dx  \int_{y_{\text{min}}(x,z)}^{y_{\text{max}}(x,z)} dy\frac{1}{\sqrt{1-x^2-y^2-z^2 + 2 x y z}}\nonumber \\
			&\times\frac{\left(1-x^2-y^2-z^2 + 2 x y z\right)}{\left(x+ i \epsilon\right)^2\left(-i y+ \epsilon\right )^2}\,.
		\end{align}
		Using the integrals listed in appendix \ref{subsection_LadderDiagram_IntegralTable} for $\textrm{Im} \; \hat{V}_{C}^{\rm IX, \, SMM}(i \mu)$ and adding the $\delta$-part, we finally end up with:
		\begin{align}
			\textrm{Im} \; V_{C}^{\rm IX, \, SMM}(i \mu) =& \frac{3 g_A^6 \mu^2}{2\left(8 \pi F_\pi^2 \right)^3}\iint \limits_{z^2<1} d \omega_{1} d \omega_{2} \bigg\{1 - \frac{1}{\sqrt{1-z^2}}\left[\pi \left(\frac{l_2}{2l_1}+z\right)-z \arccos(-z)\right]\bigg\}\;.
		\end{align}

\end{document}